\documentclass[preprintnumbers,article,amsmath,amssymb,floatfix,10pt,prd,onecolumn,superscriptaddress,nofootinbib]{revtex4-2}

\usepackage{cancel}
\usepackage{adjustbox}
\usepackage{titlesec}
\usepackage[toc]{appendix}
\usepackage[font=small,labelfont=bf,justification=justified,singlelinecheck=false]{caption}
\titleformat*{\section}{\LARGE\bfseries}
\titleformat*{\subsection}{\Large\bfseries}
\titleformat*{\subsubsection}{\large\bfseries}
\titleformat*{\paragraph}{\large\bfseries}
\titleformat*{\subparagraph}{\large\bfseries}
\usepackage{bm}
\usepackage{amsfonts}
\usepackage{latexsym}
\usepackage[utf8]{inputenc}
\usepackage{graphicx}
\usepackage{amsmath}
\usepackage{palatino}
\usepackage{mathpazo}
\usepackage{textcomp}
\usepackage{float}
\usepackage{booktabs}
\usepackage{dcolumn}
\usepackage{ragged2e}
\usepackage{hyperref}
\hypersetup{colorlinks,citecolor=blue}
\hypersetup{colorlinks=true,linkcolor=blue,filecolor=magenta,urlcolor=blue}
\usepackage{amsthm}
\usepackage{xcolor}
\usepackage{orcidlink}
\usepackage{epsfig}
\usepackage{commath}
\usepackage{cancel}
\usepackage{subcaption}
\usepackage{caption}
\usepackage{multirow}
\newcommand{\be}{\begin{equation}}
\newcommand{\ee}{\end{equation}}
\newcommand{\bea}{\begin{eqnarray}}
\newcommand{\eea}{\end{eqnarray}}
\newcommand{\eeas}{\end{eqnarray*}}
\newcommand{\beas}{\begin{eqnarray*}}

\def\jnl@style{\it}
\def\aaref@jnl#1{{\jnl@style#1}}

\def\aaref@jnl#1{{\jnl@style#1}}

\def\aj{\aaref@jnl{AJ}}                   
\def\apj{\aaref@jnl{ApJ}}                 
\def\apjl{\aaref@jnl{ApJ}}                
\def\apjs{\aaref@jnl{ApJS}}               
\def\apss{\aaref@jnl{Ap\&SS}}             
\def\aap{\aaref@jnl{A\&A}}                
\def\aapr{\aaref@jnl{A\&A~Rev.}}          
\def\aaps{\aaref@jnl{A\&AS}}              
\def\mnras{\aaref@jnl{Mon.~Not.~Roy.~Astron.~Soc.}}             
\def\prd{\aaref@jnl{Phys.~Rev.~D}}        
\def\prc{\aaref@jnl{Phys.~Rev.~C}}  
\def\prl{\aaref@jnl{Phys.~Rev.~Lett.}}    
\def\qjras{\aaref@jnl{QJRAS}}             
\def\skytel{\aaref@jnl{S\&T}}             
\def\ssr{\aaref@jnl{Space~Sci.~Rev.}}     
\def\zap{\aaref@jnl{ZAp}}                 
\def\nat{\aaref@jnl{Nature}}              
\def\aplett{\aaref@jnl{Astrophys.~Lett.}} 
\def\apspr{\aaref@jnl{Astrophys.~Space~Phys.~Res.}} 
\def\physrep{\aaref@jnl{Phys.~Rep.}}      
\def\physscr{\aaref@jnl{Phys.~Scr}}       
\def\commat{\aaref@jnl{Comm.~Math.~Phys.}}              
\def\science{\aaref@jnl{Science}}               
\def\cqg{\aaref@jnl{Classical Quant.~Grav.}}            
\def\jpcs{\aaref@jnl{JPCS}}                                     
\def\ijmpd{\aaref@jnl{Int.~J.~Mod.~Phys.~D}}                    
\def\grg{\aaref@jnl{Gen.~Relat.~Gravit.}}               
\def\rpp{\aaref@jnl{Rep.~Prog.~Phys.}}          
\def\npa{\aaref@jnl{Nucl.~Phys.~A}}        
\def\lrr{\aaref@jnl{Living Rev.~Rel.}}                   
\def\jcap{\aaref@jnl{J.~Cosmology Astropart.~Phys.}}    
\def\rmp{\aaref@jnl{Rev.~Mod.~Phys.}}   
\def\epjc{\aaref@jnl{Eur.~Phys.~J.~C}} 
\def\plb{\aaref@jnl{~Phy.~Lett.~B}} 
\def\mpla{\aaref@jnl{Mod.~Phy.~Lett.~A}} 
\def\arxiv{\aaref@jnl{arxiv.org}}

\allowdisplaybreaks[1]

\usepackage[T1]{fontenc}
\usepackage[utf8]{inputenc}
\usepackage[english]{babel}
\usepackage{newtxtext,newtxmath} 
\usepackage{microtype} 
\usepackage{graphicx}
\usepackage{xcolor}
\definecolor{prdblue}{RGB}{0,0,128} 
\usepackage{mathtools}
\usepackage{bm}

\NewDocumentCommand{\setargsaux}{mm}
{\IfNoValueTF{#2}{#1} {#1\,\delimsize|\,\mathopen{}#2}}

\newtheorem{theorem}{Theorem}[section]

\setcitestyle{numbers,square,comma} 
\usepackage{hyperref}
\hypersetup{
    colorlinks=true,
    linkcolor=prdblue,
    citecolor=prdblue,
    urlcolor=prdblue,
    breaklinks=true
}
\usepackage{quoting,xparse}
\NewDocumentCommand{\bywhom}{m}{%
  {\nobreak\hfill\penalty50\hskip1em\null\nobreak
   \hfill\mbox{\normalfont(#1)}%
   \parfillskip=0pt \finalhyphendemerits=0 \par}%
}

\NewDocumentEnvironment{pquotation}{m}
  {\begin{quoting}[
     indentfirst=true,
     leftmargin=\parindent,
     rightmargin=\parindent]\itshape
  \bywhom{#1}\end{quoting}}

\usepackage{etoolbox}
\patchcmd{\abstract}{\quotation}{\quotation\small\addtolength{\leftmargin}{2.5em}\addtolength{\rightmargin}{2.5em}}{}{}

\usepackage{tikz}
\usetikzlibrary{svg.path}
\definecolor{orcidlogocol}{HTML}{A6CE39}
\tikzset{
  orcidlogo/.style={
    yscale=-1,inner sep=0pt,
    node contents={%
      \begin{tikzpicture}[yshift=-.12ex, scale=0.06]
        \fill[orcidlogocol] svg{M256,128c0,70.7-57.3,128-128,128C57.3,256,0,198.7,0,128C0,57.3,57.3,0,128,0C198.7,0,256,57.3,256,128z};
        \fill[white] svg{M86.3,186.2H70.9V79.1h15.4v48.4V186.2z}
        svg{M108.9,79.1h21.6c24,0,32.8,19.2,32.8,31c0,22.1-18.1,31-32.8,31h-21.6V79.1z M124.3,127.4h6.8c13.9,0,17.9-10.7,17.9-17.5c0-10.5-8-17.2-17.9-17.2h-6.8V127.4z}
        svg{M108.9,186.2V141h15.4v45.2H108.9z};
      \end{tikzpicture}
    }
  }
}
\newcommand\orcid[1]{\href{https://orcid.org/#1}{\mbox{\tikz\node[orcidlogo];}}}

\usepackage{accents}

\newcommand{\utilde}[1]{\underaccent{\tilde}{#1}}

\usepackage{booktabs}
\usepackage{tabularx}
\usepackage{multirow}
\usepackage{array}

\begin{document}
\title{Hamiltonian Analysis of the Newer General Relativity}

\author{Chen-Yi Sia~\orcidlink{0009-0009-5167-202X}}
\email{siachenyi@um.edu.my}
\affiliation{Institute of Mathematical Sciences, Faculty of Science, Universiti Malaya, 50603 Kuala Lumpur, Malaysia}

\author{Tee-How Loo~\orcidlink{0000-0003-4099-9843}}
\email{looth@um.edu.my}
\affiliation{Institute of Mathematical Sciences, Faculty of Science, Universiti Malaya, 50603 Kuala Lumpur, Malaysia}

\author{Avik De~\orcidlink{0000-0001-6475-3085}}
\email{avikde@um.edu.my}
\affiliation{Institute of Mathematical Sciences, Faculty of Science, Universiti Malaya, 50603 Kuala Lumpur, Malaysia}


\begin{abstract}
We present a systematic Hamiltonian analysis of the five-parameter class of theories known as Newer General Relativity. Quadratic nonmetricity theories of this type can be divided into nine types according to the structure of their primary constraints. After reviewing the framework of symmetric teleparallel gravity, we derive the ADM formulation of Newer General Relativity, discuss the existing classification and perform the Hamiltonian formulation of the theory. For each type, we compute the conjugate momenta, identify the primary constraints, and construct both the canonical and primary Hamiltonians. We then determine the constraint structure and analyze the resulting physical degrees of freedom in each type.
\end{abstract}

\footnotetext{The research was supported by the Ministry of Higher Education (MoHE), through the Fundamental Research Grant Scheme (FRGS/1/2023/STG07/UM/02/3, project no.: FP074-2023)}

\maketitle
\tableofcontents
\newpage
\section{Introduction}
While general relativity (GR) formulated in terms of spacetime curvature serves as the standard paradigm for classical gravitation and cosmology \cite{efe_physical_implic, Carroll_2023}, conceptual challenges and cosmological tensions continue to drive the exploration of modified and extended gravitational theories \cite{Will2014, HEISENBERG20191, Stelle:1977ry, CLIFTON20121, Saridakis2021ModifiedGravity}. A particularly compelling perspective is provided by the ``geometric trinity of gravity'' \cite{BeltranJimenez:2019esp}, wherein GR admits dynamically equivalent representations within more general metric-affine geometries by relaxing the Levi-Civita condition on the connection \cite{HEHL19951}. In addition to standard curvature-based GR, gravitation can be equivalently described through torsion in the Teleparallel Equivalent of General Relativity (TEGR) or through nonmetricity in the Symmetric Teleparallel Equivalent of General Relativity (STEGR) \cite{Capozziello2022, IOSIFIDIS2024138498, three_plus_one, three_plus_one_tele_sym, TomonariMobius, PhysRevD.110.066002}.\\

Modifications beyond these equivalent limits have garnered substantial interest, including metric teleparallel gravity and its extensions \cite{tele_equiv_gr, Bahamonde_2023, Aldrovandi:2013wha, tele_theories_of_gravity, Jimenez_2018, PhysRevD.67.044016, Ong:2017xwo, energy_angularMomentum, M_Blagojevic_2000, Golovnev:2018wbh, Heisenberg_2024}, symmetric teleparallel theories \cite{Nester:1998mp, BeltranJimenez:2022azb, PhysRevD.109.044061, Tomonari2025STEGR, Adak:2008gd, Pala2022}, and non-linear extensions such as $f(R)$, $f(T)$, and $f(Q)$ gravity \cite{PhysRevD.95.104034, Joshi:2022dpz, deg_of_fT, Cai_2016, PhysRevD.88.024019, PhysRevD.97.104028, PhysRevD.102.064025, PhysRevD.109.064034, Bahamonde:2022ohm, adm_fQ_KunHu, dirac_dof_fQ, failure_fQ, cov_fQ, PhysRevD.101.103507, PhysRevD.108.124030, PhysRevLett.132.141401, HEISENBERG20241, De_2023}. Within the metric-affine framework, generalized quadratic Lagrangians provide natural avenues for extending these baseline theories. In the torsional sector, New General Relativity (NGR) constitutes a three-parameter extension of TEGR whose canonical structure has recently been characterized \cite{newGR_first, newGR_second, Tomonari:2025kgb}. In the symmetric teleparallel sector, the analog is Newer General Relativity (Newer GR), a five-parameter quadratic extension of STEGR typically formulated in the coincident gauge \cite{DAmbrosio:2020nqu, classifyNewerGR, PhysRevD.98.044048, Ferrara2026NGR}.\\

A rigorous Hamiltonian constraint analysis is paramount for establishing the physical viability of any modified gravity theory. Because generally covariant field theories are governed by singular Lagrangians, canonical formulations systematically give rise to constraint structures \cite{nonlinear_field, nonlinear_field2, hamiltonianGR, constraints, generalizedHamil, hamiltonian_field_dynamics, Dirac:1958sc, dirac_QM, PhysRev.117.1595, PhysRev.116.1322}. The Dirac--Bergmann algorithm \cite{singular_lagrangians, Saito:1989bt, Castellani:1981us, Shanmugadhasan1973, SheikhJabbari2001, PhysRevD.41.1247, ErrastiDiez:2023gme, Kaparulin:2012px, ErrastiDiez:2020dux, Sugano:1991ir, Sugano1983, TomonariBC, TomonariSymp, SWHawking_1996, Baez:1995sj, Nakahara2003, generalization_adm, cadabra, cadabra_field, apply_cadabra, xAct} provides the necessary machinery to isolate primary and secondary constraints, evaluate the constraint algebra, determine the gauge symmetries, and count the propagating physical degrees of freedom (pDoF). Such canonical analyses have previously revealed ghost instabilities, strong-coupling pathologies, or unexpected extra degrees of freedom across various teleparallel settings \cite{pri_constraints_general_tele_quadratic, gr_null_surface, hamiltonian_tele_description, PhysRevD.62.024021, PhysRevD.64.084014, PhysRevD.94.104045, Blixt:2020ekl}.\\

For Newer GR, the parameter space has been categorized into distinct classes based on the invertibility of the kinetic Hessian, and the associated primary constraints have been identified \cite{DAmbrosio:2020nqu, classifyNewerGR, PhysRevD.98.044048, Ferrara2026NGR}. However, a complete Dirac--Bergmann analysis remains absent. Crucially, the consistency conditions generated by these primary constraints, the emergence of secondary constraints, the closure of the full constraint algebra, and the definitive count of physical degrees of freedom across all parameter branches have not been systematically resolved.\\

In this work, we perform a systematic Hamiltonian constraint analysis for each class of Newer GR using the Dirac--Bergmann algorithm. By analyzing the consistency conditions of the primary constraints, we solve for the Lagrange multipliers where possible, uncover the complete chain of secondary constraints, and classify the full constraint set into first- and second-class constraints. This enables us to determine unambiguously the physical degrees of freedom propagated by each class of Newer GR.\\

The paper is structured as follows. In Sec.~\ref{sec:stegr}, we review symmetric teleparallel gravity, STEGR, and the five-parameter action of Newer GR. In Sec.~\ref{sec:dirac_bergmann}, we outline the Dirac--Bergmann algorithm for singular field theories. Sec.~\ref{sec:adm} details the $3+1$ ADM decomposition of the nonmetricity tensor and summarizes the classification of Newer GR based on the Hessian matrix. Sec.~\ref{sec:hamiltonian_analysis} presents the core canonical analysis, deriving the conjugate momenta, constraint chains, and physical degrees of freedom for each class. Finally, concluding remarks and prospects for future work are given in Sec.~\ref{sec:conclusion}.
\newpage

\section{Symmetric Teleparallel Gravity}\label{sec:stegr}
Throughout this section, we discuss the foundations of symmetric teleparallel gravity and establish the notation and conventions adopted in this paper. We first consider a four-dimensional spacetime endowed with two independent dynamical variables:
the metric tensor $g_{\mu \nu}$ and an affine connection $\Gamma^{\alpha}{}_{\mu \nu}$.  Within the geometrical trinity of gravity, General Relativity can be equivalently reformulated
in terms of three distinct geometric structures, characterized by curvature, torsion, and non-metricity, respectively \cite{BeltranJimenez:2019esp}. These are defined as
\begin{align}
	R^{\lambda}{}_{\mu \alpha \nu} &= \partial_{\alpha}\Gamma^{\lambda}{}_{\mu \nu}- \partial_{\nu}\Gamma^{\lambda}{}_{\mu \alpha}+ \Gamma^{\lambda}{}_{\sigma \alpha}\Gamma^{\sigma}{}_{\mu \nu}-\Gamma^{\lambda}{}_{\sigma \nu}\Gamma^{\sigma}{}_{\mu \alpha},\\[4pt]
	T^{\lambda}{}_{\mu \nu} &= \Gamma^{\lambda}{}_{\nu \mu} - \Gamma^{\lambda}{}_{\mu \nu},\\[4pt]
	Q_{\alpha \mu \nu}&= \nabla_{\alpha} g_{\mu \nu}= \partial_{\alpha} g_{\mu \nu}- \Gamma^{\sigma}{}_{\mu\alpha} g_{\sigma \nu}- \Gamma^{\sigma}{}_{\nu \alpha} g_{\mu \sigma}.
\end{align}
Symmetric teleparallel geometry is defined by imposing the vanishing of curvature and torsion,
\begin{equation}
R^{\lambda}{}_{\mu \alpha \nu} = 0,
\qquad 
T^{\lambda}{}_{\mu \nu} = 0,
\end{equation}
while allowing a non-trivial non-metricity tensor $Q_{\alpha \mu \nu} \neq 0$. The vanishing curvature condition implies that the affine connection is locally flat and therefore can be expressed locally as a pure gauge configuration,
\begin{equation}
\Gamma^{\lambda}{}_{\mu \nu}= (\Lambda^{-1})^{\lambda}{}_{\rho} \, \partial_{\nu} \Lambda^{\rho}{}_{\mu},
\end{equation}
for some frame field $\Lambda^{\rho}{}_{\nu}$. The torsion-free condition further restricts $\Lambda^{\rho}{}_{\nu}$ to be integrable, such that it becomes a coordinate frame field, and so it can be written in terms of scalar fields $\xi^\rho$ as
\begin{equation}
\Lambda^{\rho}{}_{\nu} = \partial_{\nu} \xi^{\rho}.
\end{equation}
The chosen gauge is known as a \emph{coincident gauge}, with which the connection takes the form
\begin{equation}
\Gamma^{\lambda}{}_{\mu \nu}= \frac{\partial x^\lambda}{\partial \xi^\rho}\, \partial_{\mu}\partial_{\nu} \xi^{\rho}, 
\end{equation}
and 
the non-metricity tensor reduces to a purely metric expression
\begin{equation}
Q_{\alpha \mu \nu}= \partial_{\alpha} g_{\mu \nu}.
\end{equation}
More generally, in an arbitrary coordinate system it can be expressed as
\begin{equation}
Q_{\alpha \mu \nu}= \partial_{\alpha} g_{\mu \nu}- \frac{\partial x^\sigma}{\partial \xi^\rho}\left(\partial_{\mu}\partial_{\alpha}\xi^\rho \, g_{\sigma \nu}+\partial_{\nu}\partial_{\alpha}\xi^\rho \, g_{\mu \sigma}\right).
\end{equation}
The affine connection can be decomposed into the Levi-Civita connection and the disformation tensor according to
\begin{equation}
\Gamma^{\lambda}{}_{\mu\nu}=\mathring{\Gamma}^{\lambda}{}_{\mu\nu}+L^{\lambda}{}_{\mu\nu},
\end{equation}
where
\begin{equation}
\mathring{\Gamma}^{\lambda}{}_{\mu\nu}=\frac{1}{2} g^{\lambda\rho} \left( \partial_\mu g_{\rho\nu}+\partial_\nu g_{\rho\mu}-\partial_\rho g_{\mu\nu} \right)
\end{equation}
is the Levi-Civita connection associated with the metric, while
\begin{equation}
L^{\lambda}{}_{\mu\nu}=\frac{1}{2} g^{\lambda\rho} \left( Q_{\rho\mu\nu} -Q_{\mu\rho\nu}-Q_{\nu\rho\mu} \right)
\end{equation}
is the disformation tensor \cite{three_plus_one}.\\

The general non-metricity scalar that is quadratic in the non-metricity tensor is given by a linear combination of five independent quadratic invariants \cite{DAmbrosio:2020nqu},
\begin{equation}
\mathbb{Q}=c_1 Q_{\alpha\mu\nu}Q^{\alpha\mu\nu}+c_2 Q_{\alpha\mu\nu}Q^{\mu\nu\alpha}+c_3 Q_\mu Q^\mu+c_4 \bar{Q}_\mu \bar{Q}^\mu+c_5 Q_\mu \bar{Q}^\mu ,
\end{equation}
where
\begin{equation}
Q_\mu \equiv Q_{\mu \alpha}{}^{\alpha},
 \qquad
\bar{Q}_\mu \equiv Q^{\alpha}{}_{\alpha\mu}
\end{equation}
denote the two independent traces of the non-metricity tensor.  The action of Newer General Relativity is then defined as
\begin{equation}
S[g,\Gamma]=\int d^4x \sqrt{-g}\mathbb{Q}.
\end{equation}
Substituting the explicit expression of the non-metricity scalar, the action can be rewritten in the form
\begin{equation}
S[g,\Gamma]=\int d^4x \sqrt{-g}P^{\alpha\beta\sigma\rho\mu\nu}Q_{\alpha\beta\sigma} Q_{\rho\mu\nu},
\end{equation}
where the tensor $P^{\alpha\beta\sigma\rho\mu\nu}$ is defined as
\begin{equation}
P^{\alpha\beta\sigma\rho\mu\nu}
=c_1 g^{\alpha\rho} g^{\beta\mu} g^{\sigma\nu}+c_2 g^{\alpha\nu} g^{\beta\mu} g^{\sigma\rho}+c_3 g^{\alpha\rho} g^{\beta\sigma} g^{\mu\nu}+c_4 g^{\alpha\beta} g^{\mu\rho} g^{\nu\sigma}+c_5 g^{\alpha\beta}g^{\mu\nu} g^{\sigma\rho}.
\end{equation}
A particular choice of the coefficients,
\begin{equation}
c_1=-\frac14,
\qquad
c_2=\frac12,
\qquad
c_3=\frac14,
\qquad
c_4=0,
\qquad
c_5=-\frac12,
\end{equation}
gives the Symmetric Teleparallel Equivalent of General Relativity (STEGR).

\section{Dirac-Bergmann Algorithm in Field Theories}\label{sec:dirac_bergmann}
This section reviews the Dirac-Bergmann algorithm in field theories, which provides the foundation for the subsequent consistency analysis within the Hamiltonian formulation. We also clarify the assumptions adopted to handle terms involving spatial derivatives of the Dirac delta distribution. Consider spacetime $\mathcal{M}$with tangent bundle $T\mathcal{M}$ and a first-order Lagrangian $L(\Phi^I,\dot{\Phi}^I)=\int \mathcal{L} d^4x$, where $\{\Phi^I\}$ are generalized coordinates and $\mathcal{L}$ is the Lagrangian density. To perform a Hamiltonian analysis, one constructs the Legendre map from the tangent bundle to the cotangent bundle \cite{Gracia:1987fy}, 
\begin{equation}
\mathcal{F}L:T\mathcal{M} \to T^*\mathcal{M},
\end{equation}
which maps $(\Phi^I, \dot{\Phi}^I)$ to $(\Phi^I, \frac{\partial L}{\partial \dot{\Phi}^I})$. The term $\pi_I = \frac{\partial L}{\partial \dot{\Phi}^I}$ is called the generalized momentum or conjugate momentum, and the canonical Hamiltonian $H_c$ is defined as the function of $(\Phi^I, \pi_I)$ whose pullback is the Lagrangian energy:
\begin{align}
H_c =& \int_{\Sigma_t} \mathcal{H}_c d^3 x,\\
\mathcal{H}_c (\Phi^I, \pi_I) =& \dot{\Phi}^I \pi_I -\mathcal{L}(\Phi^I, \dot{\Phi}^I),
\end{align}
where $\mathcal{H}_c$ is the canonical Hamiltonian density and $\Sigma_t$ represents a three-dimensional spatial hypersurface. However, the Legendre map is not always invertible; the system becomes degenerate when the Hessian matrix $\frac{\partial^2 L}{\delta \dot{\Phi}^I \delta \dot{\Phi}^J}$ is not of full rank. The Dirac-Bergmann algorithm provides a systematic method to perform a Hamiltonian analysis on such degenerate systems.\\

After determining the total number of non-invertible pairs $(\dot{\Phi}^I, \pi_I)$ by investigating the rank of the Hessian matrix, we may locally define primary constraints $\bar{\phi}_I \approx 0$, which restrict the full phase space to a subspace. This restriction must remain preserved throughout the time evolution. Therefore, the primary constraints $\bar{\phi}_I$ are required to satisfy the following consistency conditions \cite{failure_fQ}:
\begin{equation}
\dot{\bar{\phi}}_I=\{\bar{\phi}_I, H_c\}+\int_{\Sigma_t}\lambda^J \{\bar{\phi}_I, \bar{\phi}_J\} d^3x \approx 0,
\end{equation}
where $\{F(\vec{x}),G(\vec{y})\}$ is the Poisson bracket between the functions $F(\vec{x})$ and $G(\vec{y})$, defined as follows:
\begin{equation}
\{F(\vec{x}),G(\vec{y})\} = \int_{\Sigma_t} d^3z \sum_I \Big(\frac{\partial F(\vec{x})}{\partial \Phi^I (\vec{z})}\frac{\partial G(\vec{y})}{\partial \pi_I (\vec{z})}-\frac{\partial F(\vec{x})}{\partial \pi_I(\vec{z})}\frac{\partial G(\vec{y})}{\partial \Phi^I (\vec{z})} \Big),
\end{equation}
where $\vec{x}$, $\vec{y}$, and $\vec{z}$ are spatial coordinates on the spacelike hypersurface $\Sigma_t$. The Poisson brackets between the local constraints may, in general, contain terms proportional to spatial derivatives of the Dirac delta distribution \cite{failure_fQ}. By refering to Tomonari's work on coincident $f(Q)$ gravity \cite{dirac_dof_fQ}, there is the proposition said that all surface terms can be neglected freely and do not change the symplectic structure. In the present work, we also taking the same assumption that the terms involving spatial derivatives of the Dirac delta distribution can be neglected. A complete Hamiltonian analysis retaining these spatial derivatives of Dirac delta distribution is left for future work, and it might provide a direct test of the assumptions adopted in the present work. \\

The Dirac matrix $D$ is the matrix formed by the Poisson brackets between the local constraints. Therefore, the consistency conditions of the primary constraints can be expressed in matrix form:
\begin{equation}
\begin{pmatrix}
\{\bar{\phi}_I, \mathcal{H}_c\}\\
\{\bar{\phi}_J, \mathcal{H}_c\}\\
\vdots\\
\{\bar{\phi}_K, \mathcal{H}_c\}
\end{pmatrix}
+ D\begin{pmatrix}
\lambda^I\\
\lambda^J\\
\vdots\\
\lambda^K
\end{pmatrix}
\approx \begin{pmatrix}
0\\
0\\
\vdots\\
0
\end{pmatrix}.
\end{equation}
If the Dirac matrix is of full rank, we can find a unique solution for all Lagrange multipliers $\lambda^I$ and stop the Dirac-Bergmann algorithm. If the Dirac matrix is not of full rank, let us assume $\operatorname{rank}(D)=\mathsf{p}$ and that the number of primary constraints is $\mathsf{q}$. If we are able to find a unique solution for $\mathsf{p}$ Lagrange multipliers and $\mathsf{q}-\mathsf{p}$ null vectors $\undertilde{\mathsf{u}}$ such that
\begin{equation}
\undertilde{\mathsf{u}}^I\{\bar{\phi}_I, \mathcal{H}_c\}\approx 0,
\end{equation}
then we can form $\mathsf{q}-\mathsf{p}$ secondary constraints and continue the algorithm by studying their consistency conditions. This process is repeated until we obtain a full set of constraints. If we cannot find such vectors $\undertilde{\mathsf{u}}$, it means the system has no valid solution for its Lagrange multipliers. Once the full set of constraints is obtained, they are classified into first-class and second-class constraints. A constraint $\bar{\phi}_I$ is a first-class constraint if and only if $\{\bar{\phi}_I, \bar{\phi}_J\} \approx 0$ for all constraints $\bar{\phi}_J$; otherwise, it is a second-class constraint. 

\section{Classification of Newer General Relativity}\label{sec:adm}
Before performing the Hamiltonian analysis, we first apply the ADM decomposition to Newer GR and review its classification into different types. This section briefly discusses the ADM decomposition and the work of Fabio D'Ambrosio and Lavinia Heisenberg, who classify Newer GR into nine classes based on the invertibility of the Hessian matrix. Consider a spacetime $\mathcal{M}$ equipped with a metric tensor
$g_{\mu\nu}$ and an affine connection $\Gamma^\rho{}_{\mu\nu}$.
Let $\mathcal{T}$ denote the time interval associated with $\mathcal{M}$,
$f^\ast$ the pullback operator induced by a smooth map $f$,
$\mathcal{S}$ a three-dimensional spatial manifold,
and $\tau$ the canonical time coordinate on
$\mathbb{R}\times\mathcal{S}$. To decompose the spacetime $\mathcal{M}$ into a disjoint
union of spacelike hypersurfaces $\Sigma_t$, we introduce an ADM foliation
defined by a diffeomorphism \cite{three_plus_one}
\begin{equation}
\Psi : \mathcal{M} \rightarrow \mathbb{R}\times\mathcal{S}.
\end{equation}
The hypersurfaces are then given by
\begin{equation}
\Sigma_t=\{\, p\in\mathcal{M}\mid(\Psi^\ast\tau)(p)=t\},
\end{equation}
such that
\begin{equation}
\mathcal{M}=\bigsqcup_{t\in\mathcal{T}}\Sigma_t.
\end{equation}
Having foliated spacetime into a family of spatial hypersurfaces $\Sigma_t$, we perform the ADM decomposition of the four-dimensional metric $g_{\mu\nu}$ in terms of the lapse function $N$, the shift vector $N^i$, and the induced metric $\gamma_{ij}$ as follows \cite{three_plus_one}:
\begin{gather}
g_{\mu \nu} =
\begin{pmatrix}
-N^2 + N_k N^k & N_i \\
N_j & \gamma_{ij}
\end{pmatrix},
\end{gather}
where $i$, $j$, and $k$ denote spatial indices. The unit normal vector $\utilde{n}$, which is orthogonal to each hypersurface $\Sigma_t$, has components
\begin{equation}
n^\alpha=\left(\frac{1}{N},-\frac{N^1}{N}, -\frac{N^2}{N}, -\frac{N^3}{N}\right),
\end{equation}
with the associated one-form components
\begin{equation}
n_\alpha=(-N,0,0,0).
\end{equation}
The normal vector satisfies the normalization condition
\begin{equation}
n^\mu n_\mu=-1.
\end{equation}
The inverse spacetime metric can then be decomposed as
\begin{equation}
g^{\mu\nu}=\gamma^{\mu\nu}-n^\mu n^\nu,
\end{equation}
where $\gamma^{\mu\nu}$ is the four-dimensional extension of the inverse induced metric,
\begin{gather}
\gamma^{\mu\nu}=
\begin{pmatrix}
0 & 0\\
0 & \gamma^{ij}
\end{pmatrix}.
\end{gather}
Using Eqs.~$(30)$, $(33)$, and $(36)-(40)$, together with the identity $\sqrt{-g}=N\sqrt{\gamma}$, Eq.~$(16)$ can be expressed entirely in terms of the ADM variables $N$, $N^i$, and $\gamma^{ij}$. Based on the work of Fabio D'Ambrosio and Lavinia Heisenberg~\cite{classifyNewerGR}, Newer General Relativity can be classified into nine classes by analyzing the invertibility of the Hessian matrix:
\begin{equation}
H = -\int_{\Sigma_t} \frac{1}{\sqrt{\gamma}} \frac{\partial^2 S}{\partial \dot{\Phi}^{I} \partial \dot{\Phi}^{J}} d^3x,
\end{equation}
where $\dot{\Phi}^I$ denotes the generalized velocities $(\dot{N}, \dot{N}^i,\dot{\gamma}^{ij})$.
\begin{table}[h]
\centering
\caption{Classification of Newer General Relativity models}
\label{tab:NGR_classification_pro}
\begin{tabular}{clc}
\toprule
\textbf{Types} & \textbf{Assumption on coefficients} & \textbf{Number of Primary Constraints} \\
\midrule
1  & $c_4=-c_1-c_2+c_3$, $c_5=-2c_3$ & 1\\
2  & $c_4=-2c_1-c_2$ & 3 \\
3  & $c_1=0$  & 5 \\
4  & $c_1=0$, $c_4=-c2+\frac{(c_5)^2}{4c_3}$ \text{or} $c_1=0$, $c_3=0$, $c_5=0$  & 6 \\
5  & $c_3=-c_1$, $c_4=-2c_1-c_2$, $c_5=2c_1$& 4 \\
6  & $c_1=0$, $c_4=-c_2$ & 8 \\
7  & $ c_1=0$, $c_4=-c_2$, $c_5=0$ & 9 \\
8  & $c_1=0$, $c_3=0$, $c_5=0$, $c_4=-c_2$ & 10 \\
9  & No Assumption & 0 \\
\bottomrule
\end{tabular}
\end{table}

\newpage
\section{Hamiltonian Analysis of Newer General Relativity}\label{sec:hamiltonian_analysis}
In this section, we begin by constructing the expanded action of Newer GR. From the expanded action, we derive the conjugate momenta and perform the Hamiltonian analysis for each type of Newer GR. This analysis allows us to determine the full set of constraints and the physical degrees of freedom for each type. Before proceeding with the Hamiltonian analysis, we first recall the projections of the non-metricity tensor along the normal direction to the spatial hypersurfaces, as derived in Ref.~\cite{DAmbrosio:2020nqu}. These projections are given by
\begin{align}
	n^{\alpha}Q_{\alpha jk} &=\frac{\dot{\gamma}_{jk}-N^i \partial_i \gamma_{jk}}{N},\\
	n^{\alpha}Q_{i \alpha k} &= \frac{\gamma_{jk} \partial_i N^j}{N},\\
	n^{\alpha}n^{\beta}Q_{\alpha \beta k} &= \frac{\gamma_{jk}(\dot{N}^j-N^i\partial_i N^j)}{N^2},\\
	n^{\alpha} n^{\beta} Q_{i\alpha \beta} &= -\frac{2\partial_i N}{N},\\
	n^{\alpha} n^{\beta} n^{\sigma} Q_{\alpha \beta \sigma} &= \frac{2(N^i \partial_i N-\dot{N})}{N^2}.
\end{align} 
Here, Greek indices $\alpha,\beta,\sigma$ run over spacetime coordinates $\{0,1,2,3\}$, whereas Latin indices $i,j,k$ run over spatial coordinates $\{1,2,3\}$. Substituting Eqs.~(29), (30), (33), and (34), together with the above projections, into Eq.~(16), we obtains,
\begin{equation}
\begin{aligned}
S[g,\Gamma] =& \int d^4x \sqrt{\gamma}\Big[N {}^{(3)}\!Q +(c_1 + c_2 + c_3 + c_4 + c_5)\left( -\frac{4}{N^3}\right) (\dot{N}^2-2\dot{N}N^i\partial_i N+N^i N^j \partial_i N \partial_j N)\\ &+(2c_1 +c_2+c_4)\left(\frac{\gamma_{kc}}{N^3}\right) (\dot{N}^k\dot{N}^c-\dot{N}^k N^a\partial_a N^c-\dot{N}^c N^i\partial_i N^k + N^i N^a \partial_i N^k \partial_a N^c)\\ &+(2c_3+c_5)\left(\frac{2}{N^2}\right)[\gamma^{bc}(\dot{N}N^a\partial_a \gamma_{bc}-\dot{N}\dot{\gamma}_{bc})+\gamma^{jk}\dot{\gamma}_{jk}N^a\partial_a N] \\ &+(2c_4+c_5)\frac{2}{N^2}(N^a \partial_a N \partial_i N^i-\dot{N}\partial_i N^i)\\ &+(2c_2+c_5)\left(\frac{2}{N^2}\right) (N^i \partial_i N^j \partial_j N-\dot{N}^j \partial_j N)+(c_1+c_3)\left(\frac{4\gamma^{ia}\partial_i N\partial_a N}{N}\right)\\ &+(c_3+c_5)\left(-\frac{2\gamma^{bc}N^i N^a \partial_i N \partial_a \gamma_{bc}}{N^2}\right)\notag\\
&+c_1\left(\frac{N^i N^a \partial_i \gamma^{bc} \partial_a \gamma_{bc}-2\gamma^{ia}\gamma_{kc}\partial_i N^k \partial_a N^c-\gamma^{jb}\gamma^{kc}\dot{\gamma}_{jk}\dot{\gamma}_{bc}-2\dot{\gamma}_{bc}N^i \partial_i \gamma^{bc}}{N}\right)\\ &+c_2\left( \frac{2\gamma^{ic}N^a \partial_i N^b \partial_a \gamma_{bc}-\partial_i N^a \partial_a N^i -2\gamma^{ic}\dot{\gamma}_{bc}\partial_i N^b}{N}\right)\\ &+c_3\Big(4\gamma^{ia}\gamma^{jk}Q_{ijk}\partial_a N-\frac{2\gamma^{jk}N^iN^a\partial_a N\partial_i \gamma_{jk}}{N^2}\\  &-\frac{\gamma^{jk}\gamma^{bc}(N^i N^a \partial_i \gamma_{jk}\partial_a \gamma_{bc}+\dot{\gamma}_{jk}\dot{\gamma}_{bc}-\dot{\gamma}_{bc}N^i \partial_i \gamma_{jk}-\dot{\gamma}_{jk}N^a \partial_a \gamma_{bc})}{N}\Big)\notag\\
&+c_4\left(-\frac{\partial_i N^i \partial_a N^a}{N}+\frac{1}{N}\gamma^{ij}Q_{ijk}(N^a \partial_a N^k -\dot{N}^k)\right)\\ &+c_5\left( \frac{\gamma^{bc}N^a\partial_a \gamma_{bc}\partial_i N^i-\gamma^{bc}\dot{\gamma}_{bc}\partial_i N^i}{N}+2\gamma^{ij}\gamma^{ak}Q_{ijk}\partial_a N+\frac{1}{N}\gamma^{bc}Q_{abc}(N^i\partial_i N^a-\dot{N}^a)\right) \Big].
\end{aligned}
\end{equation}
By substituting the parameter choices defining each type of Newer General Relativity into Eq.~(34), we obtain the explicit form of the action for each theory. The Dirac-Bergmann algorithm can then be applied to each case individually to derive the corresponding Hamiltonian formulation and constraint structure. The Hamiltonian analysis presented below is carried out using the xAct mathematica package.

\subsection{Type 1}
Using the explicit form of the action for Type $1$ Newer General Relativity, we obtain the conjugate momenta as follows:
\begin{align}
\pi_N =& \frac{(4c_1+4c_2)}{N^2}\partial_a N^a,\\
\pi_i =& \frac{(2c_1+2c_3)}{N^3}\left(\dot{N}^a \gamma_{ia}-\gamma_{ib} \gamma^{ak} N_a \partial_k N^b \right)+\frac{(c_1+c_2-c_3)}{N}\gamma^{ab}\partial_b \gamma_{ia}+\frac{2c_3}{N}\gamma^{ab}\partial_i \gamma_{ab}-\notag \\&\frac{(4c_2-4c_3)}{N^2}\partial_i N,\\
\pi_{ij} =& -\frac{2c_3}{N}(\dot{\gamma}^{ab} \gamma_{ab} \gamma_{ij}+ \gamma_{ij} \gamma^{ak} \partial_k N_a - \gamma^{bc}\gamma_{ij} N^a \partial_a \gamma_{bc})-\frac{2c_1}{N}( \dot{\gamma}^{ab} \gamma_{ia} \gamma_{jb}+N^a \partial_a \gamma_{ij})\notag \\ &+\frac{c_2}{N}(\gamma_{ja}\partial_i N^a + \gamma_{ia} \partial_j N^a).
\end{align}
Consider the Legendre map $\mathcal{F}L:TQ \to T^*Q$ defined by $(\Phi^I,\dot{\Phi}^I) \mapsto (\Phi^I, \pi_I)$. The relations between the velocities $\dot{\gamma}^{ij}$ and $\dot{N}^i$, and their corresponding momenta $\pi_{ij}$ and $\pi_i$, can be inverted,
\begin{align}
\dot{N}^k =& \frac{N^3\gamma^{ik}}{2(c_1+c_3)}\Big(\pi_i -\frac{(c_1+c_2-c_3)}{N}\gamma^{ab}\partial_b \gamma_{ia}-\frac{2c_3}{N}\gamma^{ab}\partial_i \gamma_{ab}-\frac{4(c_3-c_2)}{N^2}\partial_i N \notag \\&+\frac{2(c_1+c_3)}{N^3}\gamma_{ib}N^a\partial_a N^b\Big)\\
\dot{\gamma}^{kl} =& -\frac{N\gamma^{ik}\gamma^{jl}}{2c_1}\Big(\pi_{ij}+\frac{2c_3}{N}(\gamma^{bc}\gamma_{ij}N^m\partial_m \gamma_{bc}+\gamma_{ij}\partial_m N^m)+\frac{2c_1}{N}N^a\partial_a \gamma_{ij}\notag \\&-\frac{c_2}{N}(\gamma_{ja}\partial_i N^a+\gamma_{ia}\partial_j N^a)-\frac{c_3}{3c_3+c_1}\gamma_{ij}\gamma^{ab}\Big[ \pi_{ab}+\frac{2c3}{N}(\gamma^{mn}\gamma_{ab}N^c\partial_c \gamma_{mn}+\gamma_{ab}\partial_c N^c)\notag \\&+\frac{2c1}{N}N^c\partial_c \gamma_{ab}-\frac{c_2}{N}(\gamma_{bm}\partial_a N^m +\gamma_{am}\partial_b N^m)\Big] \Big)
\end{align}
However, the relation between $\pi_N$ and $\dot{N}$ is not invertible, and a primary constraint arises in the form:
\begin{equation}
\bar{\phi}_N = \pi_N -  \frac{(4c_1+4c_2)\partial_a N^a}{N^2} \approx 0.
\end{equation}
The canonical Hamiltonian is,
\begin{equation}
H_c (\Sigma_t) = \int_{\Sigma_t} \mathcal{H}_c d^3x,
\end{equation}
\begin{align}
\mathcal{H}_c =& \frac{1}{N^2} ( f_1 \pi_{ab} \pi^{ab} N^3+f_2 \pi^a{}_a \pi^b{}_b N^3+f_3 N^5\pi_a \pi^a+f_4 \pi^b{}_b \gamma^{ij}N^2 N^a \partial_a \gamma_{ij} + f_5 N^3 \pi^a\partial_a N \notag \\ 
& +f_6 N^2\pi^b{}_b\partial_a N^a+f_7 N\partial_a N \partial^a N+f_8 N^2 \pi^a N^b \partial_b N_a+f_9 N^a \partial_a N \partial_b N^b \notag \\
& +f_{10} N \partial_a N^a\partial_b N^b +f_{11}\pi_{ba} N^2\partial^b N^a+f_{12} N \partial_a N_b \partial^b N^a+f_{13} N \gamma_{ac} \partial_b N^c \partial^b N^a \notag\\
& +f_{14} NN^a \gamma_{bc} \gamma^{ij}\partial_a \gamma_{ij} \partial^c N^b+f_{15}N \gamma_{bc}\partial_a N^a \partial^c N^b+f_{16}N^3 \gamma^{ab} \gamma^{ci}\gamma^{jk} \partial_c \gamma_{ab}\partial_i \gamma_{jk} \notag \\
&+f_{17}N \gamma_{ai} \gamma_{bc}\partial^b N^a \partial^i N^c +f_{18}N \gamma_{ac} \gamma_{bi}\partial^b N^a \partial^i N^c +f_{19}N\gamma_{ab} \gamma_{ci} \partial^b N^a \partial^i N^c\notag \\
&+f_{20}N^3 \gamma^{ab}\gamma^{ci} \gamma^{jk} \partial_b \gamma_{ac} \partial_k \gamma_{ij} +f_{21} N^3 \gamma^{ab}\gamma^{ci} \gamma^{jk} \partial^c \gamma_{ab}\partial_k  \gamma_{ij}),
\end{align} 
where $f_1$, ..., $f_{21}$ are functions of the coefficients $c_1$, $c_2$, ..., $c_5$. Therefore we have the primary Hamiltonian,
\begin{equation}
H_p (\Sigma_t)=H_c (\Sigma_t) + \int_{\Sigma_t} \bar{\lambda} \bar{\phi}_N d^3x,
\end{equation}
where $\bar{\lambda}$ is the Lagrange multiplier. Since the poisson bracket between primary constaint to itself vanishes,
\begin{equation}
 \{\bar{\phi}_N(\vec{x}), \bar{\phi}_N(\vec{y}) \} =0,
\end{equation}
the consistency condition reduces to:
\begin{align}
\bar{\sigma}_N(\vec{x}) =&\{\bar{\phi}_N(\vec{x}), H_p (\Sigma_t) \}, \notag \\
=&  \{\bar{\phi}_N, \mathcal{H}_c \}(\vec{x}),\notag \\
=& \frac{1}{N^3}(f_{22} N^3 \pi_{ab}\pi^{ab}+f_{23}N^3\pi^a{}_a\pi^b{}_b+f_{24}N^5\pi_a \pi^a+f_{25}N^3\pi^a \partial_a N+f_{26}N^4\partial_a \pi^a \notag \\
&+f_{27}N^2\partial_a\partial^a N+f_{28}N\partial_a N \partial^a N+f_{29} N\partial_a N^a \partial_b N^b+f_{30}N^a\partial_a N_b \partial^b N \notag \\
& +f_{31} N \partial_a N_b \partial^b N^a + f_{32} \gamma_{ac}N \partial_b N^c \partial^b N^a +f_{33} N \gamma_{bc}\gamma^{de} N^a \partial_a \gamma_{de} \partial^c N^b+f_{34}\gamma_{bc}N \partial_a N^a \partial^c N^b\notag \\
& +f_{35}\gamma^{ab}\gamma^{cd}\gamma^{ef}N^3\partial_c \gamma_{ab}\partial_d \gamma_{ef}+f_{36}N\gamma_{ad}\gamma_{bc}\partial^b N^a\partial^dN^c+f_{37}N\gamma_{ac}\gamma_{bd}\partial^b N^a \partial^d N^c\notag \\
& +f_{38}N\gamma_{ab}\gamma_{cd}\partial^bN^a\partial^dN^c+f_{39}N^3\gamma^{ab}\gamma^{cd}\gamma^{ef}\partial_b \gamma_{ac}\partial_f \gamma_{de}+f_{40}\gamma^{ab}\gamma^{cd}\gamma^{ef} N^3 \partial_c \gamma_{ab}\partial_f \gamma_{de}),
\end{align}
where $f_{22}$, ..., $f_{40}$ are functions of the coefficients $c_1$, ..., $c_5$. We then introduce the smeared constraints
\begin{align}
\phi_N [F_1] &= \int_{\Sigma_t} d^3x F_1(\vec{x}) \bar{\phi}_N (q(\vec{x}), p(\vec{x}), \partial q(\vec{x})),\\
\sigma_N [F_2] &= \int_{\Sigma_t}d^3y F_2(\vec{y}) \bar{\sigma}_N (q(\vec{y}), p(\vec{y}), \partial q(\vec{y})),
\end{align}
which in our assumption of  the terms involving spatial derivatives of the Dirac delta distribution can be neglected, and choosing $F_1 = \delta (\vec{z}-\vec{a})$ and $F_2 = \bar{\lambda}$,
\begin{align}
\{\phi_N[\delta],\sigma_N[\bar{\lambda}]\} &= \int_{\Sigma_t} d^3z \bar{\lambda}(\vec{z}) \{ \bar{ \phi }_N, \bar{\sigma}_N \}(\vec{z}) \delta (\vec{z}-\vec{a}). \notag \\
&= \bar{\lambda}(\vec{a}) \{ \bar{ \phi}_N, \bar{ \sigma}_N \}(\vec{a}),
\end{align}
with the expllicit expression for $\{ \bar{ \phi}_N, \bar{ \sigma}_N \}$, 
\begin{align}
\{ \bar{ \phi}_N, \bar{ \sigma}_N \} =& \frac{1}{N^3}(f_{41}N^4 \pi_a\pi^a+f_{42}N^2\pi^a\partial_aN+f_{43}N^3\partial_a\pi^a+f_{44}N\partial_a\partial^aN+f_{45}\partial_aN\partial^aN\notag \\
& +f_{46}\partial_aN^a\partial_b N^b+f_{47}\partial_aN_b\partial^bN^a+f_{48}\gamma_{ac}\partial_b N^c\partial^bN^a+f_{49}\gamma_{bc}\partial_aN_a\partial^cN^b\notag \\
&+f_{50}\gamma_{ad}\gamma_{bc}\partial^bN^a\partial^dN^c +f_{51}\gamma_{ac}\gamma_{bd}\partial^b N^a\partial^dN^c+f_{52}\gamma_{ab}\gamma_{cd}\partial^b N^a\partial^d N^c).
\end{align}
The consistency condition for the secondary constraint becomes
\begin{align}
\{\bar{\sigma}_N(\vec{a}), H_T(\Sigma_t) \} &= \{\bar{\sigma}_N (\vec{a}), H_c(\Sigma_t) \} +\{\sigma_N [\bar{\lambda}], \phi_N[\delta(\vec{z}-\vec{a})] \}, \notag\\
&= \{ \bar{\sigma}_N, \mathcal{H}_c\}(\vec{a})+\bar{\lambda}\{  \bar{\sigma}_N,\bar{\phi}_N \}(\vec{a}) \approx 0,
\end{align} 
with the uniquely determined lagrange multiplier $\bar{\lambda}$, 
\begin{align}
\bar{\lambda} = -\frac{\{ \bar{\sigma}_N, \mathcal{H}_c\}}{ \{  \bar{\sigma}_N,\bar{\phi}_N \}}.
\end{align}
The full set of constraints consists only of $\bar{\phi}_N$ and $\bar{\sigma}_N$, both of which are second-class constraints. Consequently, the algorithm reveals that Type 1 Newer General Relativity is a gravity theory with $10-\frac{1}{2}(2) = 9$ physical degrees of freedom.

\subsection{Type 2}
We evaluate the conjugate momenta for Type 2 Newer General Relativity as:
\begin{align}
\pi_N =&  \frac{(8c_1+4c_2-2c_5)}{N^2}\partial_a N^a+\frac{(4c_3+2c_5)}{N^2}(\dot{\gamma}^{ab} \gamma_{ab}-\gamma_{bc} N^a \partial_a \gamma^{bc})\notag \\ &+\frac{(8c_1-8c_3-8c_5)}{N^3}(\dot{N}-N^a \partial_a N),\\
\pi_i =& \frac{(2c_1+c_2)}{N}\gamma^{ab}\partial_b \gamma_{ia}-\frac{c_5}{N}\gamma^{ab}\partial_i \gamma_{ab}-\frac{(4c_2+2c_5)}{N^2}\partial_i N, \\
\pi_{ij} =& \frac{(4c_3+2c_5)}{N^2}(\dot{N} \gamma_{ij}-\gamma_{ij} N^a \partial_a N)+\frac{2c_3}{N}( -\dot{\gamma}^{ab} \gamma_{ab} \gamma_{ij} + \gamma_{bc} \gamma_{ij} N^a \partial_a \gamma^{bc}) \notag \\ &+\frac{2c_1}{N} (-\dot{\gamma}^{ab} \gamma_{ia} \gamma_{jb}+\gamma_{ib} \gamma_{jc} N^a \partial_a \gamma^{bc})+\frac{c_5}{N} \gamma_{ij}\partial_a N^a +\frac{c_2}{N} (\gamma_{ja} \partial_i N^a + \gamma_{ia} \partial_j N^a),
\end{align}
and the relation between $\pi_i$ and $\dot{N}^i$ is not invertible, leading to the primary constraints:
\begin{equation}
\bar{\phi}_i = \pi_i -\frac{(2c_1+c_2)}{N}\gamma^{ab}\partial_b \gamma_{ia}+\frac{c_5}{N}\gamma^{ab}\partial_i \gamma_{ab}-\frac{(-4c_2-2c_5)}{N^2}\partial_i N,
\end{equation}
and the following solutions of $\dot{N}$ and $\dot{\gamma}^{kl}$,
\begin{align}
\dot{N}=& \frac{N^3}{8(c_1-c_3-c_5)}\Big(\pi_N+\frac{8(c_1-c_3-c_5)}{N^3}N^a\partial_aN-\frac{2(2c_3+c_5)}{N^2}(\dot{\gamma}^{ab}\gamma_{ab}+\gamma^{bc}N^a\partial_a\gamma_{bc})\notag \\&-\frac{2(4c_1+2c_2-c_5)}{N^2}\partial_a N^a \Big)\\
\dot{\gamma}^{kl}=&-\frac{N\gamma^{ik}\gamma^{jl}}{2c_1}\Big(\pi_{ij}+\frac{2c_1}{N}N^a\partial_a \gamma_{ij}+\frac{2c_3}{N}\gamma^{bc}\gamma_{ij}N^a\partial_a\gamma_{bc}-\frac{2(2c_3+c_5)}{N^2}(\dot{N}\gamma_{ij}-\gamma_{ij}N^a\partial_aN)\notag \\&-\frac{c_5}{N}\gamma_{ij}\partial_aN^a-\frac{c_2}{N}(\gamma_{ja}\partial_i N^a+\gamma_{ia}\partial_j N^a) -\frac{c_3}{3c_3+c_1}\gamma_{ij}\gamma^{ab}\Big[\pi_{ab}+\frac{2c_1}{N}N^m \partial_m \gamma_{ab}\notag \\
&+\frac{2c_3}{N}\gamma^{mn}\gamma_{ab}N^c\partial_c\gamma_{mn}-\frac{2(2c_3+c_5)}{N^2}(\dot{N}\gamma_{ab}-\gamma_{ab}N^m\partial_mN)-\frac{c_5}{N}\gamma_{ab}\partial_m N^m\notag \\&-\frac{c_2}{N}(\gamma_{bm}\partial_a N^m+\gamma_{am}\partial_b N^m) \Big]\Big).
\end{align}
The condition $\{\bar{\phi}_i, \bar{\phi}_j \} = 0$ leads to the secondary constraints:
\begin{align}
\bar{\sigma}_i(\vec{x}) &=\{\bar{\phi}_i(\vec{x}), H_p (\Sigma_t) \}, \notag\\
&=  \{\bar{\phi}_i, \mathcal{H}_c \}(\vec{x}).
\end{align}
The consistency conditions for the secondary constraints $\bar{\sigma}_i$ are given by:
\begin{align}
\{\bar{\sigma}_i (\vec{a}), H_T(\Sigma_t) \} =&  \{\bar{\sigma}_i (\vec{a}), H_c(\Sigma_t) \} +\lambda^{j}(\vec{a}) \{\sigma_i [\bar{\lambda}], \phi_j[\delta(\vec{z}-\vec{a})] \}\notag \\ &+ \lambda'^j (\vec{a}) \{\sigma_i [\bar{\lambda}], \sigma_j[\delta(\vec{z}-\vec{a})] \}, \notag \\
=& \{ \bar{\sigma}_i, \mathcal{H}_c\}(\vec{a})+\lambda^j \{  \bar{\sigma}_i,\bar{\phi}_j \}(\vec{a}) + \lambda'^j \{\bar{\sigma}_i,\bar{\sigma}_j  \}(\vec{a}) \approx 0.
\end{align}
Setting $\{\bar{\sigma}_i(\vec{x}), \bar{\phi}_j(\vec{y}) \} = A_{ij}\delta^{(3)}(\vec{x}-\vec{y})$ for $i, j\in \{1,2,3\}$, the consistency conditions become:
\begin{equation}
\{\bar{\sigma}_i (\vec{a}), H_T(\Sigma_t) \} =  \{ \bar{\sigma}_i, \mathcal{H}_c\}(\vec{a})+\lambda^j \{  \bar{\sigma}_i,\bar{\phi}_j \}(\vec{a}) \approx 0,
\end{equation}
which has the matrix form,
\begin{equation}
\begin{pmatrix}
 \{ \bar{\sigma}_1, \mathcal{H}_c\} \\
 \{ \bar{\sigma}_2, \mathcal{H}_c\}\\
 \{ \bar{\sigma}_3, \mathcal{H}_c\}
\end{pmatrix}
+
\begin{pmatrix}
A_{11}&A_{12}&A_{13} \\
A_{21}&A_{22}&A_{23} \\
A_{31}&A_{32}&A_{33} 
\end{pmatrix}
\begin{pmatrix}
\lambda^1 \\
\lambda^2 \\
\lambda^3
\end{pmatrix}
\approx
\begin{pmatrix}
0\\
0\\
0
\end{pmatrix}.
\end{equation}
After that, we denote the matrix formed by the poisson brackets between constraints as $D^{(1)}$,
\begin{equation}
D^{(1)}=
\begin{pmatrix}
A_{11}&A_{12}&A_{13} \\
A_{21}&A_{22}&A_{23} \\
A_{31}&A_{32}&A_{33}
\end{pmatrix},
\end{equation}
and apply the LU decomposition on $D^{(1)}$ as follows,
\begin{equation}
D^{(1)}=
\begin{pmatrix}
1&0&0\\
\frac{A_{21}}{|\triangle^{(1)}_1|}&1&0\\
\frac{A_{31}}{|\triangle^{(1)}_1|}&\frac{A_{11}A_{32}-A_{31}A_{12}}{|\triangle^{(1)}_2|}&1
\end{pmatrix}
\begin{pmatrix}
|\triangle^{(1)}_1|&A_{12}&A_{13}\\
0&\frac{|\triangle^{(1)}_2|}{|\triangle^{(1)}_1|}&A_{23}-\frac{A_{21}}{|\triangle^{(1)}_1|}A_{13}\\
0&0&\frac{|\triangle^{(1)}_3|}{|\triangle^{(1)}_2|}
\end{pmatrix},
\end{equation}
where $\triangle^{(\mathcal{P})}_{\mathcal{Q}}$ denotes the $\mathcal{Q}$th leading principal submatrix of $D^{(\mathcal{P})}$ throughout the paper, and the determinant of each $\triangle^{(\mathcal{P})}_{\mathcal{Q}}$ are assumed to be nonzero. By solving the consistency conditions, we obtained the solutions of $\lambda^i$,
\begin{align}
\lambda^1 &\approx \frac{1}{|\triangle^{(1)}_1|}\Big(-\{\bar{\sigma}_1, \mathcal{H}_c\}-A_{13}\lambda^3-A_{12}\lambda^2\Big),\\
\lambda^2 &\approx \frac{|\triangle^{(1)}_1|}{|\triangle^{(1)}_2|}\Big(\frac{A_{21}}{|\triangle^{(1)}_1|}\{\bar{\sigma}_1, \mathcal{H}_c\}-\{\bar{\sigma}_2, \mathcal{H}_c\}-(A_{23}\frac{A_{21}}{|\triangle^{(1)}_1|}A_{13})\lambda^3\Big),\\
\lambda^3 &\approx \frac{|\triangle^{(1)}_2|}{|\triangle^{(1)}_3|}\Bigg(\Big(\frac{A_{31}}{|\triangle^{(1)}_1|}-\frac{A_{32}|\triangle^{(1)}_1|-A_{31}A_{12}}{A_{22}|\triangle^{(1)}_1|-A_{21}A_{12}}\Big(\frac{A_{21}}{|\triangle^{(1)}_1|}\Big)\Big)\{\bar{\sigma}_1,\mathcal{H}_c\}+\frac{A_{32}|\triangle^{(1)}_1|-A_{31}A_{12}}{A_{22}|\triangle^{(1)}_1|-A_{21}A_{12}}\{\bar{\sigma}_2,\mathcal{H}_c\}-\{\bar{\sigma}_3,\mathcal{H}_c\}\Bigg),
\end{align}
This leads to the termination of the Dirac–Bergmann algorithm. The full set of constraints consists of the second-class constraints $\bar{\sigma}_i$ and $\bar{\phi}_j$, resulting in $10-\frac{1}{2}(6)=7$ physical degrees of freedom.

\subsection{Type 3}
For Type 3 Newer General Relativity, we first consider the conjugate momenta,
\begin{align}
\pi_N =&\frac{(-4c_4-2c_5)}{N^2}\partial_a N^a+\frac{(4c_3+2c_5)}{N^2}(\dot{\gamma}^{ab} \gamma_{ab}-\gamma_{bc} N^a \partial_a \gamma^{bc})\notag \\ &+\frac{(-8c_2-8c_3-8c_4-8c_5)}{N^3}(\dot{N}-N^a \partial_a N),\\
\pi_i =& \frac{(2c_2+2c_4)}{N^3}\left(\dot{N}^a \gamma_{ia}-\gamma_{ib} N^a \partial_a N^b \right)-\frac{(4c_2+2c_5)}{N^2}\partial_i N-\frac{c_4}{N}\gamma^{ab}\partial_b \gamma_{ia}-\frac{c_5}{N}\gamma^{ab}\partial_i \gamma_{ab},\\
\pi_{ij} =& \frac{(4c_3+2c_5)}{N^2}(\dot{N} \gamma_{ij}-\gamma_{ij} N^a \partial_a N)+\frac{2c_3}{N}( -\dot{\gamma}^{ab} \gamma_{ab} \gamma_{ij} + \gamma_{bc} \gamma_{ij} N^a \partial_a \gamma^{bc}) \notag \\ &+\frac{c_5}{N} \gamma_{ij}\partial_a N^a +\frac{c_2}{N} (\gamma_{ja} \partial_i N^a + \gamma_{ia} \partial_j N^a).
\end{align}
Notice that $\dot{\gamma}^{ij}$ cannot be uniquely expressed in terms of $\pi_{ij}$ because the coefficient of the term $-\dot{\gamma}^{ab}\gamma_{ia}\gamma_{jb}$ vanishes. Consequently, the Legendre transformation is degenerate, and primary constraints arise,

\begin{align}
\bar{\phi}_{ij} =& \pi_{ij} -\frac{c2(\gamma_{ja}\partial_i N^a +\gamma_{ia} \partial_j N^a) + 2c3\gamma_{bc} \gamma_{ij} N^a \partial_a \gamma^{bc} +c_5 \gamma_{ij} \partial_a N^a}{N}\notag \\ &-\frac{(4c_3+2c_5)(\dot{N}\gamma_{ij}+\gamma_{ij} N^a \partial_a N)}{N^2} - \gamma_{ij} \gamma^{11} \Big[ \pi_{11}+\frac{(4c_3+2c_5)(\dot{N}\gamma_{11}+\gamma_{11} N^a \partial_a N)}{N^2}\notag \\ &+\frac{c2(\gamma_{1a}\partial_1 N^a +\gamma_{1a} \partial_1 N^a) + 2c3\gamma_{bc} \gamma_{11} N^a \partial_a \gamma^{bc} +c_5 \gamma_{11} \partial_a N^a}{N}\Big],
\end{align}
with the solutions of the velocities as following,
\begin{align}
\dot{N}=&-\frac{N^3}{8(c_2+c_3+c_4+c_5)}\Big(\pi_N-\frac{8(c_2+c_3+c_4+c_5)}{N^3}N^a\partial_aN+\frac{2(2c_4+c_5)}{N^2}\partial_aN^a\notag \\&-\frac{2(2c_3+c_5)}{N^2}(\dot{\gamma}^{ab}\gamma_{ab}+\gamma^{bc}N^a\partial_a\gamma_{bc}) \Big),\\
\dot{N}^k =& \frac{N^3\gamma^{ik}}{2(c_2+c_4)}\Big(\pi_i+\frac{2(c_2+c_4)}{N^3}\gamma_{ib}N^a\partial_aN^b+\frac{c_4}{N}\gamma^{ab}\partial_b\gamma_{ia}\notag \\&+\frac{c_5}{N}\gamma^{ab}\partial_i\gamma_{ab}+\frac{2(2c_2+c_5)}{N^2}\partial_iN \Big),
\end{align}
and also the consistency conditions that yield secondary constraints $\bar{\sigma}_{ij}$,\begin{align}
\bar{\sigma}_{ij}(\vec{x}) &=\{\bar{\phi}_{ij}(\vec{x}), H_p (\Sigma_t) \}, \notag \\
&=  \{\bar{\phi}_{ij}, \mathcal{H}_c \}(\vec{x}).
\end{align}
After that we compute $\{\bar{\sigma}_{ij}, \bar{\phi}_{mn} \}$ and $\{\bar{\sigma}_{ij}, \mathcal{H}_c \}$ for the consistency conditions of $\bar{\sigma}_{ij}$, 
\begin{align}
\{\bar{\sigma}_{ij} (\vec{a}), H_T(\Sigma_t) \} =&  \{\bar{\sigma}_{ij} (\vec{a}), H_c(\Sigma_t) \} +\lambda^{mn}(\vec{a}) \{\sigma_{ij} [\bar{\lambda}], \phi_{mn}[\delta(\vec{z}-\vec{a})] \}\notag \\ &+ \lambda'^{mn} (\vec{a}) \{\sigma_{ij} [\bar{\lambda}], \sigma_{mn}[\delta(\vec{z}-\vec{a})] \},\notag \\
=& \{ \bar{\sigma}_{ij}, \mathcal{H}_c\}(\vec{a})+\lambda^{mn} \{  \bar{\sigma}_{ij},\bar{\phi}_{mn} \}(\vec{a}) + \lambda'^{mn} \{\bar{\sigma}_{ij},\bar{\sigma}_{mn}  \}(\vec{a}),\notag \\
=&  \{ \bar{\sigma}_{ij}, \mathcal{H}_c\}(\vec{a})+\lambda^{mn} \{  \bar{\sigma}_{ij},\bar{\phi}_{mn} \}(\vec{a}) \approx 0.
\end{align}
Let $\{\bar{\sigma}_{ij}(\vec{x}), \bar{\phi}_{mn}(\vec{y}) \}=\mathfrak{A}_{ijmn}\delta^{(3)}(\vec{x}-\vec{y})$ with $ij, mn\in \{22,33,12,13,23\}$. In order to solve the consistency condition, we introduce some symbols to simplify the lengthly expressions as follows,
\begin{align}
\mathfrak{b}_{ij}&=\frac{\begin{vmatrix}
\mathfrak{A}_{2222}&\mathfrak{A}_{2233}\\
\mathfrak{A}_{ij22}&\mathfrak{A}_{ij33}
\end{vmatrix}
}{|\triangle^{(2)}_2|},\\
\mathfrak{c}_{ij}&=-\frac{\mathfrak{A}_{ij22}}{|\triangle^{(2)}_1|}+\frac{\mathfrak{A}_{3322}}{|\triangle^{(2)}_1|}\mathfrak{b}_{ij},\\
\mathfrak{d}_{ij}&=\frac{\begin{vmatrix}
\mathfrak{A}_{2222}&\mathfrak{A}_{2233}&\mathfrak{A}_{2212}\\
\mathfrak{A}_{3322}&\mathfrak{A}_{3333}&\mathfrak{A}_{3312}\\
\mathfrak{A}_{ij22}&\mathfrak{A}_{ij33}&\mathfrak{A}_{ij12}
\end{vmatrix}}{|\triangle^{(2)}_3|},\\
\mathfrak{f}_{ij}&=\frac{\begin{vmatrix}
\mathfrak{A}_{2222}&\mathfrak{A}_{2233}&\mathfrak{A}_{2212}&\mathfrak{A}_{2213}\\
\mathfrak{A}_{3322}&\mathfrak{A}_{3333}&\mathfrak{A}_{3312}&\mathfrak{A}_{3313}\\
\mathfrak{A}_{1222}&\mathfrak{A}_{1233}&\mathfrak{A}_{1212}&\mathfrak{A}_{1213}\\
\mathfrak{A}_{ij22}&\mathfrak{A}_{ij33}&\mathfrak{A}_{ij12}&\mathfrak{A}_{ij13}
\end{vmatrix}}{|\triangle^{(2)}_4|},\\
\bar{\mathfrak{b}}_{ij}&=\begin{vmatrix}
\mathfrak{A}_{2222}&\mathfrak{A}_{22ij}\\
\mathfrak{A}_{3322}&\mathfrak{A}_{33ij}
\end{vmatrix},\\
\bar{\mathfrak{d}}_{ij}&=\begin{vmatrix}
\mathfrak{A}_{2222}&\mathfrak{A}_{2233}&\mathfrak{A}_{22ij}\\
\mathfrak{A}_{3322}&\mathfrak{A}_{3333}&\mathfrak{A}_{33ij}\\
\mathfrak{A}_{1222}&\mathfrak{A}_{1233}&\mathfrak{A}_{12ij}
\end{vmatrix},\\
\bar{\mathfrak{f}}_{ij}&=\begin{vmatrix}
\mathfrak{A}_{2222}&\mathfrak{A}_{2233}&\mathfrak{A}_{2212}&\mathfrak{A}_{22ij}\\
\mathfrak{A}_{3322}&\mathfrak{A}_{3333}&\mathfrak{A}_{3312}&\mathfrak{A}_{33ij}\\
\mathfrak{A}_{1222}&\mathfrak{A}_{1233}&\mathfrak{A}_{1212}&\mathfrak{A}_{12ij}\\
\mathfrak{A}_{1322}&\mathfrak{A}_{1333}&\mathfrak{A}_{1312}&\mathfrak{A}_{13ij}
\end{vmatrix}.
\end{align}
The consistency condition in matrix form are shown below,
\begin{equation}
\begin{pmatrix}
 \{ \bar{\sigma}_{22}, \mathcal{H}_c\} \\
 \{ \bar{\sigma}_{33}, \mathcal{H}_c\}\\
 \{ \bar{\sigma}_{12}, \mathcal{H}_c\}\\
 \{ \bar{\sigma}_{13}, \mathcal{H}_c\}\\
 \{ \bar{\sigma}_{23}, \mathcal{H}_c\}
\end{pmatrix}
+
\begin{pmatrix}
\mathfrak{A}_{2222}&\mathfrak{A}_{2233}&\mathfrak{A}_{2212}&\mathfrak{A}_{2213}&\mathfrak{A}_{2223}\\
\mathfrak{A}_{3322}&\mathfrak{A}_{3333}&\mathfrak{A}_{3312}&\mathfrak{A}_{3313}&\mathfrak{A}_{3323}\\
\mathfrak{A}_{1222}&\mathfrak{A}_{1233}&\mathfrak{A}_{1212}&\mathfrak{A}_{1213}&\mathfrak{A}_{1223}\\
\mathfrak{A}_{1322}&\mathfrak{A}_{1333}&\mathfrak{A}_{1312}&\mathfrak{A}_{1313}&\mathfrak{A}_{1323}\\
\mathfrak{A}_{2322}&\mathfrak{A}_{2333}&\mathfrak{A}_{2312}&\mathfrak{A}_{2313}&\mathfrak{A}_{2323}
\end{pmatrix}
\begin{pmatrix}
\lambda^{22} \\
\lambda^{33} \\
\lambda^{12}\\
\lambda^{13}\\
\lambda^{23}
\end{pmatrix}
\approx
\begin{pmatrix}
0\\
0\\
0\\
0\\
0
\end{pmatrix}.
\end{equation}
By applying LU decomposition, the matrix formed by poisson bracket between constraints, $D^{(2)}$ can be decomposed into the following form, 
\begin{equation}
D^{(2)}=\begin{pmatrix}
1&0&0&0&0\\
\frac{\mathfrak{A}_{3322}}{|\triangle^{(2)}_1|}&1&0&0&0\\
\frac{\mathfrak{A}_{1222}}{|\triangle^{(2)}_1|}&\mathfrak{b}_{12}&1&0&0\\
\frac{\mathfrak{A}_{1322}}{|\triangle^{(2)}_1|}&\mathfrak{b}_{13}&\mathfrak{d}_{13}&1&0\\
\frac{\mathfrak{A}_{2322}}{|\triangle^{(2)}_1|}&\mathfrak{b}_{23}&\mathfrak{d}_{23}&\mathfrak{f}_{23}&1
\end{pmatrix}
\begin{pmatrix}
|\triangle^{(2)}_1|&\mathfrak{A}_{2233}&\mathfrak{A}_{2212}&\mathfrak{A}_{2213}&\mathfrak{A}_{2223}\\
0&\frac{|\triangle^{(2)}_2|}{|\triangle^{(2)}_1|}&\frac{\bar{\mathfrak{b}}_{12}}{|\triangle^{(2)}_1|}&\frac{\bar{\mathfrak{b}}_{13}}{|\triangle^{(2)}_1|}&\frac{\bar{\mathfrak{b}}_{23}}{|\triangle^{(2)}_1|}\\
0&0&\frac{|\triangle^{(2)}_3|}{|\triangle^{(2)}_2|}&\frac{\bar{\mathfrak{d}}_{13}}{|\triangle^{(2)}_2|}&\frac{\bar{\mathfrak{d}}_{23}}{|\triangle^{(2)}_2|}\\
0&0&0&\frac{|\triangle^{(2)}_4|}{|\triangle^{(2)}_3|}&\frac{\bar{\mathfrak{f}}_{23}}{|\triangle^{(2)}_3|}\\
0&0&0&0&\frac{|\triangle^{(2)}_5|}{|\triangle^{(2)}_4|}
\end{pmatrix},
\end{equation}
which leads to the solutions of consistency condition,
\begin{align}
\lambda^{22} =&\frac{1}{|\triangle^{(2)}_1|}\Big(-\{\bar{\sigma}_{22},\mathcal{H}_c\}-\lambda^{33}\mathfrak{A}_{2233}-\lambda^{12}\mathfrak{A}_{2212}-\lambda^{13}\mathfrak{A}_{2213}-\lambda^{23}\mathfrak{A}_{2223} \Big),\\
\lambda^{33} =& \frac{|\triangle^{(2)}_1|}{|\triangle^{(2)}_2|}\Big(\frac{\mathfrak{A}_{3322}}{|\triangle^{(2)}_1|}\{\bar{\sigma}_{22},\mathcal{H}_c\}-\{\bar{\sigma}_{33},\mathcal{H}_c\}-\lambda^{12}\frac{\bar{\mathfrak{b}}_{12}}{|\triangle^{(2)}_1|}-\lambda^{13}\frac{\bar{\mathfrak{b}}_{13}}{|\triangle^{(2)}_1|}-\lambda^{23}\frac{\bar{\mathfrak{b}}_{23}}{|\triangle^{(2)}_1|}\Big),\\
\lambda^{12} =& \frac{|\triangle^{(2)}_2|}{|\triangle^{(2)}_3|}\Big(-\mathfrak{c}_{12}\{\bar{\sigma}_{22},\mathcal{H}_c\}+\mathfrak{b}_{12}\{\bar{\sigma}_{33},\mathcal{H}_c\}-\{\bar{\sigma}_{12},\mathcal{H}_c\}-\lambda^{13}\frac{\bar{\mathfrak{d}}_{13}}{|\triangle^{(2)}_2|}-\lambda^{23}\frac{\bar{\mathfrak{d}}_{23}}{|\triangle^{(2)}_2|}\Big),\\
\lambda^{13} =& \frac{|\triangle^{(2)}_3|}{|\triangle^{(2)}_4|}\Big((-\mathfrak{c}_{13}+\mathfrak{c}_{12}\mathfrak{d}_{13})\{\bar{\sigma}_{22},\mathcal{H}_c\}+(\mathfrak{b}_{13}-\mathfrak{b}_{12}\mathfrak{d}_{13})\{\bar{\sigma}_{33},\mathcal{H}_c\}+\mathfrak{d}_{13}\{\bar{\sigma}_{12},\mathcal{H}_c\}-\{\bar{\sigma}_{13},\mathcal{H}_c\}-\lambda^{23}\frac{\bar{\mathfrak{f}}_{23}}{|\triangle^{(2)}_3|} \Big),\\
\lambda^{23} =& \frac{|\triangle^{(2)}_4|}{|\triangle^{(2)}_5|}\Big((-\mathfrak{c}_{23}+\mathfrak{c}_{12}\mathfrak{d}_{23}-\mathfrak{c}_{13}\mathfrak{f}_{23}+\mathfrak{c}_{12}\mathfrak{d}_{13}\mathfrak{f}_{23})\{\bar{\sigma}_{22},\mathcal{H}_c\}+(\mathfrak{b}_{23}-\mathfrak{b}_{12}\mathfrak{d}_{23}-\mathfrak{b}_{23}\mathfrak{f}_{23}+\mathfrak{b}_{12}\mathfrak{d}_{13}\mathfrak{f}_{23})\{\bar{\sigma}_{33},\mathcal{H}_c\}\notag\\ &+(\mathfrak{d}_{23}-\mathfrak{d}_{13}\mathfrak{f}_{23})\{\bar{\sigma}_{12},\mathcal{H}_c\}+\mathfrak{f}_{23}\{\bar{\sigma}_{13},\mathcal{H}_c\}-\{\bar{\sigma}_{23}, \mathcal{H}_c\} \Big).
\end{align}
The second-class constraints $\bar{\sigma}_{ij}$ and $\bar{\phi}_{ij}$ form the full set of constraints, yielding a total of $10-\frac{1}{2}(10)=5$ physical degrees of freedom for Type $3$ Newer GR.

\subsection{Type $4$}
The assumptions of Type $4$ Newer General Relativity is $c_1=0$ and $(2c_3+c_5)^2-4c_3(c_1+c_2+c_3+c_4+c_5)=0$. The property $(2c_3+c_5)^2-4c_3(c_1+c_2+c_3+c_4+c_5)=0$ can be satisfied within two cases, whether $c_3$ and $c_5$ equals zero or $c_4=-c_2+\frac{c_5^2}{4c_3}$.
\subsubsection{Case 1}
In this case, we assume that $c_3$ and $c_5$ are nonzero. Therefore $c_4=-c_2+\frac{c_5^2}{4c_3}$,
\begin{align}
\pi_N =&\frac{(4c_2c_3-2c_5c_3-c_5^2)}{c_3N^2}\partial_a N^a+\frac{(4c_3+2c_5)}{N^2}(\dot{\gamma}^{ab} \gamma_{ab}-\gamma_{bc} N^a \partial_a \gamma^{bc})\notag \\&+\frac{(-16c_3^2-16c_3c_5-4c_5^2)}{2c_3N^3}(\dot{N}-N^a \partial_a N), \\
\pi_i =& \frac{c_5^2}{2c_3 N^3}\left(\dot{N}^a \gamma_{ia}-\gamma_{ib} N^a \partial_a N^b \right)-\frac{(4c_2+2c_5)}{N^2}\partial_i N-\frac{c_5^2}{4c_3N}\gamma^{ab}\partial_b \gamma_{ia}+\frac{c_2}{N}\gamma^{ab}\partial_b \gamma_{ia}-\frac{c_5}{N}\gamma^{ab}\partial_i \gamma_{ab}, \\
\pi_{ij} =& \frac{(4c_3+2c_5)}{N^2}(\dot{N} \gamma_{ij}-\gamma_{ij} N^a \partial_a N)+\frac{2c_3}{N}( -\dot{\gamma}^{ab} \gamma_{ab} \gamma_{ij} + \gamma_{bc} \gamma_{ij} N^a \partial_a \gamma^{bc}) \notag \\ &+\frac{c_5}{N} \gamma_{ij}\partial_a N^a +\frac{c_2}{N} (\gamma_{ja} \partial_i N^a + \gamma_{ia} \partial_j N^a).
\end{align}
Notice that $\pi_N$ also contains the term $\dot{\gamma}^{ab}$, by expressing $\dot{\gamma}^{ab}$ in terms of $\pi_N$ and then substituting into the Eq.~(101) yields $6$ primary constraints,
\begin{align}
\bar{\phi}^{(1)}_{ij} =& \pi_{ij} -\frac{c2(\gamma_{ja}\partial_i N^a +\gamma_{ia} \partial_j N^a) + 2c3\gamma_{bc} \gamma_{ij} N^a \partial_a \gamma^{bc} +c_5 \gamma_{ij} \partial_a N^a}{N}\notag \\ &-\frac{(4c_3+2c_5)(\dot{N}\gamma_{ij}+\gamma_{ij} N^a \partial_a N)}{N^2} -  \left(\frac{c_3 \gamma_{ij} N}{2c_3+c_5}\right) [ \pi_N +\frac{2c_5^2 \dot{N}}{c_3 N^3}+\frac{(8c_3+8c_5)\dot{N}}{N^3}\notag \\ &+\frac{(4c_3+2c_5)\gamma_{bc} N^a \partial_a \gamma^{bc}-(4c_2-2c_5)\partial_a N^a-(8c_3+8c_5) N^a \partial_a N}{N^2}+\frac{c_5^2 \partial_a N^a}{c_3 N^2}-\frac{2c_5^2 N^a \partial_a N}{c_3 N^3}],
\end{align}
and the following solutions of $\dot{N}$ and$\dot{N}^k$,
\begin{align}
\dot{N} =& -\frac{N^3}{8(c_3+c_5)}\Big(\pi_N-\frac{8(c_3+c_5)}{N^3}N^a\partial_aN+\frac{2c_5^2}{c_3N^3}(\dot{N}-N^a\partial_aN)-\frac{2(2c_2-c_5)}{N^2}\partial_aN^a\notag \\&+\frac{c_5^2}{c_3N^2}\partial_aN^a-\frac{2(2c_3+c_5)}{N^2}(\dot{\gamma}^{ab}\gamma_{ab}+\gamma^{bc}N^a\partial_a\gamma_{bc}) \Big),\\
\dot{N}^k=&\frac{2c_3N^3}{c_5^2}\gamma^{ik}\Big(\pi_i+\frac{c_5^2}{2c_3N^3}\gamma_{ib}N^a\partial_aN^b+\frac{c_5^2}{4c_3N}\gamma^{ab}\partial_b \gamma_{ia}-\frac{c_2}{N}\gamma^{ab}\partial_b\gamma_{ia}\notag\\&+\frac{c_5}{N}\gamma^{ab}\partial_i\gamma_{ab}+\frac{2(2c_2+c_5)}{N^2}\partial_iN \Big),
\end{align}
Taking assumption on the consistency condition of $\bar{\phi}_{ij}$ yields the secondary constraints,
\begin{align}
\bar{\sigma}^{(1)}_{ij}(\vec{x}) &=\{\bar{\phi}^{(1)}_{ij}(\vec{x}), H_p (\Sigma_t) \}, \notag \\
&= \{\bar{\phi}^{(1)}_{ij}, \mathcal{H}_c \}(\vec{x}),
\end{align}
and its consistency condition becomes,
\begin{align}
\{\bar{\sigma}^{(1)}_{ij} (\vec{a}), H_T(\Sigma_t) \} =& \{\bar{\sigma}^{(1)}_{ij} (\vec{a}), H_c(\Sigma_t) \} +\lambda^{mn}(\vec{a}) \{\sigma^{(1)}_{ij} [\bar{\lambda}], \phi^{(1)}_{mn}[\delta(\vec{z}-\vec{a})] \}\notag \\
&+ \lambda'^{mn} (\vec{a}) \{\sigma^{(1)}_{ij} [\bar{\lambda}], \sigma^{(1)}_{mn}[\delta(\vec{z}-\vec{a})] \}, \notag \\
=& \{ \bar{\sigma}^{(1)}_{ij}, \mathcal{H}_c\}(\vec{a})+\lambda^{mn} \{  \bar{\sigma}^{(1)}_{ij},\bar{\phi}^{(1)}_{mn} \}(\vec{a}) + \lambda'^{mn} \{ \bar{\sigma}^{(1)}_{ij},  \bar{\sigma}^{(1)}_{mn} \}(\vec{a}), \notag \\
=&  \{ \bar{\sigma}^{(1)}_{ij}, \mathcal{H}_c\}(\vec{a})+\lambda^{mn} \{  \bar{\sigma}^{(1)}_{ij},\bar{\phi}^{(1)}_{mn} \}(\vec{a}) \approx 0.
\end{align} 
Let $\{\bar{\sigma}^{(1)}_{ij}(\vec{x}), \bar{\phi}^{(1)}_{mn}(\vec{y}) \}=\mathrm{A}^{(\mathfrak{n})}_{ijmn}\delta^{(3)}(\vec{x}-\vec{y})$ with $ij, mn\in \{11,22,33,12,13,23\}$, the symbol $\mathfrak{n} \in \{1,2\}$ stands for Case $1$ when $\mathfrak{n}=1$, and Case $2$ when $\mathfrak{n}=2$. Consider the following notations,
\begin{align}
\mathrm{b}^{(\mathfrak{n})}_{ij}=&\frac{\begin{vmatrix}
\mathrm{A}^{(\mathfrak{n})}_{1111}&\mathrm{A}^{(\mathfrak{n})}_{1122}\\
\mathrm{A}^{(\mathfrak{n})}_{ij11}&\mathrm{A}^{(\mathfrak{n})}_{ij22}
\end{vmatrix}
}{|\triangle^{(\mathfrak{n}+2)}_2|},\\
\mathrm{c}^{(\mathfrak{n})}_{ij}=&-\frac{\mathrm{A}^{(\mathfrak{n})}_{ij11}}{|\triangle^{(\mathfrak{n}+2)}_1|}+\frac{\mathrm{A}^{(\mathfrak{n})}_{2211}}{|\triangle^{(\mathfrak{n}+2)}_1|}\mathrm{b}^{(\mathfrak{n})}_{ij},\\
\mathrm{d}^{(\mathfrak{n})}_{ij} =&\frac{\begin{vmatrix}
\mathrm{A}^{(\mathfrak{n})}_{1111}&\mathrm{A}^{(\mathfrak{n})}_{1122}&\mathrm{A}^{(\mathfrak{n})}_{1133}\\
\mathrm{A}^{(\mathfrak{n})}_{2211}&\mathrm{A}^{(\mathfrak{n})}_{2222}&\mathrm{A}^{(\mathfrak{n})}_{2233}\\
\mathrm{A}^{(\mathfrak{n})}_{ij11}&\mathrm{A}^{(\mathfrak{n})}_{ij22}&\mathrm{A}^{(\mathfrak{n})}_{ij33}
\end{vmatrix}}{|\triangle^{(\mathfrak{n}+2)}_3|},\\
\mathrm{f}^{(\mathfrak{n})}_{ij}=&\frac{\begin{vmatrix}
\mathrm{A}^{(\mathfrak{n})}_{1111}&\mathrm{A}^{(\mathfrak{n})}_{1122}&\mathrm{A}^{(\mathfrak{n})}_{1133}&\mathrm{A}^{(\mathfrak{n})}_{1112}\\
\mathrm{A}^{(\mathfrak{n})}_{2211}&\mathrm{A}^{(\mathfrak{n})}_{2222}&\mathrm{A}^{(\mathfrak{n})}_{2233}&\mathrm{A}^{(\mathfrak{n})}_{2212}\\
\mathrm{A}^{(\mathfrak{n})}_{3311}&\mathrm{A}^{(\mathfrak{n})}_{3322}&\mathrm{A}^{(\mathfrak{n})}_{3333}&\mathrm{A}^{(\mathfrak{n})}_{3312}\\
\mathrm{A}^{(\mathfrak{n})}_{ij11}&\mathrm{A}^{(\mathfrak{n})}_{ij22}&\mathrm{A}^{(\mathfrak{n})}_{ij33}&\mathrm{A}^{(\mathfrak{n})}_{ij12}
\end{vmatrix}}{|\triangle^{(\mathfrak{n}+2)}_4|},\\
\mathrm{g}^{(\mathfrak{n})}_{ij}=&\frac{\begin{vmatrix}
\mathrm{A}^{(\mathfrak{n})}_{1111}&\mathrm{A}^{(\mathfrak{n})}_{1122}&\mathrm{A}^{(\mathfrak{n})}_{1133}&\mathrm{A}^{(\mathfrak{n})}_{1112}&\mathrm{A}^{(\mathfrak{n})}_{1113}\\
\mathrm{A}^{(\mathfrak{n})}_{2211}&\mathrm{A}^{(\mathfrak{n})}_{2222}&\mathrm{A}^{(\mathfrak{n})}_{2233}&\mathrm{A}^{(\mathfrak{n})}_{2212}&\mathrm{A}^{(\mathfrak{n})}_{2213}\\
\mathrm{A}^{(\mathfrak{n})}_{3311}&\mathrm{A}^{(\mathfrak{n})}_{3322}&\mathrm{A}^{(\mathfrak{n})}_{3333}&\mathrm{A}^{(\mathfrak{n})}_{3312}&\mathrm{A}^{(\mathfrak{n})}_{3313}\\
\mathrm{A}^{(\mathfrak{n})}_{1211}&\mathrm{A}^{(\mathfrak{n})}_{1222}&\mathrm{A}^{(\mathfrak{n})}_{1233}&\mathrm{A}^{(\mathfrak{n})}_{1212}&\mathrm{A}^{(\mathfrak{n})}_{1213}\\
\mathrm{A}^{(\mathfrak{n})}_{ij11}&\mathrm{A}^{(\mathfrak{n})}_{ij22}&\mathrm{A}^{(\mathfrak{n})}_{ij33}&\mathrm{A}^{(\mathfrak{n})}_{ij12}&\mathrm{A}^{(\mathfrak{n})}_{ij13}
\end{vmatrix}}{|\triangle^{(\mathfrak{n}+2)}_5|},\\
\bar{\mathrm{b}}^{(\mathfrak{n})}_{ij}=&\begin{vmatrix}
\mathrm{A}^{(\mathfrak{n})}_{1111}&\mathrm{A}^{(\mathfrak{n})}_{11ij}\\
\mathrm{A}^{(\mathfrak{n})}_{2211}&\mathrm{A}^{(\mathfrak{n})}_{22ij}
\end{vmatrix},\\
\bar{\mathrm{d}}^{(\mathfrak{n})}_{ij}=&\begin{vmatrix}
\mathrm{A}^{(\mathfrak{n})}_{1111}&\mathrm{A}^{(\mathfrak{n})}_{1122}&\mathrm{A}^{(\mathfrak{n})}_{11ij}\\
\mathrm{A}^{(\mathfrak{n})}_{2211}&\mathrm{A}^{(\mathfrak{n})}_{2222}&\mathrm{A}^{(\mathfrak{n})}_{22ij}\\
\mathrm{A}^{(\mathfrak{n})}_{3311}&\mathrm{A}^{(\mathfrak{n})}_{3322}&\mathrm{A}^{(\mathfrak{n})}_{33ij}
\end{vmatrix},\\
\bar{\mathrm{f}}^{(\mathfrak{n})}_{ij}=&\begin{vmatrix}
\mathrm{A}^{(\mathfrak{n})}_{1111}&\mathrm{A}^{(\mathfrak{n})}_{1122}&\mathrm{A}^{(\mathfrak{n})}_{1133}&\mathrm{A}^{(\mathfrak{n})}_{11ij}\\
\mathrm{A}^{(\mathfrak{n})}_{2211}&\mathrm{A}^{(\mathfrak{n})}_{2222}&\mathrm{A}^{(\mathfrak{n})}_{2233}&\mathrm{A}^{(\mathfrak{n})}_{22ij}\\
\mathrm{A}^{(\mathfrak{n})}_{3311}&\mathrm{A}^{(\mathfrak{n})}_{3322}&\mathrm{A}^{(\mathfrak{n})}_{3333}&\mathrm{A}^{(\mathfrak{n})}_{33ij}\\
\mathrm{A}^{(\mathfrak{n})}_{1211}&\mathrm{A}^{(\mathfrak{n})}_{1222}&\mathrm{A}^{(\mathfrak{n})}_{1233}&\mathrm{A}^{(\mathfrak{n})}_{12ij}
\end{vmatrix},\\
\bar{\mathrm{g}}^{(\mathfrak{n})}_{ij}=&\begin{vmatrix}
\mathrm{A}^{(\mathfrak{n})}_{1111}&\mathrm{A}^{(\mathfrak{n})}_{1122}&\mathrm{A}^{(\mathfrak{n})}_{1133}&\mathrm{A}^{(\mathfrak{n})}_{1112}&\mathrm{A}^{(\mathfrak{n})}_{11ij}\\
\mathrm{A}^{(\mathfrak{n})}_{2211}&\mathrm{A}^{(\mathfrak{n})}_{2222}&\mathrm{A}^{(\mathfrak{n})}_{2233}&\mathrm{A}^{(\mathfrak{n})}_{2212}&\mathrm{A}^{(\mathfrak{n})}_{22ij}\\
\mathrm{A}^{(\mathfrak{n})}_{3311}&\mathrm{A}^{(\mathfrak{n})}_{3322}&\mathrm{A}^{(\mathfrak{n})}_{3333}&\mathrm{A}^{(\mathfrak{n})}_{3312}&\mathrm{A}^{(\mathfrak{n})}_{33ij}\\
\mathrm{A}^{(\mathfrak{n})}_{1211}&\mathrm{A}^{(\mathfrak{n})}_{1222}&\mathrm{A}^{(\mathfrak{n})}_{1233}&\mathrm{A}^{(\mathfrak{n})}_{1212}&\mathrm{A}^{(\mathfrak{n})}_{12ij}\\
\mathrm{A}^{(\mathfrak{n})}_{1311}&\mathrm{A}^{(\mathfrak{n})}_{1322}&\mathrm{A}^{(\mathfrak{n})}_{1333}&\mathrm{A}^{(\mathfrak{n})}_{1312}&\mathrm{A}^{(\mathfrak{n})}_{13ij}
\end{vmatrix},\\
\mathrm{h}^{(\mathfrak{n})}_{ij}=&\mathrm{c}^{(\mathfrak{n})}_{ij}-\mathrm{c}^{(\mathfrak{n})}_{33}\mathrm{d}^{(\mathfrak{n})}_{ij}-\mathrm{c}^{(\mathfrak{n})}_{12}\mathrm{f}^{(\mathfrak{n})}_{ij}+\mathrm{c}^{(\mathfrak{n})}_{33}\mathrm{d}^{(\mathfrak{n})}_{12}\mathrm{f}^{(\mathfrak{n})}_{ij},\\
\mathrm{k}^{(\mathfrak{n})}_{ij}=&\mathrm{b}^{(\mathfrak{n})}_{ij}-\mathrm{b}^{(\mathfrak{n})}_{33}\mathrm{d}^{(\mathfrak{n})}_{ij}-\mathrm{b}^{(\mathfrak{n})}_{12}\mathrm{f}^{(\mathfrak{n})}_{ij}+\mathrm{b}^{(\mathfrak{n})}_{33}\mathrm{d}^{(\mathfrak{n})}_{12}\mathrm{f}^{(\mathfrak{n})}_{ij}.
\end{align}
The consistency conditions in matrix form is as follows,
\begin{equation}
\begin{pmatrix}
 \{ \bar{\sigma}^{(1)}_{11}, \mathcal{H}_c\}\\
 \{ \bar{\sigma}^{(1)}_{22}, \mathcal{H}_c\}\\
 \{ \bar{\sigma}^{(1)}_{33}, \mathcal{H}_c\}\\
 \{ \bar{\sigma}^{(1)}_{12}, \mathcal{H}_c\}\\
 \{ \bar{\sigma}^{(1)}_{13}, \mathcal{H}_c\}\\
 \{ \bar{\sigma}^{(1)}_{23}, \mathcal{H}_c\}
\end{pmatrix}
+
\begin{pmatrix}
\mathrm{A}^{(\mathfrak{n})}_{1111}&\mathrm{A}^{(\mathfrak{n})}_{1122}&\mathrm{A}^{(\mathfrak{n})}_{1133}&\mathrm{A}^{(\mathfrak{n})}_{1112}&\mathrm{A}^{(\mathfrak{n})}_{1113}&\mathrm{A}^{(\mathfrak{n})}_{1123}\\
\mathrm{A}^{(\mathfrak{n})}_{2211}&\mathrm{A}^{(\mathfrak{n})}_{2222}&\mathrm{A}^{(\mathfrak{n})}_{2233}&\mathrm{A}^{(\mathfrak{n})}_{2212}&\mathrm{A}^{(\mathfrak{n})}_{2213}&\mathrm{A}^{(\mathfrak{n})}_{2223}\\
\mathrm{A}^{(\mathfrak{n})}_{3311}&\mathrm{A}^{(\mathfrak{n})}_{3322}&\mathrm{A}^{(\mathfrak{n})}_{3333}&\mathrm{A}^{(\mathfrak{n})}_{3312}&\mathrm{A}^{(\mathfrak{n})}_{3313}&\mathrm{A}^{(\mathfrak{n})}_{3323}\\
\mathrm{A}^{(\mathfrak{n})}_{1211}&\mathrm{A}^{(\mathfrak{n})}_{1222}&\mathrm{A}^{(\mathfrak{n})}_{1233}&\mathrm{A}^{(\mathfrak{n})}_{1212}&\mathrm{A}^{(\mathfrak{n})}_{1213}&\mathrm{A}^{(\mathfrak{n})}_{1223}\\
\mathrm{A}^{(\mathfrak{n})}_{1311}&\mathrm{A}^{(\mathfrak{n})}_{1322}&\mathrm{A}^{(\mathfrak{n})}_{1333}&\mathrm{A}^{(\mathfrak{n})}_{1312}&\mathrm{A}^{(\mathfrak{n})}_{1313}&\mathrm{A}^{(\mathfrak{n})}_{1323}\\
\mathrm{A}^{(\mathfrak{n})}_{2311}&\mathrm{A}^{(\mathfrak{n})}_{2322}&\mathrm{A}^{(\mathfrak{n})}_{2333}&\mathrm{A}^{(\mathfrak{n})}_{2312}&\mathrm{A}^{(\mathfrak{n})}_{2313}&\mathrm{A}^{(\mathfrak{n})}_{2323}
\end{pmatrix}
\begin{pmatrix}
\lambda^{11}\\
\lambda^{22}\\
\lambda^{33}\\
\lambda^{12}\\
\lambda^{13}\\
\lambda^{23}
\end{pmatrix}
\approx
\begin{pmatrix}
0\\
0\\
0\\
0\\
0\\
0
\end{pmatrix}.
\end{equation}
The matrix of the possion brackets among constriants, $D^{(\mathfrak{n}+2)}$ have following LU decomposition, 
\begin{equation}
D^{(\mathfrak{n}+2)}=\begin{pmatrix}
1&0&0&0&0&0\\
\frac{\mathrm{A}^{(\mathfrak{n})}_{2211}}{|\triangle^{(\mathfrak{n}+2)}_1|}&1&0&0&0&0\\
\frac{\mathrm{A}^{(\mathfrak{n})}_{3311}}{|\triangle^{(\mathfrak{n}+2)}_1|}&\mathrm{b}^{(\mathfrak{n})}_{33}&1&0&0&0\\
\frac{\mathrm{A}^{(\mathfrak{n})}_{1211}}{|\triangle^{(\mathfrak{n}+2)}_1|}&\mathrm{b}^{(\mathfrak{n})}_{12}&\mathrm{d}^{(\mathfrak{n})}_{12}&1&0&0\\
\frac{\mathrm{A}^{(\mathfrak{n})}_{1311}}{|\triangle^{(\mathfrak{n}+2)}_1|}&\mathrm{b}^{(\mathfrak{n})}_{13}&\mathrm{d}^{(\mathfrak{n})}_{13}&\mathrm{f}^{(\mathfrak{n})}_{13}&1&0\\
\frac{\mathrm{A}^{(\mathfrak{n})}_{2311}}{|\triangle^{(\mathfrak{n}+2)}_1|}&\mathrm{b}^{(\mathfrak{n})}_{23}&\mathrm{d}^{(\mathfrak{n})}_{23}&\mathrm{f}^{(\mathfrak{n})}_{23}&\mathrm{g}^{(\mathfrak{n})}_{23}&1
\end{pmatrix}
\begin{pmatrix}
|\triangle^{(\mathfrak{n}+2)}_1|&\mathrm{A}^{(\mathfrak{n})}_{1122}&\mathrm{A}^{(\mathfrak{n})}_{1133}&\mathrm{A}^{(\mathfrak{n})}_{1112}&\mathrm{A}^{(\mathfrak{n})}_{1113}&\mathrm{A}^{(\mathfrak{n})}_{1123}\\
0&\frac{|\triangle^{(\mathfrak{n}+2)}_2|}{|\triangle^{(\mathfrak{n}+2)}_1|}&\frac{\bar{\mathrm{b}}^{(\mathfrak{n})}_{33}}{|\triangle^{(\mathfrak{n}+2)}_1|}&\frac{\bar{\mathrm{b}}^{(\mathfrak{n})}_{12}}{|\triangle^{(\mathfrak{n}+2)}_1|}&\frac{\bar{\mathrm{b}}^{(\mathfrak{n})}_{13}}{|\triangle^{(\mathfrak{n}+2)}_1|}&\frac{\bar{\mathrm{b}}^{(\mathfrak{n})}_{23}}{|\triangle^{(\mathfrak{n}+2)}_1|}\\
0&0&\frac{|\triangle^{(\mathfrak{n}+2)}_3|}{|\triangle^{(\mathfrak{n}+2)}_2|}&\frac{\bar{\mathrm{d}}^{(\mathfrak{n})}_{12}}{|\triangle^{(\mathfrak{n}+2)}_2|}&\frac{\bar{\mathrm{d}}^{(\mathfrak{n})}_{13}}{|\triangle^{(\mathfrak{n}+2)}_2|}&\frac{\bar{\mathrm{d}}^{(\mathfrak{n})}_{23}}{|\triangle^{(\mathfrak{n}+2)}_2|}\\
0&0&0&\frac{|\triangle^{(\mathfrak{n}+2)}_4|}{|\triangle^{(\mathfrak{n}+2)}_3|}&\frac{\bar{\mathrm{f}}^{(\mathfrak{n})}_{13}}{|\triangle^{(\mathfrak{n}+2)}_3|}&\frac{\bar{\mathrm{f}}^{(\mathfrak{n})}_{23}}{|\triangle^{(\mathfrak{n}+2)}_3|}\\
0&0&0&0&\frac{|\triangle^{(\mathfrak{n}+2)}_5|}{|\triangle^{(\mathfrak{n}+2)}_4|}&\frac{\bar{\mathrm{g}}^{(\mathfrak{n})}_{23}}{|\triangle^{(\mathfrak{n}+2)}_4|}\\
0&0&0&0&0&\frac{|\triangle^{(\mathfrak{n}+2)}_6|}{|\triangle^{(\mathfrak{n}+2)}_5|}
\end{pmatrix},
\end{equation}
which yields the following solutions of lagrange multipliers in Case $1$,
\begin{align}
\lambda^{11} =&\frac{1}{|\triangle^{(3)}_1|}\Big(-\{\bar{\sigma}^{(1)}_{11},\mathcal{H}_c\}-\lambda^{22}\mathrm{A}^{(1)}_{1122}-\lambda^{33}\mathrm{A}^{(1)}_{1133}-\lambda^{12}\mathrm{A}^{(1)}_{1112}-\lambda^{13}\mathrm{A}^{(1)}_{1113}-\lambda^{23}\mathrm{A}^{(1)}_{1123} \Big),\\
\lambda^{22} =& \frac{|\triangle^{(3)}_1|}{|\triangle^{(3)}_2|}\Big(\frac{\mathrm{A}^{(1)}_{2211}}{|\triangle^{(3)}_1|}\{\bar{\sigma}^{(1)}_{11},\mathcal{H}_c\}-\{\bar{\sigma}^{(1)}_{22},\mathcal{H}_c\}-\lambda^{33}\frac{\bar{\mathrm{b}}^{(1)}_{33}}{|\triangle^{(3)}_1|}-\lambda^{12}\frac{\bar{\mathrm{b}}^{(1)}_{12}}{|\triangle^{(3)}_1|}-\lambda^{13}\frac{\bar{\mathrm{b}}^{(1)}_{13}}{|\triangle^{(3)}_1|}-\lambda^{23}\frac{\bar{\mathrm{b}}^{(1)}_{23}}{|\triangle^{(3)}_1|}\Big),\\
\lambda^{33} =& \frac{|\triangle^{(3)}_2|}{|\triangle^{(3)}_3|}\Big(-\mathrm{c}^{(1)}_{33}\{\bar{\sigma}^{(1)}_{11},\mathcal{H}_c\}+\mathrm{b}^{(1)}_{33}\{\bar{\sigma}^{(1)}_{22},\mathcal{H}_c\}-\{\bar{\sigma}^{(1)}_{33},\mathcal{H}_c\}-\lambda^{12}\frac{\bar{\mathrm{d}}^{(1)}_{12}}{|\triangle^{(3)}_2|}-\lambda^{13}\frac{\bar{\mathrm{d}}^{(1)}_{13}}{|\triangle^{(3)}_2|}-\lambda^{23}\frac{\bar{\mathrm{d}}^{(1)}_{23}}{|\triangle^{(3)}_2|}\Big),\\
\lambda^{12} =& \frac{|\triangle^{(3)}_3|}{|\triangle^{(3)}_4|}\Big((-\mathrm{c}^{(1)}_{12}+\mathrm{c}^{(1)}_{33}\mathrm{d}^{(1)}_{12})\{\bar{\sigma}^{(1)}_{11},\mathcal{H}_c\}+(\mathrm{b}^{(1)}_{12}-\mathrm{b}^{(1)}_{33}\mathrm{d}^{(1)}_{12})\{\bar{\sigma}^{(1)}_{22},\mathcal{H}_c\}+\mathrm{d}^{(1)}_{12}\{\bar{\sigma}^{(1)}_{33},\mathcal{H}_c\}\notag\\ &-\{\bar{\sigma}^{(1)}_{12},\mathcal{H}_c\}-\lambda^{13}\frac{\bar{\mathrm{f}}^{(1)}_{13}}{|\triangle^{(3)}_3|}-\lambda^{23}\frac{\bar{\mathrm{f}}^{(1)}_{23}}{|\triangle^{(3)}_3|} \Big),\\
\lambda^{13} =& \frac{|\triangle^{(3)}_4|}{|\triangle^{(3)}_5|}\Big(-\mathrm{h}^{(1)}_{13}\{\bar{\sigma}^{(1)}_{11},\mathcal{H}_c\}+\mathrm{k}^{(1)}_{13}\{\bar{\sigma}^{(1)}_{22}, \mathcal{H}_c\}+(\mathrm{d}^{(1)}_{13}-\mathrm{d}^{(1)}_{12}\mathrm{f}^{(1)}_{13})\{\bar{\sigma}^{(1)}_{33},\mathcal{H}_c\}+\mathrm{f}^{(1)}_{13}\{\bar{\sigma}^{(1)}_{12},\mathcal{H}_c\}\notag \\ &-\{\bar{\sigma}^{(1)}_{13},\mathcal{H}_c\}-\lambda^{23}\frac{\bar{\mathrm{g}}^{(1)}_{23}}{|\triangle^{(3)}_4|}\Big),\\
\lambda^{23} =& \frac{|\triangle^{(3)}_5|}{|\triangle^{(3)}_6|}\Big(-(\mathrm{h}^{(1)}_{23}-\mathrm{h}^{(1)}_{13}\mathrm{g}^{(1)}_{23})\{\bar{\sigma}^{(1)}_{11},\mathcal{H}_c\}+(\mathrm{k}^{(1)}_{23}-\mathrm{k}^{(1)}_{13}\mathrm{g}^{(1)}_{23})\{\bar{\sigma}^{(1)}_{22}, \mathcal{H}_c\}\notag \\ &+(\mathrm{d}^{(1)}_{23}-\mathrm{d}^{(1)}_{12}\mathrm{f}^{(1)}_{23}-\mathrm{d}^{(1)}_{13}\mathrm{g}^{(1)}_{23}+\mathrm{d}^{(1)}_{12}\mathrm{f}^{(1)}_{13}\mathrm{g}^{(1)}_{23})\{\bar{\sigma}^{(1)}_{33},\mathcal{H}_c\}+(\mathrm{f}^{(1)}_{23}-\mathrm{f}^{(1)}_{13}\mathrm{g}^{(1)}_{23})\{\bar{\sigma}^{(1)}_{12},\mathcal{H}_c\}\notag \\&+\mathrm{g}^{(1)}_{23}\{\bar{\sigma}^{(1)}_{13},\mathcal{H}_c\}-\{\bar{\sigma}^{(1)}_{23},\mathcal{H}_c\}\Big).
\end{align}
The full set of constraints here contains only second-class constraints $\bar{\sigma}^{(1)}_{ij}$ and $\bar{\phi}^{(1)}_{ij}$, then Type $4$ Newer General Relativity have a total of $10-\frac{1}{2}(12)=4$ physical degrees of freedom when $c_4 = -c_2+\frac{c_5^2}{4c_3}$, $c_3$ and $c_5$ are nonzero.

\subsubsection{Case $2$}
If $c_1$, $c_3$ and $c_5$ are zero, the conjugate momenta are given as below:
\begin{align}
\pi_N =&\frac{(-4c_4)}{N^2}\partial_a N^a+\frac{(-8c_2-8c_4)}{N^3}(\dot{N}-N^a \partial_a N), \\
\pi_i =& \frac{(2c_2+2c_4)}{N^3}\left(\dot{N}^a \gamma_{ia}-\gamma_{ib} N^a \partial_a N^b \right)-\frac{4c_2}{N^2}\partial_i N-\frac{c_4}{N}\gamma^{ab}\partial_b \gamma_{ia}, \\
\pi_{ij} =& \frac{c_2}{N} (\gamma_{ja} \partial_i N^a + \gamma_{ia} \partial_j N^a).
\end{align}
By using the conjugate momenta, construct the primary constraints of Type $4$ Case $2$ Newer General Relativity,
\begin{align}
\bar{\phi}^{(2)}_{ij} =& \pi_{ij} -\frac{c_2 (\gamma_{ja} \partial_i N^a + \gamma_{ia} \partial_j N^a)}{N},
\end{align}
and the solutions of velocities,
\begin{align}
\dot{N}=&-\frac{N^3}{8(c_2+c_4)}\Big(\pi_N-\frac{8(c_2+c_4)}{N^3}N^a\partial_aN+\frac{4c_4}{N^2}\partial_aN^a\Big),\\
\dot{N}^k=&\frac{N^3}{2(c_2+c_4)}\gamma^{ik}\Big(\pi_i+\frac{2(c_2+c_4)}{N^3}\gamma_{ib}N^a\partial_aN^b+\frac{c_4}{N}\gamma^{ab}\partial_b\gamma_{ia}+\frac{4c_2}{N^2}\partial_iN \Big),
\end{align}
which have the consistency conditions,
\begin{align}
\bar{\sigma}^{(2)}_{ij}(\vec{x}) &=\{\bar{\phi}^{(2)}_{ij}(\vec{x}), H_p (\Sigma_t) \}, \notag \\
&=  \{\bar{\phi}^{(2)}_{ij}, \mathcal{H}_c \}(\vec{x}),
\end{align}
and yields the secondary constraints $\bar{\sigma}^{(2)}_{ij} (\vec{x})$. Similarly, the consistency conditions of secondary constraints are expressed as below,
\begin{align}
\{\bar{\sigma}^{(2)}_{ij} (\vec{a}), H_T(\Sigma_t)\} =& \{\bar{\sigma}^{(2)}_{ij}(\vec{a}), H_c(\Sigma_t) \} +\lambda^{mn}(\vec{a}) \{\sigma^{(2)}_{ij} [\bar{\lambda}], \phi^{(2)}_{mn}[\delta(\vec{z}-\vec{a})] \}\notag \\
&+ \lambda'^{mn} (\vec{a}) \{\sigma^{(2)}_{ij} [\bar{\lambda}], \sigma^{(2)}_{mn}[\delta(\vec{z}-\vec{a})] \},\notag \\
=&  \{ \bar{\sigma}^{(2)}_{ij}, \mathcal{H}_c\}(\vec{a})+\lambda^{mn} \{  \bar{\sigma}^{(2)}_{ij},\bar{\phi}^{(2)}_{mn} \}(\vec{a}) \approx 0.
\end{align} 
The process of solving the consistency conditions for Case $2$ is similar to that for Case $1$. The solutions for the Lagrange multipliers are listed below,
\begin{align}
\lambda^{11} =&\frac{1}{|\triangle_1|}\Big(-\{\bar{\sigma}^{(2)}_{11},\mathcal{H}_c\}-\lambda^{22}\mathrm{A}^{(2)}_{1122}-\lambda^{33}\mathrm{A}^{(2)}_{1133}-\lambda^{12}\mathrm{A}^{(2)}_{1112}-\lambda^{13}\mathrm{A}^{(1)}_{1113}-\lambda^{23}\mathrm{A}^{(2)}_{1123} \Big),\\
\lambda^{22} =& \frac{|\triangle_1|}{|\triangle_2|}\Big(\frac{\mathrm{A}^{(2)}_{2211}}{|\triangle_1|}\{\bar{\sigma}^{(2)}_{11},\mathcal{H}_c\}-\{\bar{\sigma}^{(2)}_{22},\mathcal{H}_c\}-\lambda^{33}\frac{\bar{\mathrm{b}}^{(2)}_{33}}{|\triangle_1|}-\lambda^{12}\frac{\bar{\mathrm{b}}^{(2)}_{12}}{|\triangle_1|}-\lambda^{13}\frac{\bar{\mathrm{b}}^{(2)}_{13}}{|\triangle_1|}-\lambda^{23}\frac{\bar{\mathrm{b}}^{(2)}_{23}}{|\triangle_1|}\Big),\\
\lambda^{33} =& \frac{|\triangle_2|}{|\triangle_3|}\Big(-\mathrm{c}^{(2)}_{33}\{\bar{\sigma}^{(2)}_{11},\mathcal{H}_c\}+\mathrm{b}^{(2)}_{33}\{\bar{\sigma}^{(2)}_{22},\mathcal{H}_c\}-\{\bar{\sigma}^{(2)}_{33},\mathcal{H}_c\}-\lambda^{12}\frac{\bar{\mathrm{d}}^{(2)}_{12}}{|\triangle_2|}-\lambda^{13}\frac{\bar{\mathrm{d}}^{(2)}_{13}}{|\triangle_2|}-\lambda^{23}\frac{\bar{\mathrm{d}}^{(2)}_{23}}{|\triangle_2|}\Big),
\end{align}
\begin{align}
\lambda^{12} =& \frac{|\triangle_3|}{|\triangle_4|}\Big((-\mathrm{c}^{(2)}_{12}+\mathrm{c}^{(2)}_{33}\mathrm{d}^{(2)}_{12})\{\bar{\sigma}^{(2)}_{11},\mathcal{H}_c\}+(\mathrm{b}^{(2)}_{12}-\mathrm{b}^{(2)}_{33}\mathrm{d}^{(2)}_{12})\{\bar{\sigma}^{(2)}_{22},\mathcal{H}_c\}+\mathrm{d}^{(2)}_{12}\{\bar{\sigma}^{(2)}_{33},\mathcal{H}_c\}\notag\\ &-\{\bar{\sigma}^{(2)}_{12},\mathcal{H}_c\}-\lambda^{13}\frac{\bar{\mathrm{f}}^{(2)}_{13}}{|\triangle_3|}-\lambda^{23}\frac{\bar{\mathrm{f}}^{(2)}_{23}}{|\triangle_3|} \Big),\\
\lambda^{13} =& \frac{|\triangle_4|}{|\triangle_5|}\Big(-\mathrm{h}^{(2)}_{13}\{\bar{\sigma}^{(2)}_{11},\mathcal{H}_c\}+\mathrm{k}^{(2)}_{13}\{\bar{\sigma}^{(2)}_{22}, \mathcal{H}_c\}+(\mathrm{d}^{(2)}_{13}-\mathrm{d}^{(2)}_{12}\mathrm{f}^{(2)}_{13})\{\bar{\sigma}^{(2)}_{33},\mathcal{H}_c\}+\mathrm{f}^{(2)}_{13}\{\bar{\sigma}^{(2)}_{12},\mathcal{H}_c\}\notag \\ &-\{\bar{\sigma}^{(2)}_{13},\mathcal{H}_c\}-\lambda^{23}\frac{\bar{\mathrm{g}}^{(2)}_{23}}{|\triangle_4|}\Big),\\
\lambda^{23} =& \frac{|\triangle_5|}{|\triangle_6|}\Big(-(\mathrm{h}^{(2)}_{23}-\mathrm{h}^{(2)}_{13}\mathrm{g}^{(2)}_{23})\{\bar{\sigma}^{(2)}_{11},\mathcal{H}_c\}+(\mathrm{k}^{(2)}_{23}-\mathrm{k}^{(2)}_{13}\mathrm{g}^{(2)}_{23})\{\bar{\sigma}^{(2)}_{22}, \mathcal{H}_c\}\notag \\ &+(\mathrm{d}^{(2)}_{23}-\mathrm{d}^{(2)}_{12}\mathrm{f}^{(2)}_{23}-\mathrm{d}^{(2)}_{13}\mathrm{g}^{(2)}_{23}+\mathrm{d}^{(2)}_{12}\mathrm{f}^{(2)}_{13}\mathrm{g}^{(2)}_{23})\{\bar{\sigma}^{(2)}_{33},\mathcal{H}_c\}+(\mathrm{f}^{(2)}_{23}-\mathrm{f}^{(2)}_{13}\mathrm{g}^{(2)}_{23})\{\bar{\sigma}^{(2)}_{12},\mathcal{H}_c\}\notag \\&+\mathrm{g}^{(2)}_{23}\{\bar{\sigma}^{(2)}_{13},\mathcal{H}_c\}-\{\bar{\sigma}^{(2)}_{23},\mathcal{H}_c\}\Big).
\end{align}
Type $4$ Case $2$ Newer General Relativity has the same physical degrees of freedom as Case $1$, which is a total of $10-\frac{1}{2}(12)=4$.

\subsection{Type $5$}
Type $5$ Newer General Relativity is the intersection between Type $1$ and Type $2$, therefore the only velocities which can be expressed in terms of corresponding conjugate momenta is $\dot{\gamma}^{ij}$, 
\begin{align}
\pi_N =& \frac{(4c_1+4c_2)}{N^2}\partial_a N^a,\\
\pi_i =&\frac{(2c_1+c_2)}{N}\gamma^{ab}\partial_b \gamma_{ia}-\frac{2c_1}{N}\gamma^{ab}\partial_i \gamma_{ab}-\frac{(4c_1+4c_2)}{N^2}\partial_i N, \\
\pi_{ij} =&\frac{2c_1}{N}( \dot{\gamma}^{ab} \gamma_{ab} \gamma_{ij}-\dot{\gamma}^{ab} \gamma_{ia} \gamma_{jb} - \gamma_{bc} \gamma_{ij} N^a \partial_a \gamma^{bc}+\gamma_{ib} \gamma_{jc} N^a \partial_a \gamma^{bc})\notag \\ &+\frac{2c_1}{N} \gamma_{ij}\partial_a N^a +\frac{c_2}{N} (\gamma_{ja} \partial_i N^a + \gamma_{ia} \partial_j N^a),
\end{align}
and yields four primary constraints,
\begin{align}
	\bar{\phi}_N &= \pi_N -  \frac{(4c_1+4c_2)}{N^2}\partial_a N^a\approx 0,\\
	\bar{\phi}_i &= \pi_i -\frac{(2c_1+c_2)}{N}\gamma^{ab}\partial_b \gamma_{ia}+\frac{2c_1}{N}\gamma^{ab}\partial_i \gamma_{ab}+\frac{(4c_1+4c_2)}{N^2}\partial_i N\approx 0,
\end{align}
and the following solution of $\dot{\gamma}^{kl}$,
\begin{align}
\dot{\gamma}^{kl}=&-\frac{N\gamma^{ik}\gamma^{jl}}{2c_1}\Big(\pi_{ij}-\frac{2c_1}{N}(\gamma^{bc}\gamma_{ij}N^a\partial_a \gamma_{bc}-N^a\partial_a\gamma_{ij}+\gamma_{ij}\partial_aN^a)\notag \\&-\frac{c_2}{N}(\gamma_{ja}\partial_i N^a+\gamma_{ia}\partial_j N^a)-\frac{1}{2}\gamma_{ij}\gamma^{mn}\Big[\pi_{mn}-\frac{2c_1}{N}(\gamma^{bc}\gamma_{mn}N^a\partial_a \gamma_{bc}-N^a\partial_a \gamma_{mn}+\gamma_{mn}\partial_a N^a)\notag \\&-\frac{c_2}{N}(\gamma_{na}\partial_mN^a+\gamma_{ma}\partial_nN^a) \Big] \Big).
\end{align}
The nonzero Poisson brackets between the primary constraints are shown below,
\begin{align}
\{\phi_N[\delta],\phi_i[\bar{\lambda}]\} &= \int_{\Sigma_t} d^3z \bar{\lambda}(\vec{z}) \{ \bar{ \phi }_N, \bar{\phi}_i \}(\vec{z}) \delta (\vec{z}-\vec{a}),\notag \\
&= \bar{\lambda}(\vec{a}) \{ \bar{ \phi}_N, \bar{ \phi}_i \}(\vec{a}),
\end{align}
which can be applied to the consistency conditions of $\bar{\phi}_N$,
\begin{align}
\{\bar{\phi}_N (\vec{x}), H_p (\Sigma_t)\} =& \{\bar{\phi}_N, \mathcal{H}_c \}(\vec{x}) + \{\phi_N[\delta(\vec{z}-\vec{x})], \phi_i[\lambda^i] \}, \notag \\
=&  \{\bar{\phi}_N, \mathcal{H}_c \}(\vec{x}) +\lambda^i (\vec{x})\{\bar{\phi}_N, \bar{\phi}_i \}(\vec{x})\approx 0,
\end{align}
and also the consistency conditions of $\bar{\phi}_i$,
\begin{align}
\{\bar{\phi}_i (\vec{x}), H_p (\Sigma_t)\} =& \{\bar{\phi}_i, \mathcal{H}_c \}(\vec{x}) + \{\phi_i[\delta(\vec{z}-\vec{x})], \phi_N[\lambda^N] \},\notag \\
=&  \{\bar{\phi}_i, \mathcal{H}_c \}(\vec{x}) +\lambda^N (\vec{x})\{\bar{\phi}_i, \bar{\phi}_N \}(\vec{x})\approx 0.
\end{align}
The explicit expression for $\{\bar{\phi}_N, \bar{\phi}_i\}$ is shown below,
\begin{align}
\{\bar{\phi}_N (\vec{x}), \bar{\phi}_i (\vec{y})\} =& \frac{1}{N^2}[(-2c_1-c_2)\gamma^{ab}\partial_b \gamma_{ia}+2c_1 \gamma^{ab}\partial_i \gamma_{ab}] \delta^{(3)}(\vec{x}-\vec{y}), \notag \\
=&\mathsf{A}_i \delta^{(3)}(\vec{x}-\vec{y}),
\end{align}
which form the matrix $D^{(5)}$,
\begin{equation}
D^{(5)} =\begin{pmatrix}
0&\mathsf{A}_1&\mathsf{A}_2&\mathsf{A}_3\\
-\mathsf{A}_1&0&0&0\\
-\mathsf{A}_2&0&0&0\\
-\mathsf{A}_3&0&0&0
\end{pmatrix}.
\end{equation}
We may construct matrix $\bar{D}^{(5)}$ which satisfies,
\begin{align}
\bar{D}^{(5)}=& P^{(5)}D^{(5)}[P^{(5)}]^{\top}, \notag \\
=& \begin{pmatrix}
0&0&0&\mathsf{A}_3\\
0&0&0&0\\
0&0&0&0\\
-\mathsf{A}_3&0&0&0
\end{pmatrix},
\end{align}
where the invertible matrix $P^{(5)}$ are, 
\begin{equation}
P^{(5)} = \begin{pmatrix}
1&0&0&0\\
0&1&0&-\frac{\mathsf{A}_1}{\mathsf{A}_3}\\
0&0&1&-\frac{\mathsf{A}_2}{\mathsf{A}_3}\\
0&0&0&1
\end{pmatrix}.
\end{equation}
These matrix decomposition properties can be apply to solve the following consistency conditions of primary constraints,
\begin{align}
 \begin{pmatrix}
\{\bar{\phi}_N, \mathcal{H}_c\}\\
\{\bar{\phi}_1, \mathcal{H}_c\}\\
\{\bar{\phi}_2, \mathcal{H}_c\}\\
\{\bar{\phi}_3, \mathcal{H}_c\}
\end{pmatrix}
+ D^{(5)}\begin{pmatrix}
\lambda^N\\
\lambda^1\\
\lambda^2\\
\lambda^3
\end{pmatrix}
\approx & \begin{pmatrix}
0\\
0\\
0\\
0
\end{pmatrix},\notag \\
 P^{(5)}\begin{pmatrix}
\{\bar{\phi}_N, \mathcal{H}_c\}\\
\{\bar{\phi}_1, \mathcal{H}_c\}\\
\{\bar{\phi}_2, \mathcal{H}_c\}\\
\{\bar{\phi}_3, \mathcal{H}_c\}
\end{pmatrix}
+ \bar{D}^{(5)}[(P^{(5)})^{\top}]^{-1}\begin{pmatrix}
\lambda^N\\
\lambda^1\\
\lambda^2\\
\lambda^3
\end{pmatrix}
\approx & \begin{pmatrix}
0\\
0\\
0\\
0
\end{pmatrix},
\end{align}
which yields the unique solution for $\lambda^N$ and $\lambda^3$,
\begin{align}
\lambda^N=&\frac{1}{\mathsf{A}_3}\{\bar{\phi}_3, \mathcal{H}_c\},\\
\lambda^3=&-\frac{1}{\mathsf{A}_3}\{\bar{\phi}_N, \mathcal{H}_c\}-\lambda^1\frac{\mathsf{A}_1}{\mathsf{A}_3}-\lambda^2\frac{\mathsf{A}_2}{\mathsf{A}_3},
\end{align}
and also the following secondary constraints,
\begin{align}
\bar{\sigma}_1 =& \{\bar{\phi}_1, \mathcal{H}_c\}-\frac{\mathsf{A}_1}{\mathsf{A}_3}\{\bar{\phi}_3, \mathcal{H}_c\} \approx 0,\\
\bar{\sigma}_2 =& \{\bar{\phi}_2, \mathcal{H}_c\}-\frac{\mathsf{A}_2}{\mathsf{A}_3}\{\bar{\phi}_3, \mathcal{H}_c\} \approx 0.
\end{align}
Setting $\{\bar{\sigma}_k, \bar{\phi}_l\}=\mathsf{B}_{kl}\delta^{(3)}(\vec{x}-\vec{y})$, and form the matrix $D^{(6)}$ with LU decomposition,
\begin{align}
D^{(6)}=& \begin{pmatrix}
\mathsf{B}_{11}&\mathsf{B}_{12}\\
\mathsf{B}_{21}&\mathsf{B}_{22}
\end{pmatrix}, \notag \\
=& \begin{pmatrix}
1&0\\
\frac{\mathsf{B}_{21}}{|\triangle^{(6)}_1|}&1
\end{pmatrix} \begin{pmatrix}
|\triangle^{(6)}_1|&\mathsf{B}_{12}\\
0&\frac{|\triangle^{(6)}_2|}{|\triangle^{(6)}_1|}
\end{pmatrix}.
\end{align}
The consistency condtition for the secondary constraints are shown below,
\begin{equation}
\begin{pmatrix}
\{\bar{\sigma}_1, \mathcal{H}^{(2)}_c\}\\
\{\bar{\sigma}_1, \mathcal{H}^{(2)}_c\}
\end{pmatrix}+D^{(6)}
\begin{pmatrix}
\lambda^1_{(2)}\\
\lambda^2_{(2)}
\end{pmatrix}\approx \begin{pmatrix}
0\\
0
\end{pmatrix},
\end{equation}
where $\mathcal{H}^{(2)}_c=\mathcal{H}_c+\lambda^N\bar{\phi}_N+\lambda^i\bar{\phi}_i$. By the decomposition of matrix $D^{(6)}$, we obtained the solutions of lagrange multipliers $\lambda^1_{(2)}$ and $\lambda^2_{(2)}$,
\begin{align}
\lambda^2_{(2)}=& \frac{|\triangle^{(6)}_1|}{|\triangle^{(6)}_2|}\Big(\frac{\mathsf{B}_{21}}{|\triangle^{(6)}_1|}\{\bar{\sigma}_1, \mathcal{H}^{(2)}_c\}-\{\bar{\sigma}_2, \mathcal{H}^{(2)}_c\}\Big),\\
\lambda^1_{(2)}=& \frac{1}{|\triangle^{(6)}_1|}\Big(-\{\bar{\sigma}_1, \mathcal{H}^{(2)}_c\}-\mathsf{B}_{12}\lambda^2_{(2)} \Big).
\end{align}
The Dirac–Bergmann algorithm terminates, and the full set of constraints consists of the second-class constraints $\bar{\phi}_N, \bar{\phi}_i, \bar{\sigma}_1$, and $\bar{\sigma}_2$. Therefore, the physical degrees of freedom of Type $5$ Newer General Relativity are $10-\frac{1}{2}(6)=7$.

\subsection{Type $6$}
The Hamiltonian analysis of Type $6$ Newer General Relativity consider $c_1=0$ and also $2c_1+c_2+c_4$, which is also the intersection between Type $2$ and Type $3$. Its primary constraints are listed below,
\begin{align}
\pi_N =&\frac{(4c_2-2c_5)}{N^2}\partial_a N^a+\frac{(4c_3+2c_5)}{N^2}(\dot{\gamma}^{ab} \gamma_{ab}-\gamma_{bc} N^a \partial_a \gamma^{bc})\notag \\ &+\frac{(-8c_3-8c_5)}{N^3}(\dot{N}-N^a \partial_a N), \\
\pi_i =&\frac{c_2}{N}\gamma^{ab}\partial_b \gamma_{ia}-\frac{c_5}{N}\gamma^{ab}\partial_i \gamma_{ab}-\frac{2(2c_2+c_5)}{N^2}\partial_i N, \\
\pi_{ij} =& \frac{(4c_3+2c_5)}{N^2}(\dot{N} \gamma_{ij}-\gamma_{ij} N^a \partial_a N)+\frac{2c_3}{N}( -\dot{\gamma}^{ab} \gamma_{ab} \gamma_{ij} + \gamma_{bc} \gamma_{ij} N^a \partial_a \gamma^{bc}) \notag \\ &+\frac{c_5}{N}\gamma_{ij}\partial_a N^a +\frac{c_2}{N}(\gamma_{ja} \partial_i N^a + \gamma_{ia} \partial_j N^a),
\end{align}
and this time we have primary constraints from the conjugate momenta $\pi_i$ and $\pi_{ij}$,
\begin{align}
\bar{\phi}_i &= \pi_i-\frac{c_2}{N}\gamma^{ab}\partial_b \gamma_{ia}+\frac{c_5}{N}\gamma^{ab}\partial_i \gamma_{ab}+\frac{2(2c_2+c_5)}{N^2}\partial_i N,\\
\bar{\phi}_{ij} =& \pi_{ij} -\frac{c2(\gamma_{ja}\partial_i N^a +\gamma_{ia} \partial_j N^a) + 2c3\gamma_{bc} \gamma_{ij} N^a \partial_a \gamma^{bc} +c_5 \gamma_{ij} \partial_a N^a}{N}\notag \\ &-\frac{(4c_3+2c_5)(\dot{N}\gamma_{ij}+\gamma_{ij} N^a \partial_a N)}{N^2} - \gamma_{ij} \gamma^{11} \Big[ \pi_{11}+\frac{(4c_3+2c_5)(\dot{N}\gamma_{11}+\gamma_{11} N^a \partial_a N)}{N^2}\notag \\ &+\frac{c2(\gamma_{1a}\partial_1 N^a +\gamma_{1a} \partial_1 N^a) + 2c3\gamma_{bc} \gamma_{11} N^a \partial_a \gamma^{bc} +c_5 \gamma_{11} \partial_a N^a}{N}\Big],
\end{align}
and also the solution of $\dot{N}$,
\begin{equation}
\dot{N}=-\frac{N^3}{8(c_3+c_5)}\Big(\pi_N-\frac{8(c_3+c_5)}{N^3}N^a\partial_aN-\frac{2(2c_2-c_5)}{N^2}\partial_aN^a-\frac{2(2c_3+c_5)}{N^2}(\dot{\gamma}^{ab}\gamma_{ab}+\gamma^{bc}N^a\partial_a\gamma_{bc}) \Big).
\end{equation}
Since we have two types of primary constraints, the nonzero poisson brackets between constraints is calculated as below,
\begin{align}
\{\phi_i[\delta],\phi_{kl}[\bar{\lambda}]\} =& \int_{\Sigma_t} d^3z \bar{\lambda}(\vec{z}) \{ \bar{ \phi }_i, \bar{\phi}_{kl} \}(\vec{z}) \delta (\vec{z}-\vec{a}),\\
=& \bar{\lambda}(\vec{a}) \{ \bar{ \phi}_i, \bar{ \phi}_{kl} \}(\vec{a}),\\
=&\frac{1}{2N}\bar{\lambda}[c_2 \gamma^{ab}(\gamma_{li}\partial_b \gamma_{ka}+\gamma_{ki}\partial_b \gamma_{la})-2c_5\partial_i \gamma_{kl}]+\frac{4c_3+2c_5}{N^2}\bar{\lambda}(\gamma_{kl}\partial_i N - \gamma_{kl}\gamma_{11}\gamma^{11}\partial_i N)\notag \\
&-\frac{2c_3}{N}\bar{\lambda}(\gamma^{ab}\gamma_{kl}\partial_i \gamma_{ab} + \gamma^{ab}\gamma_{kl}\gamma_{11}\gamma^{11}\partial_i \gamma_{ab}),
\end{align}
which is useful to investigate the consistency conditions,
\begin{align}
\{\bar{\phi}_i (\vec{x}), H_p (\Sigma_t)\} =&  \{\bar{\phi}_i, \mathcal{H}_c \}(\vec{x}) +\lambda^{kl} (\vec{x})\{\bar{\phi}_i, \bar{\phi}_{kl} \}(\vec{x})\approx 0,\\
\{\bar{\phi}_{kl} (\vec{x}), H_p (\Sigma_t)\} =&  \{\bar{\phi}_{kl}, \mathcal{H}_c \}(\vec{x}) +\lambda^{i} (\vec{x})\{\bar{\phi}_{kl}, \bar{\phi}_{i} \}(\vec{x})\approx 0.
\end{align}
For Type $6$, Type $7$ and Type $8$, we use the notation $\{\bar{\phi}_i (\vec{x}), \bar{\phi}_{kl}(\vec{y})\}=\mathtt{A}^{(\mathtt{N})}_{ikl}\delta^{(3)(\vec{x}-\vec{y})}$, where $\mathtt{N}\in \{6, 7, 8\}$. $\mathtt{A}^{(\mathtt{N})}_{ikl}$ represents the poisson brackets between $\bar{\phi}_i $ and $\bar{\phi}_{kl}$ for Type $\mathtt{N}$. The poisson brackets form the following matrix $D^{(7)}$,
\begin{equation}
D^{(7)} = \begin{pmatrix}
0&0&0&\mathtt{A}^{(6)}_{122}&\mathtt{A}^{(6)}_{133}&\mathtt{A}^{(6)}_{112}&\mathtt{A}^{(6)}_{113}&\mathtt{A}^{(6)}_{123}\\
0&0&0&\mathtt{A}^{(6)}_{222}&\mathtt{A}^{(6)}_{233}&\mathtt{A}^{(6)}_{212}&\mathtt{A}^{(6)}_{213}&\mathtt{A}^{(6)}_{223}\\
0&0&0&\mathtt{A}^{(6)}_{322}&\mathtt{A}^{(6)}_{333}&\mathtt{A}^{(6)}_{312}&\mathtt{A}^{(6)}_{313}&\mathtt{A}^{(6)}_{323}\\
-\mathtt{A}^{(6)}_{122}&-\mathtt{A}^{(6)}_{222}&-\mathtt{A}^{(6)}_{322}&0&0&0&0&0\\
-\mathtt{A}^{(6)}_{133}&-\mathtt{A}^{(6)}_{233}&-\mathtt{A}^{(6)}_{333}&0&0&0&0&0\\
-\mathtt{A}^{(6)}_{112}&-\mathtt{A}^{(6)}_{212}&-\mathtt{A}^{(6)}_{312}&0&0&0&0&0\\
-\mathtt{A}^{(6)}_{113}&-\mathtt{A}^{(6)}_{213}&-\mathtt{A}^{(6)}_{313}&0&0&0&0&0\\
-\mathtt{A}^{(6)}_{123}&-\mathtt{A}^{(6)}_{223}&-\mathtt{A}^{(6)}_{323}&0&0&0&0&0
\end{pmatrix},
\end{equation}
which can be used to construct the matrix $\bar{D}^{(7)}=P^{(6)}D^{(7)}[P^{(6)}]^{\top}$,
\begin{align}
\bar{D}^{(7)} =& \begin{pmatrix}
0&0&0&0&0&0&0&\mathtt{A}^{(6)}_{123}\\
0&0&0&0&0&0&-\frac{\mathtt{C}^{(6)}_{213}}{\mathtt{A}^{(6)}_{123}}&0\\
0&0&0&0&0&\frac{\mathtt{d}^{(6)}_{12}}{\mathtt{C}^{(6)}_{213}}&0&0\\
0&0&0&0&0&0&0&0\\
0&0&0&0&0&0&0&0\\
0&0&-\frac{\mathtt{d}^{(6)}_{12}}{\mathtt{C}^{(6)}_{213}}&0&0&0&0&0\\
0&\frac{\mathtt{C}^{(6)}_{213}}{\mathtt{A}^{(6)}_{123}}&0&0&0&0&0&0\\
-\mathtt{A}^{(6)}_{123}&0&0&0&0&0&0&0
\end{pmatrix},\\
P^{(6)}=&\begin{pmatrix}
1&0&0&0&0&0&0&0\\
0&1&0&0&0&0&0&0\\
0&0&1&0&0&0&0&0\\
0&0&0&1&0&-\frac{\mathtt{d}^{(6)}_{22}}{\mathtt{d}^{(6)}_{12}}&\mathtt{m}^{(6)}_{22}&\mathtt{n}^{(6)}_{22}\\
0&0&0&0&1&-\frac{\mathtt{d}^{(6)}_{33}}{\mathtt{d}^{(6)}_{12}}&\mathtt{m}^{(6)}_{33}&\mathtt{n}^{(6)}_{33}\\
0&0&0&0&0&1&-\frac{\mathtt{C}^{(6)}_{212}}{\mathtt{C}^{(6)}_{213}}&\mathtt{k}^{(6)}_{12}\\
0&0&0&0&0&\mathtt{x}^{(6)}_{13}&\mathtt{y}^{(6)}_{13}&\mathtt{z}^{(6)}_{13}\\
0&0&0&0&0&\mathtt{x}^{(6)}_{23}&\mathtt{y}^{(6)}_{23}&\mathtt{z}^{(6)}_{23}
\end{pmatrix},
\end{align}
where,
\begin{align}
\mathtt{C}^{(\mathtt{N})}_{2jk} =& \begin{vmatrix}
\mathtt{A}^{(N)}_{1jk}&\mathtt{A}^{(N)}_{2jk}\\
\mathtt{A}^{(N)}_{123}&\mathtt{A}^{(N)}_{223}
\end{vmatrix},\\
\mathtt{C}^{(\mathtt{N})}_{3jk} =& \begin{vmatrix}
\mathtt{A}^{(N)}_{1jk}&\mathtt{A}^{(N)}_{3jk}\\
\mathtt{A}^{(N)}_{123}&\mathtt{A}^{(N)}_{323}
\end{vmatrix},\\
\mathtt{d}^{(\mathtt{N})}_{jk} =& \begin{vmatrix}
\mathtt{A}^{(N)}_{1jk}&\mathtt{A}^{(N)}_{2jk}&\mathtt{A}^{(N)}_{3jk}\\
\mathtt{A}^{(N)}_{113}&\mathtt{A}^{(N)}_{213}&\mathtt{A}^{(N)}_{313}\\
\mathtt{A}^{(N)}_{123}&\mathtt{A}^{(N)}_{223}&\mathtt{A}^{(N)}_{323}
\end{vmatrix},\\
\mathtt{m}^{(\mathtt{N})}_{ij} =& -\frac{\mathtt{C}^{(\mathtt{N})}_{2ij}}{\mathtt{C}^{(\mathtt{N})}_{213}}+\frac{\mathtt{C}^{(\mathtt{N})}_{212}}{\mathtt{C}^{(\mathtt{N})}_{213}}\frac{\mathtt{d}^{(\mathtt{N})}_{ij}}{\mathtt{d}^{(\mathtt{N})}_{12}},\\
\mathtt{k}^{(\mathtt{N})}_{ij} =& -\frac{\mathtt{A}^{(\mathtt{N})}_{1ij}}{\mathtt{A}^{(\mathtt{N})}_{123}}+\frac{\mathtt{A}^{(\mathtt{N})}_{113}}{\mathtt{A}^{(\mathtt{N})}_{123}}\frac{\mathtt{C}^{(\mathtt{N})}_{2ij}}{\mathtt{C}^{(\mathtt{N})}_{213}},\\
\mathtt{n}^{(\mathtt{N})}_{ij} =& \mathtt{k}^{(\mathtt{N})}_{ij}-\frac{\mathtt{d}^{(\mathtt{N})}_{ij}}{\mathtt{d}^{(\mathtt{N})}_{12}}\mathtt{k}^{(\mathtt{N})}_{12},\\
\mathtt{x}^{(\mathtt{N})}_{13} =& \frac{\mathtt{C}^{(\mathtt{N})}_{213}\mathtt{C}^{(\mathtt{N})}_{313}}{\mathtt{A}^{(\mathtt{N})}_{123}\mathtt{d}^{(\mathtt{N})}_{12}},\\
\mathtt{y}^{(\mathtt{N})}_{13} =&1- \frac{\mathtt{C}^{(\mathtt{N})}_{212}\mathtt{C}^{(\mathtt{N})}_{313}}{\mathtt{A}^{(\mathtt{N})}_{123}\mathtt{d}^{(\mathtt{N})}_{12}},\\
\mathtt{z}^{(\mathtt{N})}_{13} =& \frac{\mathtt{C}^{(\mathtt{N})}_{213}\mathtt{C}^{(\mathtt{N})}_{313}}{\mathtt{A}^{(\mathtt{N})}_{123}\mathtt{d}^{(\mathtt{N})}_{12}}\mathtt{k}^{(\mathtt{N})}_{12}-\frac{\mathtt{A}^{(\mathtt{N})}_{113}}{\mathtt{A}^{(\mathtt{N})}_{123}},\\
\mathtt{x}^{(\mathtt{N})}_{23}=&\frac{\mathtt{A}^{(\mathtt{N})}_{223}\mathtt{C}^{(\mathtt{N})}_{313}-\mathtt{A}^{(\mathtt{N})}_{323}\mathtt{C}^{(\mathtt{N})}_{213}}{\mathtt{d}^{(\mathtt{N})}_{12}},\\
\mathtt{y}^{(\mathtt{N})}_{23}=&\frac{\mathtt{C}^{(\mathtt{N})}_{212}}{\mathtt{C}^{(\mathtt{N})}_{213}}\Big(\frac{-\mathtt{A}^{(\mathtt{N})}_{223}\mathtt{C}^{(\mathtt{N})}_{313}+\mathtt{A}^{(\mathtt{N})}_{323}\mathtt{C}^{(\mathtt{N})}_{213}}{\mathtt{d}^{(\mathtt{N})}_{12}}\Big)+\frac{\mathtt{A}^{(\mathtt{N})}_{223}\mathtt{A}^{(\mathtt{N})}_{123}}{\mathtt{C}^{(\mathtt{N})}_{213}},\\
\mathtt{z}^{(\mathtt{N})}_{23}=&\frac{\mathtt{A}^{(\mathtt{N})}_{223}\mathtt{C}^{(\mathtt{N})}_{313}-\mathtt{A}^{(\mathtt{N})}_{323}\mathtt{C}^{(\mathtt{N})}_{213}}{\mathtt{d}^{(\mathtt{N})}_{12}}\mathtt{k}^{(\mathtt{N})}_{12}-\frac{\mathtt{A}^{(\mathtt{N})}_{223}\mathtt{A}^{(\mathtt{N})}_{113}}{\mathtt{C}^{(\mathtt{N})}_{213}}+1,\\
\bar{\mathtt{B}}^{(\mathtt{N})}=&\begin{pmatrix}
1&-\frac{\mathtt{C}^{(\mathtt{N})}_{212}}{\mathtt{C}^{(\mathtt{N})}_{213}}&\mathtt{k}^{(\mathtt{N})}_{12}\\
\mathtt{x}^{(\mathtt{N})}_{13}&\mathtt{y}^{(\mathtt{N})}_{13}&\mathtt{z}^{(\mathtt{N})}_{13}\\
\mathtt{x}^{(\mathtt{N})}_{23}&\mathtt{y}^{(\mathtt{N})}_{23}&\mathtt{z}^{(\mathtt{N})}_{23}
\end{pmatrix}.
\end{align}
By applying the construction of the matrix $D^{(7)}$ to solve the following consistency condition,
\begin{align}
 \begin{pmatrix}
\{\bar{\phi}_1, \mathcal{H}_c\}\\
\{\bar{\phi}_2, \mathcal{H}_c\}\\
\{\bar{\phi}_3, \mathcal{H}_c\}\\
\{\bar{\phi}_{22}, \mathcal{H}_c\}\\
\{\bar{\phi}_{33}, \mathcal{H}_c\}\\
\{\bar{\phi}_{12}, \mathcal{H}_c\}\\
\{\bar{\phi}_{13}, \mathcal{H}_c\}\\
\{\bar{\phi}_{23}, \mathcal{H}_c\}
\end{pmatrix}
+ D^{(7)}\begin{pmatrix}
\lambda^1\\
\lambda^2\\
\lambda^3\\
\lambda^{22}\\
\lambda^{33}\\
\lambda^{12}\\
\lambda^{13}\\
\lambda^{23}
\end{pmatrix}
\approx & \begin{pmatrix}
0\\
0\\
0\\
0\\
0\\
0\\
0\\
0
\end{pmatrix}, \notag \\
 P^{(6)}\begin{pmatrix}
\{\bar{\phi}_1, \mathcal{H}_c\}\\
\{\bar{\phi}_2, \mathcal{H}_c\}\\
\{\bar{\phi}_3, \mathcal{H}_c\}\\
\{\bar{\phi}_{22}, \mathcal{H}_c\}\\
\{\bar{\phi}_{33}, \mathcal{H}_c\}\\
\{\bar{\phi}_{12}, \mathcal{H}_c\}\\
\{\bar{\phi}_{13}, \mathcal{H}_c\}\\
\{\bar{\phi}_{23}, \mathcal{H}_c\}
\end{pmatrix}
+ \bar{D}^{(7)}[(P^{(6)})^{\top}]^{-1}\begin{pmatrix}
\lambda^1\\
\lambda^2\\
\lambda^3\\
\lambda^{22}\\
\lambda^{33}\\
\lambda^{12}\\
\lambda^{13}\\
\lambda^{23}
\end{pmatrix}
\approx & \begin{pmatrix}
0\\
0\\
0\\
0\\
0\\
0\\
0\\
0
\end{pmatrix},
\end{align}
which gives the following solutions of lagrange multipliers and secondary constraints,
\begin{align}
\lambda^{1}=& \frac{1}{\mathtt{A}^{(6)}_{123}}\Big(\mathtt{x}^{(6)}_{23}\{\bar{\phi}_{12}, \mathcal{H}_c\}+\mathtt{y}^{(6)}_{23}\{\bar{\phi}_{13}, \mathcal{H}_c\}+\mathtt{z}^{(6)}_{23}\{\bar{\phi}_{23}, \mathcal{H}_c\} \Big),\\
\lambda^{2}=& \frac{\mathtt{A}^{(6)}_{123}}{\mathtt{C}^{(6)}_{213}}\Big(-\mathtt{x}^{(6)}_{13}\{\bar{\phi}_{12}, \mathcal{H}_c\}-\mathtt{y}^{(6)}_{13}\{\bar{\phi}_{13}, \mathcal{H}_c\}-\mathtt{z}^{(6)}_{13}\{\bar{\phi}_{23}, \mathcal{H}_c\} \Big),\\
\lambda^{3}=& \frac{\mathtt{C}^{(6)}_{213}}{\mathtt{d}^{(6)}_{12}}\Big(\{\bar{\phi}_{12}, \mathcal{H}_c\}-\frac{\mathtt{C}^{(6)}_{212}}{\mathtt{C}^{(6)}_{213}}\{\bar{\phi}_{13}, \mathcal{H}_c\}+\mathtt{k}^{(6)}_{12}\{\bar{\phi}_{23}, \mathcal{H}_c\} \Big),\\
\bar{\sigma}_{22}=& \frac{\mathtt{d}^{(6)}_{22}}{\mathtt{d}^{(6)}_{12}}\{\bar{\phi}_{12}, \mathcal{H}_c\}-\mathtt{m}^{(6)}_{22}\{\bar{\phi}_{13}, \mathcal{H}_c\}-\mathtt{n}^{(6)}_{22}\{\bar{\phi}_{23}, \mathcal{H}_c\},\\
\bar{\sigma}_{33}=& \frac{\mathtt{d}^{(6)}_{33}}{\mathtt{d}^{(6)}_{12}}\{\bar{\phi}_{12}, \mathcal{H}_c\}-\mathtt{m}^{(6)}_{33}\{\bar{\phi}_{13}, \mathcal{H}_c\}-\mathtt{n}^{(6)}_{33}\{\bar{\phi}_{23}, \mathcal{H}_c\},
\end{align}
and the solutions of $\lambda^{12}, \lambda^{13},$ and $\lambda^{23}$ can be solved by the following equations,
\begin{align}
\{\bar{\phi}_{1}, \mathcal{H}_c\}+\frac{\mathtt{A}^{(\mathtt{6})}_{123}}{det \bar{\mathtt{B}}^{(\mathtt{6})}}\Big(\lambda^{12}(\mathtt{x}^{(6)}_{13}\mathtt{y}^{(6)}_{23}-\mathtt{x}^{(6)}_{23}\mathtt{y}^{(6)}_{13})-\frac{\lambda^{13}}{\mathtt{C}^{(6)}_{213}}(\mathtt{C}^{(6)}_{213}\mathtt{y}^{(6)}_{23}+\mathtt{C}^{(6)}_{212}\mathtt{x}^{(6)}_{23})&\notag\\+\frac{\lambda^{23}}{\mathtt{C}^{(6)}_{213}}(\mathtt{C}^{(6)}_{212}\mathtt{x}^{(6)}_{13}+\mathtt{C}^{(6)}_{213}\mathtt{y}^{(6)}_{13}) \Big)&\approx 0,\\
\{\bar{\phi}_{2}, \mathcal{H}_c\}+\frac{\mathtt{C}^{(\mathtt{6})}_{213}}{\mathtt{A}^{(\mathtt{6})}_{213}det \bar{\mathtt{B}}^{(\mathtt{6})}}\Big(\lambda^{12}(\mathtt{x}^{(6)}_{13}\mathtt{z}^{(6)}_{23}-\mathtt{x}^{(6)}_{23}\mathtt{z}^{(6)}_{13})+\lambda^{13}(\mathtt{k}^{(6)}_{12}\mathtt{x}^{(6)}_{23}-\mathtt{z}^{(6)}_{23})&\notag\\+\lambda^{23}(\mathtt{z}^{(6)}_{13}-\mathtt{k}^{(6)}_{12}\mathtt{x}^{(6)}_{13}) \Big)&\approx 0,\\
\{\bar{\phi}_{3}, \mathcal{H}_c\}+\frac{\mathtt{d}^{(\mathtt{6})}_{12}}{\mathtt{C}^{(\mathtt{6})}_{213}det \bar{\mathtt{B}}^{(\mathtt{6})}}\Big(\lambda^{12}(\mathtt{y}^{(6)}_{13}\mathtt{z}^{(6)}_{23}-\mathtt{y}^{(6)}_{23}\mathtt{z}^{(6)}_{13})+\frac{\lambda^{13}}{\mathtt{C}^{(6)}_{213}}(\mathtt{C}^{(6)}_{213}\mathtt{k}^{(6)}_{12}\mathtt{y}^{(6)}_{23}+\mathtt{C}^{(6)}_{212}\mathtt{z}^{(6)}_{23})&\notag \\-\frac{\lambda^{23}}{\mathtt{C}^{(6)}_{213}}(\mathtt{C}^{(6)}_{213}\mathtt{k}^{(6)}_{12}\mathtt{y}^{(6)}_{13}-\mathtt{C}^{(6)}_{212}\mathtt{z}^{(6)}_{13}) \Big)&\approx 0.
\end{align}
After that we define $\mathcal{H}^{(2)}_c=\mathcal{H}_c+\lambda^i\bar{\phi}_i+\lambda^{12}\bar{\phi}_{12}+\lambda^{13}\bar{\phi}_{13}+\lambda^{23}\bar{\phi}_{23}$ and also $\{\bar{\sigma}_{ij},\bar{\phi}_{mn}\}=\mathtt{B}^{(6)}_{ijmn}\delta^{(3)}(\vec{x}-\vec{y})$. The matrix $D^{(8)}$ is defined as follows,
\begin{equation}
D^{(8)}=\begin{pmatrix}
\mathtt{B}^{(6)}_{2222}&\mathtt{B}^{(6)}_{2233}\\
\mathtt{B}^{(6)}_{3322}&\mathtt{B}^{(6)}_{3333}
\end{pmatrix},
\end{equation}
and study the consistency conditions for secondary constraints,
\begin{equation}
 \begin{pmatrix}
\{\bar{\sigma}_{22}, \mathcal{H}^{(2)}_c\}\\
\{\bar{\sigma}_{33}, \mathcal{H}^{(2)}_c\}
\end{pmatrix}
+ \begin{pmatrix}
\mathtt{B}^{(6)}_{2222}&\mathtt{B}^{(6)}_{2233}\\
\mathtt{B}^{(6)}_{3322}&\mathtt{B}^{(6)}_{3333}
\end{pmatrix}\begin{pmatrix}
\lambda^{22}\\
\lambda^{33}
\end{pmatrix}\delta^{(3)}(\vec{x}-\vec{y})
\approx \begin{pmatrix}
0\\
0
\end{pmatrix},
\end{equation}
which gives the solutions of $\lambda^{22}$ and $\lambda^{33}$,
\begin{align}
\lambda^{22} =& \frac{1}{|\triangle^{(8)}_1|}(-\{\bar{\sigma}_{22}, \mathcal{H}^{(2)}_c\}-\lambda^{33}\mathtt{B}^{(6)}_{2233}),\\
\lambda^{33} =& \frac{|\triangle^{(8)}_1|}{|\triangle^{(8)}_2|}(\frac{\mathtt{B}^{(6)}_{3322}}{|\triangle^{(8)}_1|}\{\bar{\sigma}_{22}, \mathcal{H}^{(2)}_c\}-\{\bar{\sigma}_{33}, \mathcal{H}^{(2)}_c\}).
\end{align}
The Dirac-Bergmann algorithm stops with a total of $10$ second class constraints, $\bar{\sigma}_{22}, \bar{\sigma}_{33}, \bar{\phi}_{ij},$ and $\bar{\phi}_{i}$. Therefore Type $6$ Newer General Relativity have a total of $10-\frac{1}{2}(10)=5$ physical degrees of freedoms.

\subsection{Type $7$}
Type $7$ Newer General Relativity consider $c_1=0$, $2c_1+c_2+c_4=0$ and $(2c_3+c_5)^2-4c_3(c_1+c_2+c_3+c_4+c_5)=0$, however it is not the intersection between Type 2 and Type 4. This is because Type 4 consider whether $c_3$ and $c_5$ nonzero or $c_1=c_3=c_5=0$, Type 7 is neither both cases and only consider $c_1=0$, $c_3$ and $c_5$ remaining nonzero, 
\begin{align}
\pi_N =&\frac{4c_2}{N^2}\partial_a N^a+\frac{4c_3}{N^2}(\dot{\gamma}^{ab} \gamma_{ab}-\gamma_{bc} N^a \partial_a \gamma^{bc})+\frac{-8c_3}{N^3}(\dot{N}-N^a \partial_a N) , \\
\pi_i =& \frac{c_2}{N}\gamma^{ab}\partial_b \gamma_{ia}-\frac{4c_2}{N^2}\partial_i N, \\
\pi_{ij} =& \frac{(4c_3)}{N^2}(\dot{N} \gamma_{ij}-\gamma_{ij} N^a \partial_a N)+\frac{2c_3}{N}( -\dot{\gamma}^{ab} \gamma_{ab} \gamma_{ij} + \gamma_{bc} \gamma_{ij} N^a \partial_a \gamma^{bc}) \notag \\ &+\frac{c_2 }{N}(\gamma_{ja} \partial_i N^a + \gamma_{ia} \partial_j N^a).
\end{align}
Comparing to Type $4$, with additional $c_1=0$ simplify $\bar{\phi}_{ij}$ to be much more simple than $\bar{\phi}_{ij}$ in Type $4$ Case $1$,
\begin{align}
\bar{\phi}_i =& \pi_i-\frac{c_2}{N}\gamma^{ab}\partial_b \gamma_{ia}+\frac{4c_2}{N^2}\partial_i N,\\
\bar{\phi}_{ij} =& \pi_{ij} +\frac{-2c_2 N( \gamma_{jk} \partial_i N^k+\gamma_{ik} \partial_j N^k)+\gamma_{ij} (N^3 \pi_N - 4c_2 N \gamma^{mk} \partial_m N_k-16c_3 N_k \gamma^{km} \partial_m N)}{2N^2},
\end{align}
and the solution of $\dot{N}$ is presented as below,
\begin{align}
\dot{N}=& -\frac{N^3}{8c_3}\Big(\pi_N-\frac{8c_3}{N^3}N^a\partial_aN-\frac{4c_3}{N^2}(\dot{\gamma}^{ab}\gamma_{ab}+\gamma^{bc}N^a\partial_a\gamma_{bc})-\frac{4c_2}{N^2}\partial_aN^a \Big).
\end{align}
After that the poisson bracket between two types of constraints,
\begin{align}
\{\phi_i[\delta],\phi_{kl}[\bar{\lambda}]\} =& \int_{\Sigma_t} d^3z \bar{\lambda}(\vec{z}) \{ \bar{ \phi }_i, \bar{\phi}_{kl} \}(\vec{z}) \delta (\vec{z}-\vec{a}),\notag \\
=& \bar{\lambda}(\vec{a}) \{ \bar{ \phi}_i, \bar{ \phi}_{kl} \}(\vec{a}),\notag \\
=&\frac{(-4c_2+8c_3) \bar{\lambda}}{N^2}\gamma_{kl}\partial_i N + \frac{c_2 \bar{\lambda}}{N^2}[(\gamma_{ki}\gamma_{la}+\gamma_{ka}\gamma_{li})\partial^aN-\gamma_{li}\partial_k N-\gamma_{ki}\partial_l N]\notag \\
& +\frac{c_2 \bar{\lambda}}{2N}\gamma^{ab}(\gamma_{li}\partial_b \gamma_{ka}+\gamma_{ki}\partial_b \gamma_{la}+\gamma_{kl}\partial_b \gamma_{ia})-\frac{c_2 \bar{\lambda}}{N}(\partial_k \gamma_{li}+\partial_l \gamma_{ki}+\partial_i \gamma_{kl}),
\end{align}
will be substitute into the consistency conditions,
\begin{align}
\{\bar{\phi}_i (\vec{x}), H_p (\Sigma_t)\} =&  \{\bar{\phi}_i, \mathcal{H}_c \}(\vec{x}) +\lambda^{kl} (\vec{x})\{\bar{\phi}_i, \bar{\phi}_{kl} \}(\vec{x})\approx 0,\\
\{\bar{\phi}_{kl} (\vec{x}), H_p (\Sigma_t)\} =&  \{\bar{\phi}_{kl}, \mathcal{H}_c \}(\vec{x}) +\lambda^{i} (\vec{x})\{\bar{\phi}_{kl}, \bar{\phi}_{i} \}(\vec{x})\approx 0,
\end{align}
which has the matrix form,
\begin{equation}
 \begin{pmatrix}
\{\bar{\phi}_1, \mathcal{H}_c\}\\
\{\bar{\phi}_2, \mathcal{H}_c\}\\
\{\bar{\phi}_3, \mathcal{H}_c\}\\
\{\bar{\phi}_{11}, \mathcal{H}_c\}\\
\{\bar{\phi}_{22}, \mathcal{H}_c\}\\
\{\bar{\phi}_{33}, \mathcal{H}_c\}\\
\{\bar{\phi}_{12}, \mathcal{H}_c\}\\
\{\bar{\phi}_{13}, \mathcal{H}_c\}\\
\{\bar{\phi}_{23}, \mathcal{H}_c\}
\end{pmatrix}
+ D^{(9)}\begin{pmatrix}
\lambda^1\\
\lambda^2\\
\lambda^3\\
\lambda^{11}\\
\lambda^{22}\\
\lambda^{33}\\
\lambda^{12}\\
\lambda^{13}\\
\lambda^{23}
\end{pmatrix}
\approx  \begin{pmatrix}
0\\
0\\
0\\
0\\
0\\
0\\
0\\
0\\
0
\end{pmatrix}.
\end{equation}
The matrix $D^{(9)}$, $\bar{D}^{(9)}$ and $P^{(7)}$ is defined as follows,
\begin{align}
D^{(9)} =& \begin{pmatrix}
0&0&0&\mathtt{A}^{(7)}_{111}&\mathtt{A}^{(7)}_{122}&\mathtt{A}^{(7)}_{133}&\mathtt{A}^{(7)}_{112}&\mathtt{A}^{(7)}_{113}&\mathtt{A}^{(7)}_{123}\\
0&0&0&\mathtt{A}^{(7)}_{211}&\mathtt{A}^{(7)}_{222}&\mathtt{A}^{(7)}_{233}&\mathtt{A}^{(7)}_{212}&\mathtt{A}^{(7)}_{213}&\mathtt{A}^{(7)}_{223}\\
0&0&0&\mathtt{A}^{(7)}_{311}&\mathtt{A}^{(7)}_{322}&\mathtt{A}^{(7)}_{333}&\mathtt{A}^{(7)}_{312}&\mathtt{A}^{(7)}_{313}&\mathtt{A}^{(7)}_{323}\\
-\mathtt{A}^{(7)}_{111}&-\mathtt{A}^{(7)}_{211}&-\mathtt{A}^{(7)}_{311}&0&0&0&0&0&0\\
-\mathtt{A}^{(7)}_{122}&-\mathtt{A}^{(7)}_{222}&-\mathtt{A}^{(7)}_{322}&0&0&0&0&0&0\\
-\mathtt{A}^{(7)}_{133}&-\mathtt{A}^{(7)}_{233}&-\mathtt{A}^{(7)}_{333}&0&0&0&0&0&0\\
-\mathtt{A}^{(7)}_{112}&-\mathtt{A}^{(7)}_{212}&-\mathtt{A}^{(7)}_{312}&0&0&0&0&0&0\\
-\mathtt{A}^{(7)}_{113}&-\mathtt{A}^{(7)}_{213}&-\mathtt{A}^{(7)}_{313}&0&0&0&0&0&0\\
-\mathtt{A}^{(7)}_{123}&-\mathtt{A}^{(7)}_{223}&-\mathtt{A}^{(7)}_{323}&0&0&0&0&0&0
\end{pmatrix},\\
\bar{D}^{(9)} =& \begin{pmatrix}
0&0&0&0&0&0&0&0&\mathtt{A}^{(7)}_{123}\\
0&0&0&0&0&0&0&-\frac{\mathtt{C}^{(7)}_{213}}{\mathtt{A}^{(7)}_{123}}&0\\
0&0&0&0&0&0&\frac{\mathtt{d}^{(7)}_{12}}{\mathtt{C}^{(7)}_{213}}&0&0\\
0&0&0&0&0&0&0&0&0\\
0&0&0&0&0&0&0&0&0\\
0&0&0&0&0&0&0&0&0\\
0&0&-\frac{\mathtt{d}^{(7)}_{12}}{\mathtt{C}^{(7)}_{213}}&0&0&0&0&0&0\\
0&\frac{\mathtt{C}^{(7)}_{213}}{\mathtt{A}^{(7)}_{123}}&0&0&0&0&0&0&0\\
-\mathtt{A}^{(7)}_{123}&0&0&0&0&0&0&0&0
\end{pmatrix},\\
P^{(7)}=&\begin{pmatrix}
1&0&0&0&0&0&0&0&0\\
0&1&0&0&0&0&0&0&0\\
0&0&1&0&0&0&0&0&0\\
0&0&0&1&0&0&-\frac{\mathtt{d}^{(7)}_{11}}{\mathtt{d}^{(7)}_{12}}&\mathtt{m}^{(7)}_{11}&\mathtt{n}^{(7)}_{11}\\
0&0&0&0&1&0&-\frac{\mathtt{d}^{(7)}_{22}}{\mathtt{d}^{(7)}_{12}}&\mathtt{m}^{(7)}_{22}&\mathtt{n}^{(7)}_{22}\\
0&0&0&0&0&1&-\frac{\mathtt{d}^{(7)}_{33}}{\mathtt{d}^{(7)}_{12}}&\mathtt{m}^{(7)}_{33}&\mathtt{n}^{(7)}_{33}\\
0&0&0&0&0&0&1&-\frac{\mathtt{C}^{(7)}_{212}}{\mathtt{C}^{(7)}_{213}}&\mathtt{k}^{(7)}_{12}\\
0&0&0&0&0&0&\mathtt{x}^{(7)}_{13}&\mathtt{y}^{(7)}_{13}&\mathtt{z}^{(7)}_{13}\\
0&0&0&0&0&0&\mathtt{x}^{(7)}_{23}&\mathtt{y}^{(7)}_{23}&\mathtt{z}^{(7)}_{23}
\end{pmatrix},
\end{align}
and the solutions of lagrange multipliers are shown below,
\begin{align}
\lambda^{1}=& \frac{1}{\mathtt{A}^{(7)}_{123}}\Big(\mathtt{x}^{(7)}_{23}\{\bar{\phi}_{12}, \mathcal{H}_c\}+\mathtt{y}^{(7)}_{23}\{\bar{\phi}_{13}, \mathcal{H}_c\}+\mathtt{z}^{(7)}_{23}\{\bar{\phi}_{23}, \mathcal{H}_c\} \Big),\\
\lambda^{2}=& \frac{\mathtt{A}^{(7)}_{123}}{\mathtt{C}^{(7)}_{213}}\Big(-\mathtt{x}^{(7)}_{13}\{\bar{\phi}_{12}, \mathcal{H}_c\}-\mathtt{y}^{(7)}_{13}\{\bar{\phi}_{13}, \mathcal{H}_c\}-\mathtt{z}^{(7)}_{13}\{\bar{\phi}_{23}, \mathcal{H}_c\} \Big),\\
\lambda^{3}=& \frac{\mathtt{C}^{(7)}_{213}}{\mathtt{d}^{(7)}_{12}}\Big(\{\bar{\phi}_{12}, \mathcal{H}_c\}-\frac{\mathtt{C}^{(7)}_{212}}{\mathtt{C}^{(7)}_{213}}\{\bar{\phi}_{13}, \mathcal{H}_c\}+\mathtt{k}^{(7)}_{12}\{\bar{\phi}_{23}, \mathcal{H}_c\} \Big),\\
\bar{\sigma}_{11}=& \frac{\mathtt{d}^{(7)}_{11}}{\mathtt{d}^{(7)}_{12}}\{\bar{\phi}_{12}, \mathcal{H}_c\}-\mathtt{m}^{(7)}_{11}\{\bar{\phi}_{13}, \mathcal{H}_c\}-\mathtt{n}^{(7)}_{11}\{\bar{\phi}_{23}, \mathcal{H}_c\},\\
\bar{\sigma}_{22}=& \frac{\mathtt{d}^{(7)}_{22}}{\mathtt{d}^{(7)}_{12}}\{\bar{\phi}_{12}, \mathcal{H}_c\}-\mathtt{m}^{(7)}_{22}\{\bar{\phi}_{13}, \mathcal{H}_c\}-\mathtt{n}^{(7)}_{22}\{\bar{\phi}_{23}, \mathcal{H}_c\},\\
\bar{\sigma}_{33}=& \frac{\mathtt{d}^{(7)}_{33}}{\mathtt{d}^{(7)}_{12}}\{\bar{\phi}_{12}, \mathcal{H}_c\}-\mathtt{m}^{(7)}_{33}\{\bar{\phi}_{13}, \mathcal{H}_c\}-\mathtt{n}^{(7)}_{33}\{\bar{\phi}_{23}, \mathcal{H}_c\},
\end{align}
and the explicit expressions for $\lambda^{12}, \lambda^{13}$ and $\lambda^{23}$ are similar to Type $6$, can be solved from the equations below,
\begin{align}
\{\bar{\phi}_{1}, \mathcal{H}_c\}+\frac{\mathtt{A}^{(7)}_{123}}{det \bar{\mathtt{B}}^{(7)}}\Big(\lambda^{12}(\mathtt{x}^{(7)}_{13}\mathtt{y}^{(7)}_{23}-\mathtt{x}^{(7)}_{23}\mathtt{y}^{(7)}_{13})-\frac{\lambda^{13}}{\mathtt{C}^{(7)}_{213}}(\mathtt{C}^{(7)}_{213}\mathtt{y}^{(7)}_{23}+\mathtt{C}^{(7)}_{212}\mathtt{x}^{(7)}_{23})&\notag\\+\frac{\lambda^{23}}{\mathtt{C}^{(7)}_{213}}(\mathtt{C}^{(7)}_{212}\mathtt{x}^{(7)}_{13}+\mathtt{C}^{(7)}_{213}\mathtt{y}^{(7)}_{13}) \Big)&\approx 0,\\
\{\bar{\phi}_{2}, \mathcal{H}_c\}+\frac{\mathtt{C}^{(7)}_{213}}{\mathtt{A}^{(7)}_{213}det \bar{\mathtt{B}}^{(7)}}\Big(\lambda^{12}(\mathtt{x}^{(7)}_{13}\mathtt{z}^{(7)}_{23}-\mathtt{x}^{(7)}_{23}\mathtt{z}^{(7)}_{13})+\lambda^{13}(\mathtt{k}^{(7)}_{12}\mathtt{x}^{(7)}_{23}-\mathtt{z}^{(7)}_{23})&\notag\\+\lambda^{23}(\mathtt{z}^{(7)}_{13}-\mathtt{k}^{(7)}_{12}\mathtt{x}^{(7)}_{13}) \Big)&\approx 0,\\
\{\bar{\phi}_{3}, \mathcal{H}_c\}+\frac{\mathtt{d}^{(7)}_{12}}{\mathtt{C}^{(7)}_{213}det \bar{\mathtt{B}}^{(7)}}\Big(\lambda^{12}(\mathtt{y}^{(7)}_{13}\mathtt{z}^{(7)}_{23}-\mathtt{y}^{(7)}_{23}\mathtt{z}^{(7)}_{13})+\frac{\lambda^{13}}{\mathtt{C}^{(7)}_{213}}(\mathtt{C}^{(7)}_{213}\mathtt{k}^{(7)}_{12}\mathtt{y}^{(7)}_{23}+\mathtt{C}^{(7)}_{212}\mathtt{z}^{(7)}_{23})&\notag \\-\frac{\lambda^{23}}{\mathtt{C}^{(7)}_{213}}(\mathtt{C}^{(7)}_{213}\mathtt{k}^{(7)}_{12}\mathtt{y}^{(7)}_{13}-\mathtt{C}^{(7)}_{212}\mathtt{z}^{(7)}_{13}) \Big)&\approx 0.
\end{align}
Let $\mathcal{H}^{(2)}_c=\mathcal{H}_c+\lambda^i\bar{\phi}_i+\lambda^{12}\bar{\phi}_{12}+\lambda^{13}\bar{\phi}_{13}+\lambda^{23}\bar{\phi}_{23}$ and $\{\bar{\sigma}_{ij},\bar{\phi}_{mn}=\mathtt{B}^{(7)}_{ijmn}\delta^{(3)}(\vec{x}-\vec{y}) \}$. The matrix $D^{(10)}$ is defined as follows,
\begin{equation}
D^{(10)}= \begin{pmatrix}
\mathtt{B}^{(7)}_{1111}&\mathtt{B}^{(7)}_{1122}&\mathtt{B}^{(7)}_{1133}\\
\mathtt{B}^{(7)}_{2211}&\mathtt{B}^{(7)}_{2222}&\mathtt{B}^{(7)}_{2233}\\
\mathtt{B}^{(7)}_{3311}&\mathtt{B}^{(7)}_{3322}&\mathtt{B}^{(7)}_{3333}
\end{pmatrix}.
\end{equation}
The consistency conditions of secondary constraints,
\begin{equation}
 \begin{pmatrix}
\{\bar{\sigma}_{11}, \mathcal{H}^{(2)}_c\}\\
\{\bar{\sigma}_{22}, \mathcal{H}^{(2)}_c\}\\
\{\bar{\sigma}_{33}, \mathcal{H}^{(2)}_c\}
\end{pmatrix}
+ \begin{pmatrix}
\mathtt{B}^{(7)}_{1111}&\mathtt{B}^{(7)}_{1122}&\mathtt{B}^{(7)}_{1133}\\
\mathtt{B}^{(7)}_{2211}&\mathtt{B}^{(7)}_{2222}&\mathtt{B}^{(7)}_{2233}\\
\mathtt{B}^{(7)}_{3311}&\mathtt{B}^{(7)}_{3322}&\mathtt{B}^{(7)}_{3333}
\end{pmatrix}\begin{pmatrix}
\lambda^{11}\\
\lambda^{22}\\
\lambda^{33}
\end{pmatrix}
\approx \begin{pmatrix}
0\\
0\\
0
\end{pmatrix},
\end{equation}
which gives the following solutions of $\lambda^{11}, \lambda^{22}$, and $\lambda^{33}$,
\begin{align}
\lambda^{11} =& \frac{1}{|\triangle^{(10)}_1|}\Big(-\{\bar{\sigma}_{11}, \mathcal{H}^{(2)}_c \}-\lambda^{22}\mathtt{B}^{(7)}_{1122}-\lambda^{33}\mathtt{B}^{(7)}_{1133} \Big),\\
\lambda^{22} =& \frac{|\triangle^{(10)}_1|}{|\triangle^{(10)}_2|}\Big( \frac{\mathtt{B}^{(7)}_{2211}}{|\triangle^{(10)}_1|}\{\bar{\sigma}_{11}, \mathcal{H}^{(2)}_c\}-\{\bar{\sigma}_{22}, \mathcal{H}^{(2)}_c\}-\lambda^{33}\frac{\begin{vmatrix}
\mathtt{B}^{(7)}_{1111}&\mathtt{B}^{(7)}_{1133}\\
\mathtt{B}^{(7)}_{2211}&\mathtt{B}^{(7)}_{2233}
\end{vmatrix}}{|\triangle^{(10)}_1|}\Big),\\
\lambda^{33}=&  \frac{|\triangle^{(10)}_2|}{|\triangle^{(10)}_3|}\Big( \frac{\mathtt{B}^{(7)}_{3311}}{|\triangle^{(10)}_1|}-\frac{\mathtt{B}^{(7)}_{2211}}{|\triangle^{(10)}_1|}\frac{\begin{vmatrix}
\mathtt{B}^{(7)}_{1111}&\mathtt{B}^{(7)}_{1122}\\
\mathtt{B}^{(7)}_{3311}&\mathtt{B}^{(7)}_{3322}
\end{vmatrix}}{|\triangle^{(10)}_2|}\{\bar{\sigma}_{11}, \mathcal{H}^{(2)}_c\}+\frac{\begin{vmatrix}
\mathtt{B}^{(7)}_{1111}&\mathtt{B}^{(7)}_{1122}\\
\mathtt{B}^{(7)}_{3311}&\mathtt{B}^{(7)}_{3322}
\end{vmatrix}}{|\triangle^{(10)}_2|}\{\bar{\sigma}_{22}, \mathcal{H}^{(2)}_c\}-\{\bar{\sigma}_{33}, \mathcal{H}^{(2)}_c\}\Big).
\end{align}
Since we have a total of $12$ second-class constraints, Type $7$ Newer General Relativity has a total of $10-\frac{1}{2}(12)=4$ physical degrees of freedoms.

\subsection{Type $8$}
As the intersection between Type $1$, Type $2$ and Type $3$, relations between all conjugate momenta of Type $8$ Newer General Relativity and their corresponding velocities are not invertible,
\begin{align}
\pi_N =& \frac{4c_2}{N^2}\partial_a N^a,\\
\pi_i =&\frac{c_2}{N}\gamma^{ab}\partial_b \gamma_{ia}-\frac{4c_2}{N^2} \partial_i N,\\
\pi_{ij} =& \frac{c_2}{N}(\gamma_{ja} \partial_i N^a + \gamma_{ia} \partial_j N^a),
\end{align}
and it is the theory with the most number of primary constraints,
\begin{align}
\bar{\phi}_N =& \pi_N -  \frac{4c_2\partial_a N^a}{N^2},\\
\bar{\phi}_i =& \pi_i-\frac{c_2}{N}\gamma^{ab}\partial_b \gamma_{ia}+\frac{4c_2}{N^2} \partial_i N, \\
\bar{\phi}_{ij} =& \pi_{ij} -\frac{c_2 (\gamma_{ja} \partial_i N^a + \gamma_{ia} \partial_j N^a)}{N}.
\end{align}
We first list out the poisson bracket between all density constraints and canonical Hamiltonian density, 
\begin{align}
\{\bar{\phi}_N, \mathcal{H}_c \}(\vec{x}) =& \frac{c_2}{N^2}(3\partial_a N^a \partial_b N^b -3\gamma^{bc} \partial_a N_b \partial_c N^a + N^a \partial_a N^b \partial^c \gamma_{bc}-4N^a\partial_b \partial_a N^b)(\vec{x}),\\
\{\bar{\phi}_i, \mathcal{H}_c \}(\vec{x}) =& -\frac{c_2}{2N^2}[-N \gamma_{ib} \gamma^{bk} \partial_a \partial_k N^a - N \partial_a \partial_i N^a -N \gamma^{bk}\partial_a N^a \partial_k \gamma_{ib}+N^a (4\gamma_{ib}\gamma^{bk} \partial_a \partial_k N \notag \\
&+\gamma^{bk}\partial_a N \partial_k \gamma_{ib}+N\gamma^{bc} \gamma^{dk}\partial_a \gamma_{cd}\partial_k \gamma_{ib})+6\partial_a N^a\partial_i N
-6\gamma^{ak}\partial_k N\partial_i N_a \notag \\
&+N\gamma^{bk}\partial_k \gamma_{ab}\partial_i N^a](\vec{x}),\\
\{\bar{\phi}_{ij}, \mathcal{H}_c \}(\vec{x}) =&-\frac{c_2}{N^2}[N\partial_a N^b (\gamma_{jb}\partial_i N^a+\gamma_{ib}\partial_j N^a)+N^a(-N\gamma_{ib}\gamma^{ck}\partial_a N^b \partial_k \gamma_{jc}\notag \\
&+N\partial_a N^b \partial_i \gamma_{jb} -\gamma_{jb}(N\gamma^{ck}\partial_a N^b \partial_k \gamma_{ic} +\partial_a N^b \partial_i N)\notag \\
&+N\partial_a N^b \partial_j \gamma_{ib} -\gamma_{ib}\partial_a N^b \partial_j N)](\vec{x}),
\end{align}
and also the poisson bracket between the density constraints,
\begin{align}
\{\bar{\phi}_N(\vec{x}), \bar{\phi}_i (\vec{y})\} =& -\frac{2c_2}{N^2}\gamma^{ab}\partial_b \gamma_{ia}\delta^{(3)}(\vec{x}-\vec{y}),\\
=&\mathtt{A}^{(8)}_i \delta^{(3)}(\vec{x}-\vec{y}),\\
\{\bar{\phi}_N(\vec{x}), \bar{\phi}_{ij}(\vec{y}) \} =& -\frac{c_2}{N^2}\gamma^{bk}\partial_k N^a (\gamma_{ib}\gamma_{ja}+\gamma_{ia}\gamma_{jb})\delta^{(3)}(\vec{x}-\vec{y}),\\
=&\mathtt{A}^{(8)}_{ij}\delta^{(3)}(\vec{x}-\vec{y}),\\
\{\bar{\phi}_l(\vec{x}), \bar{\phi}_{ij}(\vec{y}) \} =& \frac{c_2}{N^2}[\gamma^{ak}\partial_k N(\gamma_{il}\gamma_{ja}+\gamma_{ia}\gamma_{jl})-\gamma_{jl}\partial_i N -\gamma_{il}\partial_i N\notag \\
&+N(\gamma^{ab}\gamma_{jl}\partial_b \gamma_{ia}+\gamma^{ab}\gamma_{il}\partial_b \gamma_{ja}-\partial_i \gamma_{jl}-\partial_j \gamma_{il})]\delta^{(3)}(\vec{x}-\vec{y}),\\
=&\mathtt{A}^{(8)}_{lij}\delta^{(3)}(\vec{x}-\vec{y}),
\end{align}
The consistency conditions for each constraints are listed as below,
\begin{align}
\{\bar{\phi}_N (\vec{x}), H_p (\Sigma_t)\} =&  \{\bar{\phi}_N, \mathcal{H}_c \}(\vec{x}) +\lambda^{p} (\vec{x})\{\bar{\phi}_N, \bar{\phi}_{p}\}+\lambda^{kl} (\vec{x})\{\bar{\phi}_N, \bar{\phi}_{kl} \}(\vec{x})\approx 0,\\
\{\bar{\phi}_i (\vec{x}), H_p (\Sigma_t)\} =&  \{\bar{\phi}_i, \mathcal{H}_c \}(\vec{x}) +\lambda^N (\vec{x})\{\bar{\phi}_i, \bar{\phi}_N\}+\lambda^{kl} (\vec{x})\{\bar{\phi}_i, \bar{\phi}_{kl} \}(\vec{x})\approx 0,\\
\{\bar{\phi}_{kl} (\vec{x}), H_p (\Sigma_t)\} =&  \{\bar{\phi}_{kl}, \mathcal{H}_c \}(\vec{x})+\lambda^N (\vec{x})\{\bar{\phi}_{kl}, \bar{\phi}_N\} +\lambda^{p} (\vec{x})\{\bar{\phi}_{kl}, \bar{\phi}_{p} \}(\vec{x})\approx 0,
\end{align}
which forms the matrix $D^{(11)}$,
\begin{equation}
D^{(11)} =\begin{pmatrix}
0&\mathtt{A}^{(8)}_{1}&\mathtt{A}^{(8)}_{2}&\mathtt{A}^{(8)}_{3}&\mathtt{A}^{(8)}_{11}&\mathtt{A}^{(8)}_{22}&\mathtt{A}^{(8)}_{33}&\mathtt{A}^{(8)}_{12}&\mathtt{A}^{(8)}_{13}&\mathtt{A}^{(8)}_{23}\\
-\mathtt{A}^{(8)}_{1}&0&0&0&\mathtt{A}^{(8)}_{111}&\mathtt{A}^{(8)}_{122}&\mathtt{A}^{(8)}_{133}&\mathtt{A}^{(8)}_{112}&\mathtt{A}^{(8)}_{113}&\mathtt{A}^{(8)}_{123}\\
-\mathtt{A}^{(8)}_{2}&0&0&0&\mathtt{A}^{(8)}_{211}&\mathtt{A}^{(8)}_{222}&\mathtt{A}^{(8)}_{233}&\mathtt{A}^{(8)}_{212}&\mathtt{A}^{(8)}_{213}&\mathtt{A}^{(8)}_{223}\\
-\mathtt{A}^{(8)}_{3}&0&0&0&\mathtt{A}^{(8)}_{311}&\mathtt{A}^{(8)}_{322}&\mathtt{A}^{(8)}_{333}&\mathtt{A}^{(8)}_{312}&\mathtt{A}^{(8)}_{313}&\mathtt{A}^{(8)}_{323}\\
-\mathtt{A}^{(8)}_{11}&-\mathtt{A}^{(8)}_{111}&-\mathtt{A}^{(8)}_{211}&-\mathtt{A}^{(8)}_{311}&0&0&0&0&0&0\\
-\mathtt{A}^{(8)}_{22}&-\mathtt{A}^{(8)}_{122}&-\mathtt{A}^{(8)}_{222}&-\mathtt{A}^{(8)}_{322}&0&0&0&0&0&0\\
-\mathtt{A}^{(8)}_{33}&-\mathtt{A}^{(8)}_{133}&-\mathtt{A}^{(8)}_{233}&-\mathtt{A}^{(8)}_{333}&0&0&0&0&0&0\\
-\mathtt{A}^{(8)}_{12}&-\mathtt{A}^{(8)}_{112}&-\mathtt{A}^{(8)}_{212}&-\mathtt{A}^{(8)}_{312}&0&0&0&0&0&0\\
-\mathtt{A}^{(8)}_{13}&-\mathtt{A}^{(8)}_{113}&-\mathtt{A}^{(8)}_{213}&-\mathtt{A}^{(8)}_{313}&0&0&0&0&0&0\\
-\mathtt{A}^{(8)}_{23}&-\mathtt{A}^{(8)}_{123}&-\mathtt{A}^{(8)}_{223}&-\mathtt{A}^{(8)}_{323}&0&0&0&0&0&0
\end{pmatrix}. 
\end{equation}
After that we construct $\bar{D}^{(11)}=P^{(8)}D^{(11)}(P^{(8)})^{\top}$,
\begin{align}
P^{(8)}=& \begin{pmatrix}
1&0&0&0&0&0&0&0&0&0\\
0&1&0&0&0&0&-\frac{\mathtt{A}^{(8)}_{1}}{\mathtt{F}^{(8)}_{33}}&\frac{\mathtt{A}^{(8)}_{1}}{\mathtt{F}^{(8)}_{33}}\frac{\mathtt{d}^{(8)}_{33}}{\mathtt{d}^{(8)}_{12}}&-\frac{\mathtt{A}^{(8)}_{1}}{\mathtt{F}^{(8)}_{33}}\mathtt{m}^{(8)}_{33}&-\frac{\mathtt{A}^{(8)}_{1}}{\mathtt{F}^{(8)}_{33}}\mathtt{n}^{(8)}_{33}\\
0&0&1&0&0&0&-\frac{\mathtt{A}^{(8)}_{2}}{\mathtt{F}^{(8)}_{33}}&\frac{\mathtt{A}^{(8)}_{2}}{\mathtt{F}^{(8)}_{33}}\frac{\mathtt{d}^{(8)}_{33}}{\mathtt{d}^{(8)}_{12}}&-\frac{\mathtt{A}^{(8)}_{2}}{\mathtt{F}^{(8)}_{33}}\mathtt{m}^{(8)}_{33}&-\frac{\mathtt{A}^{(8)}_{2}}{\mathtt{F}^{(8)}_{33}}\mathtt{n}^{(8)}_{33}\\
0&0&0&1&0&0&-\frac{\mathtt{A}^{(8)}_{3}}{\mathtt{F}^{(8)}_{33}}&\frac{\mathtt{A}^{(8)}_{3}}{\mathtt{F}^{(8)}_{33}}\frac{\mathtt{d}^{(8)}_{33}}{\mathtt{d}^{(8)}_{12}}&-\frac{\mathtt{A}^{(8)}_{3}}{\mathtt{F}^{(8)}_{33}}\mathtt{m}^{(8)}_{33}&-\frac{\mathtt{A}^{(8)}_{3}}{\mathtt{F}^{(8)}_{33}}\mathtt{n}^{(8)}_{33}\\
0&0&0&0&1&0&-\frac{\mathtt{F}^{(8)}_{11}}{\mathtt{F}^{(8)}_{33}}&-\frac{\mathtt{d}^{(8)}_{11}}{\mathtt{d}^{(8)}_{12}}+\frac{\mathtt{F}^{(8)}_{11}}{\mathtt{F}^{(8)}_{33}}\frac{\mathtt{d}^{(8)}_{33}}{\mathtt{d}^{(8)}_{12}}&\mathtt{m}^{(8)}_{11}-\frac{\mathtt{F}^{(8)}_{11}}{\mathtt{F}^{(8)}_{33}}\mathtt{m}^{(8)}_{33}&\mathtt{n}^{(8)}_{11}-\frac{\mathtt{F}^{(8)}_{11}}{\mathtt{F}^{(8)}_{33}}\mathtt{n}^{(8)}_{33}\\
0&0&0&0&0&1&-\frac{\mathtt{F}^{(8)}_{22}}{\mathtt{F}^{(8)}_{33}}&-\frac{\mathtt{d}^{(8)}_{22}}{\mathtt{d}^{(8)}_{12}}+\frac{\mathtt{F}^{(8)}_{22}}{\mathtt{F}^{(8)}_{33}}\frac{\mathtt{d}^{(8)}_{33}}{\mathtt{d}^{(8)}_{12}}&\mathtt{m}^{(8)}_{22}-\frac{\mathtt{F}^{(8)}_{22}}{\mathtt{F}^{(8)}_{33}}\mathtt{m}^{(8)}_{33}&\mathtt{n}^{(8)}_{22}-\frac{\mathtt{F}^{(8)}_{22}}{\mathtt{F}^{(8)}_{33}}\mathtt{n}^{(8)}_{33}\\
0&0&0&0&0&0&-\frac{\mathtt{F}^{(8)}_{12}}{\mathtt{F}^{(8)}_{33}}&1+\frac{\mathtt{F}^{(8)}_{12}}{\mathtt{F}^{(8)}_{33}}\frac{\mathtt{d}^{(8)}_{33}}{\mathtt{d}^{(8)}_{12}}&-\frac{\mathtt{C}^{(8)}_{212}}{\mathtt{C}^{(8)}_{213}}-\frac{\mathtt{F}^{(8)}_{12}}{\mathtt{F}^{(8)}_{33}}\mathtt{m}^{(8)}_{33}&\mathtt{k}^{(8)}_{12}-\frac{\mathtt{F}^{(8)}_{12}}{\mathtt{F}^{(8)}_{33}}\mathtt{n}^{(8)}_{33}\\
0&0&0&0&0&0&-\frac{\mathtt{F}^{(8)}_{13}}{\mathtt{F}^{(8)}_{33}}&\mathtt{x}^{(8)}_{13}+\frac{\mathtt{F}^{(8)}_{13}}{\mathtt{F}^{(8)}_{33}}\frac{\mathtt{d}^{(8)}_{33}}{\mathtt{d}^{(8)}_{12}}&\mathtt{y}^{(8)}_{13}-\frac{\mathtt{F}^{(8)}_{13}}{\mathtt{F}^{(8)}_{33}}\mathtt{m}^{(8)}_{33}&\mathtt{z}^{(8)}_{13}-\frac{\mathtt{F}^{(8)}_{13}}{\mathtt{F}^{(8)}_{33}}\mathtt{n}^{(8)}_{33}\\
0&0&0&0&0&0&-\frac{\mathtt{F}^{(8)}_{23}}{\mathtt{F}^{(8)}_{33}}&\mathtt{x}^{(8)}_{23}+\frac{\mathtt{F}^{(8)}_{23}}{\mathtt{F}^{(8)}_{33}}\frac{\mathtt{d}^{(8)}_{33}}{\mathtt{d}^{(8)}_{12}}&\mathtt{y}^{(8)}_{23}-\frac{\mathtt{F}^{(8)}_{23}}{\mathtt{F}^{(8)}_{33}}\mathtt{m}^{(8)}_{33}&\mathtt{z}^{(8)}_{23}-\frac{\mathtt{F}^{(8)}_{23}}{\mathtt{F}^{(8)}_{33}}\mathtt{n}^{(8)}_{33}\\
0&0&0&0&0&0&1&-\frac{\mathtt{d}^{(8)}_{33}}{\mathtt{d}^{(8)}_{12}}&\mathtt{m}^{(8)}_{33}&\mathtt{n}^{(8)}_{33}
\end{pmatrix},\\
\bar{D}^{(11)}=&\begin{pmatrix}
0&0&0&0&0&0&0&0&0&\mathtt{F}^{(8)}_{33}\\
0&0&0&0&0&0&0&0&\mathtt{A}^{(8)}_{123}&0\\
0&0&0&0&0&0&0&-\frac{\mathtt{C}^{(8)}_{213}}{\mathtt{A}^{(8)}_{213}}&0&0\\
0&0&0&0&0&0&\frac{\mathtt{d}^{(8)}_{12}}{\mathtt{C}^{(8)}_{213}}&0&0&0\\
0&0&0&0&0&0&0&0&0&0\\
0&0&0&0&0&0&0&0&0&0\\
0&0&0&-\frac{\mathtt{d}^{(8)}_{12}}{\mathtt{C}^{(8)}_{213}}&0&0&0&0&0&0\\
0&0&\frac{\mathtt{C}^{(8)}_{213}}{\mathtt{A}^{(8)}_{213}}&0&0&0&0&0&0&0\\
0&-\mathtt{A}^{(8)}_{123}&0&0&0&0&0&0&0&0\\
-\mathtt{F}^{(8)}_{33}&0&0&0&0&0&0&0&0&0
\end{pmatrix},
\end{align}
where $\mathtt{F}^{(8)}_{ij}$ are defined as follows,
\begin{align}
\mathtt{F}^{(8)}_{11}=& \mathtt{A}^{(8)}_{11}-\frac{\mathtt{d}^{(8)}_{11}}{\mathtt{d}^{(8)}_{12}}\mathtt{A}^{(8)}_{12}+\mathtt{m}^{(8)}_{11}\mathtt{A}^{(8)}_{13}+\mathtt{n}^{(8)}_{11}\mathtt{A}^{(8)}_{23},\\
\mathtt{F}^{(8)}_{22}=& \mathtt{A}^{(8)}_{22}-\frac{\mathtt{d}^{(8)}_{22}}{\mathtt{d}^{(8)}_{12}}\mathtt{A}^{(8)}_{12}+\mathtt{m}^{(8)}_{22}\mathtt{A}^{(8)}_{13}+\mathtt{n}^{(8)}_{22}\mathtt{A}^{(8)}_{23},\\
\mathtt{F}^{(8)}_{33}=& \mathtt{A}^{(8)}_{33}-\frac{\mathtt{d}^{(8)}_{33}}{\mathtt{d}^{(8)}_{12}}\mathtt{A}^{(8)}_{12}+\mathtt{m}^{(8)}_{33}\mathtt{A}^{(8)}_{13}+\mathtt{n}^{(8)}_{33}\mathtt{A}^{(8)}_{23},\\
\mathtt{F}^{(8)}_{12}=& \mathtt{A}^{(8)}_{12}-\frac{\mathtt{C}^{(8)}_{212}}{\mathtt{C}^{(8)}_{213}}\mathtt{A}^{(8)}_{13}+\mathtt{k}^{(8)}_{12}\mathtt{A}^{(8)}_{23},\\
\mathtt{F}^{(8)}_{13}=& \mathtt{x}^{(8)}_{13}\mathtt{A}^{(8)}_{12}+\mathtt{y}^{(8)}_{13}\mathtt{A}^{(8)}_{13}+\mathtt{z}^{(8)}_{13}\mathtt{A}^{(8)}_{23},\\
\mathtt{F}^{(8)}_{23}=& \mathtt{x}^{(8)}_{23}\mathtt{A}^{(8)}_{12}+\mathtt{y}^{(8)}_{23}\mathtt{A}^{(8)}_{13}+\mathtt{z}^{(8)}_{23}\mathtt{A}^{(8)}_{23}.
\end{align}
The consistency condition of primary constraints in matrix form,
\begin{equation}
 \begin{pmatrix}
\{\bar{\phi}_N, \mathcal{H}_c\}\\
\{\bar{\phi}_1, \mathcal{H}_c\}\\
\{\bar{\phi}_2, \mathcal{H}_c\}\\
\{\bar{\phi}_3, \mathcal{H}_c\}\\
\{\bar{\phi}_{11}, \mathcal{H}_c\}\\
\{\bar{\phi}_{22}, \mathcal{H}_c\}\\
\{\bar{\phi}_{33}, \mathcal{H}_c\}\\
\{\bar{\phi}_{12}, \mathcal{H}_c\}\\
\{\bar{\phi}_{13}, \mathcal{H}_c\}\\
\{\bar{\phi}_{23}, \mathcal{H}_c\}
\end{pmatrix}
+ D^{(11)}\begin{pmatrix}
\lambda^N\\
\lambda^1\\
\lambda^2\\
\lambda^3\\
\lambda^{11}\\
\lambda^{22}\\
\lambda^{33}\\
\lambda^{12}\\
\lambda^{13}\\
\lambda^{23}
\end{pmatrix}
\approx  \begin{pmatrix}
0\\
0\\
0\\
0\\
0\\
0\\
0\\
0\\
0\\
0
\end{pmatrix},
\end{equation}
which can be solved by applying the construction of $\bar{D}^{(11)}$ and yields the following solutions of lagrange multipliers and secondary constriants,
\begin{align}
\lambda^1 =&-\frac{1}{\mathtt{A}^{(8)}_{123}}\Big(\frac{\mathtt{F}^{(8)}_{23}}{\mathtt{F}^{(8)}_{33}}\{\bar{\phi}_{33}, \mathcal{H}_c\}-(\mathtt{x}^{(8)}_{23}+\frac{\mathtt{F}^{(8)}_{23}}{\mathtt{F}^{(8)}_{33}}\frac{\mathtt{d}^{(8)}_{33}}{\mathtt{d}^{(8)}_{12}})\{\bar{\phi}_{12}, \mathcal{H}_c\}-(\mathtt{y}^{(8)}_{23}-\frac{\mathtt{F}^{(8)}_{23}}{\mathtt{F}^{(8)}_{33}}\mathtt{m}^{(8)}_{33})\{\bar{\phi}_{13}, \mathcal{H}_c\}\notag \\&-(\mathtt{z}^{(8)}_{23}-\frac{\mathtt{F}^{(8)}_{23}}{\mathtt{F}^{(8)}_{33}}\mathtt{n}^{(8)}_{33})\{\bar{\phi}_{23}, \mathcal{H}_c\} \Big),\\
\lambda^2 =&\frac{\mathtt{A}^{(8)}_{123}}{\mathtt{C}^{(8)}_{213}}\Big(\frac{\mathtt{F}^{(8)}_{13}}{\mathtt{F}^{(8)}_{33}}\{\bar{\phi}_{33}, \mathcal{H}_c\}-(\mathtt{x}^{(8)}_{13}+\frac{\mathtt{F}^{(8)}_{13}}{\mathtt{F}^{(8)}_{33}}\frac{\mathtt{d}^{(8)}_{33}}{\mathtt{d}^{(8)}_{12}})\{\bar{\phi}_{12}, \mathcal{H}_c\}-(\mathtt{y}^{(8)}_{13}-\frac{\mathtt{F}^{(8)}_{13}}{\mathtt{F}^{(8)}_{33}}\mathtt{m}^{(8)}_{33})\{\bar{\phi}_{13}, \mathcal{H}_c\}\notag \\&-(\mathtt{z}^{(8)}_{13}-\frac{\mathtt{F}^{(8)}_{13}}{\mathtt{F}^{(8)}_{33}}\mathtt{n}^{(8)}_{33})\{\bar{\phi}_{23}, \mathcal{H}_c\} \Big),\\
\lambda^3 =&-\frac{\mathtt{C}^{(8)}_{213}}{\mathtt{d}^{(8)}_{12}}\Big(\frac{\mathtt{F}^{(8)}_{12}}{\mathtt{F}^{(8)}_{33}}\{\bar{\phi}_{33}, \mathcal{H}_c\}-(1+\frac{\mathtt{F}^{(8)}_{12}}{\mathtt{F}^{(8)}_{33}}\frac{\mathtt{d}^{(8)}_{33}}{\mathtt{d}^{(8)}_{12}})\{\bar{\phi}_{12}, \mathcal{H}_c\}+(\frac{\mathtt{C}^{(8)}_{212}}{\mathtt{C}^{(8)}_{213}}+\frac{\mathtt{F}^{(8)}_{12}}{\mathtt{F}^{(8)}_{33}}\mathtt{m}^{(8)}_{33})\{\bar{\phi}_{13}, \mathcal{H}_c\}\notag \\&-(\mathtt{k}^{(8)}_{12}-\frac{\mathtt{F}^{(8)}_{12}}{\mathtt{F}^{(8)}_{33}}\mathtt{n}^{(8)}_{33})\{\bar{\phi}_{23}, \mathcal{H}_c\} \Big),\\
\lambda^N =&-\frac{1}{\mathtt{F}^{(8)}_{33}}\Big(-\{\bar{\phi}_{33}, \mathcal{H}_c\}+\frac{\mathtt{d}^{(8)}_{33}}{\mathtt{d}^{(8)}_{12}}\{\bar{\phi}_{12}, \mathcal{H}_c\}-\mathtt{m}^{(8)}_{33}\{\bar{\phi}_{13}, \mathcal{H}_c\}-\mathtt{n}^{(8)}_{33}\{\bar{\phi}_{23}, \mathcal{H}_c\} \Big),\\
\bar{\sigma}_{11} =& -\{\bar{\phi}_{11}, \mathcal{H}_c\}+\frac{\mathtt{F}^{(8)}_{11}}{\mathtt{F}^{(8)}_{33}}\{\bar{\phi}_{33}, \mathcal{H}_c\}+(\frac{\mathtt{d}^{(8)}_{11}}{\mathtt{d}^{(8)}_{12}}-\frac{\mathtt{F}^{(8)}_{11}}{\mathtt{F}^{(8)}_{33}}\frac{\mathtt{d}^{(8)}_{33}}{\mathtt{d}^{(8)}_{12}})\{\bar{\phi}_{12}, \mathcal{H}_c\}-(\mathtt{m}^{(8)}_{11}-\frac{\mathtt{F}^{(8)}_{11}}{\mathtt{F}^{(8)}_{33}}\mathtt{m}^{(8)}_{33})\{\bar{\phi}_{13}, \mathcal{H}_c\}\notag \\&-(\mathtt{n}^{(8)}_{11}-\frac{\mathtt{F}^{(8)}_{11}}{\mathtt{F}^{(8)}_{33}}\mathtt{n}^{(8)}_{33})\{\bar{\phi}_{23}, \mathcal{H}_c\} \approx 0,\\
\bar{\sigma}_{22} =& -\{\bar{\phi}_{22}, \mathcal{H}_c\}+\frac{\mathtt{F}^{(8)}_{22}}{\mathtt{F}^{(8)}_{33}}\{\bar{\phi}_{33}, \mathcal{H}_c\}+(\frac{\mathtt{d}^{(8)}_{22}}{\mathtt{d}^{(8)}_{12}}-\frac{\mathtt{F}^{(8)}_{22}}{\mathtt{F}^{(8)}_{33}}\frac{\mathtt{d}^{(8)}_{33}}{\mathtt{d}^{(8)}_{12}})\{\bar{\phi}_{12}, \mathcal{H}_c\}-(\mathtt{m}^{(8)}_{22}-\frac{\mathtt{F}^{(8)}_{22}}{\mathtt{F}^{(8)}_{33}}\mathtt{m}^{(8)}_{33})\{\bar{\phi}_{13}, \mathcal{H}_c\}\notag \\&-(\mathtt{n}^{(8)}_{22}-\frac{\mathtt{F}^{(8)}_{22}}{\mathtt{F}^{(8)}_{33}}\mathtt{n}^{(8)}_{33})\{\bar{\phi}_{23}, \mathcal{H}_c\} \approx 0,
\end{align}
the lagrange multipliers $\lambda^{33}, \lambda^{12}, \lambda^{13}$ and $\lambda^{23}$ also can be solved uniquely. Let $\mathcal{H}^{(2)}_c=\mathcal{H}_c+\lambda^N \bar{\phi}_N+\lambda^i\bar{\phi}_i+\lambda^{33}\bar{\phi}_{33}+\lambda^{12}\bar{\phi}_{12}+\lambda^{13}\bar{\phi}_{13}+\lambda^{23}\bar{\phi}_{23}$ and $\{\bar{\sigma}_{ij},\bar{\phi}_{mn}\}=\mathtt{B}^{(8)}_{ijmn}\delta^{(3)}(\vec{x}-\vec{y}) $. The dirac matrix $D^{(12)}$ are defined as follows,
\begin{equation}
D^{(12)}=\begin{pmatrix}
\mathtt{B}^{(8)}_{1111}&\mathtt{B}^{(8)}_{1122}\\
\mathtt{B}^{(8)}_{2211}&\mathtt{B}^{(8)}_{2222}
\end{pmatrix}.
\end{equation}
The consistency conditions of secondary constraints,
\begin{equation}
 \begin{pmatrix}
\{\bar{\sigma}_{11}, \mathcal{H}^{(2)}_c\}\\
\{\bar{\sigma}_{22}, \mathcal{H}^{(2)}_c\}
\end{pmatrix}
+ \begin{pmatrix}
\mathtt{B}^{(8)}_{1111}&\mathtt{B}^{(8)}_{1122}\\
\mathtt{B}^{(8)}_{2211}&\mathtt{B}^{(8)}_{2222}
\end{pmatrix}\begin{pmatrix}
\lambda^{11}\\
\lambda^{22}
\end{pmatrix}
\approx \begin{pmatrix}
0\\
0
\end{pmatrix},
\end{equation}
have the following solutions of lagrange multipliers,

\begin{align}
\lambda^{22} =& \frac{1}{|\triangle^{(12)}_1|}(-\{\bar{\sigma}_{22}, \mathcal{H}^{(2)}_c\}-\lambda^{33}\mathtt{B}^{(6)}_{2233}),\\
\lambda^{33} =& \frac{|\triangle^{(12)}_1|}{|\triangle^{(12)}_2|}(\frac{\mathtt{B}^{(6)}_{3322}}{|\triangle^{(12)}_1|}\{\bar{\sigma}_{22}, \mathcal{H}^{(2)}_c\}-\{\bar{\sigma}_{33}, \mathcal{H}^{(2)}_c\}).
\end{align}
Since each of the lagrange multipliers can be solved uniquely, the Dirac-Bergmann algorithm terminated. All constriants in Type $8$ Newer General Relativity are second-class, with the total number of physical degrees of freedom $10-\frac{1}{2}(12)=4$.

\subsection{Type $9$}
For Type $9$ Newer General Relativity, each of the conjugate momenta have invertible relation with corresponding velocities term. There is no primary constraint in this gravity theory, it is eligible for the complete Legendre transformation, and therefore it has $10$ physical degrees of freedom.
\begin{align}
\pi_N =&\frac{(-4c_4-2c_5)}{N^2}\partial_a N^a+\frac{(4c_3+2c_5)}{N^2}(\dot{\gamma}^{ab} \gamma_{ab}-\gamma_{bc} N^a \partial_a \gamma^{bc})\notag \\ &+\frac{(-8c_1-8c_2-8c_3-8c_4-8c_5)}{N^3}(\dot{N}-N^a \partial_a N), \\
\pi_i =& \frac{(4c_1+2c_2+2c_4)}{N^3}\left(\dot{N}^a \gamma_{ia}-\gamma_{ib} N^a \partial_a N^b \right)-\frac{(4c_2+2c_5)}{N^2}\partial_i N-\frac{c_4}{N}\gamma^{ab}\partial_b \gamma_{ia}-\frac{c_5}{N}\gamma^{ab}\partial_i \gamma_{ab}, \\
\pi_{ij} =& \frac{(4c_3+2c_5)}{N^2}(\dot{N} \gamma_{ij}-\gamma_{ij} N^a \partial_a N)+\frac{2c_3}{N}( -\dot{\gamma}^{ab} \gamma_{ab} \gamma_{ij} + \gamma_{bc} \gamma_{ij} N^a \partial_a \gamma^{bc}) \notag \\ &+\frac{c_5}{N}\gamma_{ij}\partial_a N^a +\frac{c_2}{N} (\gamma_{ja}\partial_i N^a + \gamma_{ia} \partial_j N^a)+\frac{2c_1}{N}(N^a \partial_a \gamma_{ij}-\dot{\gamma}^{ab} \gamma_{ia} \gamma_{jb}).
\end{align}
The solutions of the velocities are listed as following,
\begin{align}
\dot{N}=& -\frac{N^3}{8(c_1+c_2+c_3+c_4+c_5)}\Big(\pi_N-\frac{8(c_1+c_2+c_3+c_4+c_5)}{N^3}N^a\partial_aN\notag \\&-\frac{2(2c_3+c_5)}{N^2}(\dot{\gamma}^{ab}\gamma_{ab}+\gamma^{bc}N^a\partial_a\gamma_{bc})+\frac{2(2c_4+c_5)}{N^2}\partial_aN^a \Big),\\
\dot{N}^k=&\frac{N^3\gamma^{ik}}{2(2c_1+c_2+c_4)}\Big(\pi_i+\frac{2(2c_1+c_2+c_4)}{N^3}\gamma_{ib}N^a\partial_aN^b+\frac{c_4}{N}\gamma^{ab}\partial_b\gamma_{ia}+\frac{c_5}{N}\gamma^{ab}\partial_i\gamma_{ab}\notag \\&+\frac{2(2c_2+c_5)}{N^2}\partial_iN \Big),\\
\dot{\gamma}^{kl}=&-\frac{N\gamma^{ik}\gamma^{jl}}{2c_1}\Big(\pi_{ij}+\frac{2c_3}{N}\gamma^{bc}\gamma_{ij}N^a\partial_a\gamma_{bc}+\frac{2c_1}{N}N^a\partial_a\gamma_{ij}-\frac{2(2c_3+c_5)}{N^2}(\dot{N}\gamma_{ij}-\gamma_{ij}N^a\partial_aN)\notag \\&-\frac{c_5}{N}\gamma_{ij}\partial_aN^a-\frac{c_2}{N}(\gamma_{ja}\partial_iN^a+\gamma_{ia}\partial_j N^a)-\frac{c_3}{3c_3+c_1}\gamma_{ij}\gamma^{ab}\Big[ \pi_{ab}\notag \\&+\frac{2c_3}{N}\gamma^{mn}\gamma_{ab}N^c\partial_c\gamma_{mn}+\frac{2c_1}{N}N^m\partial_m \gamma_{ab}-\frac{2(2c_3+c_5)}{N^2}(\dot{N}\gamma_{ab}-\gamma_{ab}N^m\partial_mN)\notag\\&-\frac{c_5}{N}\gamma_{ab}\partial_mN^m-\frac{c_2}{N}(\gamma_{bm}\partial_aN^m+\gamma_{am}\partial_bN^m)\Big] \Big).
\end{align}

\section{Conclusions}\label{sec:conclusion}
The Hamiltonian analysis developed in this work investigates the constraint structure of the different types of Newer General Relativity and determines their physical degrees of freedom based on the numbers of first-class and second-class constraints associated with the corresponding classes of quadratic nonmetricity theories.\\

The study begins with a review of symmetric teleparallel gravity and the Dirac–Bergmann algorithm in the context of field theories. We adopt the same assumption used by Tomonari in the Hamiltonian analysis of Coincident $f(Q)$ gravity, namely that spatial boundary terms can be neglected under suitable boundary conditions and therefore do not modify the symplectic structure \cite{dirac_dof_fQ}. We then review the classification of Newer General Relativity by Fabio D'Ambrosio and Lavinia Heisenberg, in which the theory is divided into nine types according to different assumptions on the coefficients of the quadratic nonmetricity terms \cite{classifyNewerGR}.\\

For each type, the corresponding assumptions on the coefficients are imposed, and the conjugate momenta are calculated using the xAct package for Mathematica. The resulting primary constraints are then identified from the degeneracy of the Legendre transformation. The consistency conditions of the primary constraints are subsequently investigated for each type. Different techniques are employed either to determine the unique values of the corresponding Lagrange multipliers or to identify secondary constraints when the consistency conditions cannot be solved solely for the multipliers. The complete set of constraints is obtained for each type, and the corresponding numbers of first-class and second-class constraints are used to determine the physical degrees of freedom. The results are summarized in the following table.
 \begin{table}[h]
\centering
\caption{Physical Degrees of Freedom of each Newer General Relativity models}
\label{tab:NGR_dof}
\begin{tabular}{ccc}
\toprule
\textbf{Types} & \textbf{Total number of second-class constraints} & \textbf{pDoF} \\
\midrule
1  & 2 & 9\\
2  & 6& 7 \\
3  & 10  & 5 \\
4 (Case $1$)  & 12 & 4 \\
4 (Case $2$) & 12 & 4 \\
5  & 6& 7 \\
6  & 10& 5 \\
7  & 12& 4\\
8  &12& 4\\
9  & 0 & 10 \\
\bottomrule
\end{tabular}
\end{table}
\\
\newpage
For further research, a complete Hamiltonian analysis of Newer General Relativity without neglecting the spatial boundary terms would provide a direct test of the assumptions adopted in the present work. If the assumption that these boundary terms do not affect the symplectic structure is valid, the resulting constraint structure and physical degrees of freedom should agree with those obtained in the present analysis. If discrepancies arise, it would then be important to investigate how the spatial boundary terms modify the constraint structure and whether the Dirac–Bergmann analysis needs to be extended or reformulated to consistently incorporate their contributions. Another potential direction is to investigate Newer General Relativity without imposing the coincident gauge. Relaxing this gauge condition would allow the constraint structure to be examined in a more general setting and could provide further insight into the role of the affine connection in the Hamiltonian formulation.\\

\bibliographystyle{apsrev4-2}
\bibliography{MyRef}      

\end{document}